\pdfoutput=1

\documentclass[10pt]{siamltex}
\usepackage[]{amsmath,amssymb,epsfig}
\usepackage[top=18mm, bottom=18mm, left=20mm, right=20mm]{geometry}

\usepackage{subfigure}
\usepackage{amsmath}
\usepackage{amssymb}
\usepackage{graphicx}
\usepackage{comment}
\usepackage{array}
\usepackage{algorithm}
\usepackage{algorithmic}
\usepackage{url}

\usepackage{setspace}
\usepackage{multicol}
\usepackage{multirow}

\usepackage{color}

\usepackage{colortbl}
\usepackage{xcolor}

\usepackage{hyperref}

\hypersetup{
    colorlinks=true,
    citecolor=red,
    linkcolor=blue,
    urlcolor=blue
  }

\newcommand{\mb}[1]{\mbox{\boldmath$#1$}}
\newcounter{ale}

\newenvironment{liste}{\begin{itemize}}{\end{itemize}}
\newcommand{\aliste}{\begin{liste} \setcounter{ale}{1}}
\newcommand{\zliste}{\end{liste}}

\begin{document}



\title{Synchronization over $Z_2$ and community detection in multiplex signed networks with constraints}
\author{Mihai~Cucuringu \footnotemark[1] \ \footnotemark[2]}



\maketitle

\renewcommand{\thefootnote}{\fnsymbol{footnote}}
\footnotetext[1]{Department of Mathematics, UCLA, 520 Portola Plaza, Mathematical Sciences Building 6363, Los Angeles, CA 90095-1555, email: mihai@math.ucla.edu}
\footnotetext[2]{Program in Applied and Computational Mathematics (PACM), Princeton University, Fine Hall, Washington Road, Princeton, NJ, 08544-1000 USA. Part of this work was undertaken while the author was a Ph.D. student supported by PACM.}
\renewcommand{\thefootnote}{\arabic{footnote}}


\begin{abstract}
Finding group elements from noisy measurements of their pairwise ratios is also known as the group synchronization problem, first introduced in the context of the group SO(2) of planar rotations.
The usefulness of synchronization over the group $\mathbb{Z}_2$ has been demonstrated in recent algorithms for localization of sensor networks and three-dimensional structuring of molecules.
In this paper, we focus on synchronization over $\mathbb{Z}_2$, and consider the problem of identifying communities in a multiplex network when the interaction between the nodes is described by a signed (and possibly weighted) measure of similarity, and when the multiplex network has a natural partition into two communities, of possibly different sizes.
In the setting where one has the additional information that certain subsets of nodes represent the same (unknown) group element, we consider and compare several algorithms for synchronization over $\mathbb{Z}_2$, based on spectral and semidefinite programming relaxations (SDP), and message passing algorithms.   In other words, all nodes within such a subset represent the same unknown group element,
and one has available noisy pairwise measurements between pairs of nodes that belong to different non-overlapping subsets.
Following a recent analysis of the eigenvector method for synchronization over SO(2), we analyze the robustness to noise of the eigenvector method for synchronization over $\mathbb{Z}_2$, when the underlying graph of pairwise measurements is the Erd\H{o}s-R\'{e}nyi random graph, using results from the random matrix theory literature on the largest eigenvalue of rank-1 deformation of large random matrices.
We also propose a message passing synchronization algorithm, inspired by the standard belief propagation algorithm,  
 that outperforms the existing eigenvector synchronization algorithm only for certain classes of graphs and noise models, and enjoys the flexibility of incorporating additional 
constraints that may not be easily accommodated by any of the other spectral or SDP-based methods.
We apply the synchronization methods both to several synthetic models and a real data set of roll call voting patterns in the U.S. Congress across time, to identify the two existing communities, i.e., the Democratic and Republican parties. Finally, we discuss a number of related open problems and future research directions. 
\end{abstract}



\section{Introduction}	\label{intro}

During the last decade, the emerging area of network science has witnessed an explosive growth, with virtually thousands of papers written on the topic \cite{booknewman,booknewmanpton}. Much  of this work has focused on the detection of a mesoscale structure known as \textit{community structure}, where subgroups called \textit{communities} are composed of nodes densely connected with each other, while the connection between nodes across different communities is relatively sparse. There is already a vast literature on community detection \cite{masonams,fortunato,mason6,mason7}, which includes methods  that allow for the detection of overlapping communities \cite{mason8,mason9,mason10}. The above results have found numerous applications in areas including committee and voting networks in political science \cite{mason11, multiplex_Mucha}, friendship networks \cite{Traud2011,mason14}, protein-protein interaction networks \cite{mason15}, financial networks \cite{masonExchangeRates,pottsHeimo}, mobile phone networks 
\cite{louvainMethod,localizationBelgium}, transportation networks \cite{Barthelemysubway}, and brain networks from the neuroscience community \cite{masonbrain}.  In terms of the underlying (possibly weighted) graph associated to the network, the mathematical task is that of identifying clusters of highly interconnected nodes, or  subgraphs whose internal edge density is large compared to the rest of the graph.

In this paper, we consider the related problem of identifying communities in a graph, for the particular case when the interaction between the nodes is described by a signed measure of similarity or correlation, and when the network has a natural separation into two  (not necessarily equally sized)  communities. The goal is to  recover the two subgroups of nodes whose internal pairwise similarity or correlation is significantly stronger when compared to the rest of the network. Note that in the ground truth solution, each node belongs to exactly one of the two communities, and the task is to recover these two communities given a noisy set of pairwise signed interactions. 
Signed networks have also been recently considered in the context of identifying  community structure in a voting network in the United States General Assembly \cite{masonUnited}. 
In a dynamical setting, community detection algorithms have been used to analyze multiple time series data, such as rates in the foreign exchange market \cite{masonExchangeRates}. In \cite{pottsHeimo}, the authors introduced a multiresolution module detection approach for dense weighted networks, and successfully applied it to stock price correlations data. A random matrix theory based technique has been recently proposed for the particular task of clustering correlation data \cite{comDecCorrMtx}, and was shown to be able to capture well-known structural properties of the financial stock market.
        

In this work, the graphs we consider have a special structure, in the form of a \textit{multiplex network}, in the sense that each graph can be decomposed into a sequence of subgraphs, each of which corresponds to a \textit{layer} of the network, and there exist interconnections linking nodes across different layers.
We refer the reader to \cite{Mathematical_Multilayer} for a mathematical formulation of multilayer networks, of which multiplex networks are a subset. Unlike a multilayer network, a multiplex network only allows for a single type of inter-layer connections via which any given node is connected only to its counterpart nodes in the other layers. The rich variety of interlayer connections in a multilayer network can be mathematically captured by the \textit{multilayer adjacency tensor} introduced in \cite{Mathematical_Multilayer}.   
The very recent survey \cite{reviewMultiplex} discusses the rich history of multilayer networks, and provides a thorough review of the growing literature on this topic. 
The structure of such multiplex networks motivates our work, in the sense that the interconnections often correspond to certain structural constraints, which on one hand make the problem harder to solve, but on the other hand, if properly exploited, may enhance the quality and robustness of the community detection process.


The approach we consider in this paper is motivated by the \textit{group synchronization} problem and the recently introduced spectral and semidefinite programming based algorithms \cite{sync}. The remainder of this section describes the group synchronization problem, and its many recent applications.  As a word of caution to the reader, we remark that throughout this paper, the group synchronization problem is not be confused with the classical synchronization phenomena in large ensembles of coupled oscillators. To this end, we refer the reader to \cite{ArenasSync} for an  extensive review of numerical and analytical techniques underlying the interplay between  synchronization phenomena and the topology of complex networks.

Finding group elements from noisy measurements of their ratios is also known as the group synchronization problem \cite{timeoneSync,timetwoSync}.  
In this paper, we focus exclusively on synchronization over the group $\mathbb{Z}_2$, which can be formulated as follows.  Consider an undirected graph $G=(V,E)$, with the node set $V$ of size $|V|=n$ corresponding to a set of $n$ unknown elements $z_1,\ldots,z_n \in \mathbb{Z}_2$,  and the edge set $E$ of size $m \leq {n\choose 2} $ (though often $m \ll {n\choose 2} $) corresponding to an incomplete set of $m$ (possibly noisy) pairwise measurements $z_i z_j^{-1}$ available to the user.
   For mathematical convenience, we choose to work with the following representation of the group $\mathbb{Z}_2$, with elements $\{-1, +1\}$ and the usual multiplication operation, for which it holds true that $z_i z_j^{-1} = z_i z_j$ since an element is its own inverse.
Note that the above defined group is isomorphic to the group of integers modulo $2$, often written as $\mathbb{Z}/ 2 \mathbb{Z}$.
One may interpret the unavailable group element $z_i \in \mathbb{Z}_2, i=1,\ldots,n$ as the true (unknown) sign or polarity of node $i$ in $G$. Throughout the paper, we use the notation $z = (z_1, \ldots, z_n)$ to represent the unavailable vector of size $n$ with entries $ z_i \in \mathbb{Z}_2$, a notation useful in the context of the optimization problems from subsequent sections.

We denote by $Z$ the symmetric matrix of size $n \times n$, whose entries $Z_{ij}=Z_{ji}$ represent the pairwise measurement available for each edge $(i,j)$ in $G$, which encodes the similarity measure between the pair of nodes $i$ and $j$. In the noise free case, the measurement $Z_{ij}$ precisely equals the product of its endpoints $Z_{ij} = z_i z_j$; however, in most scenarios, many of the edge measurements are corrupted by noise with some given probability, in which case $Z_{ij} = -z_i z_j$. Given the pairwise measurement matrix $Z$, the goal is to recover the unknown elements $z_1,\ldots,z_n \in \mathbb{Z}_2$, in other words, to recover the polarity of each node of the network, such that we satisfy as many constraints $ Z_{ij} = z_i z_j$ as possible. 
It is also often the case that the information available on the edges does not necessarily take only $\pm1$ values, but rather takes on continuous values in the range $[-1,1]$ (as is the case with correlation data). In this setup, one can interpret the magnitude $w_{ij} \in [0,1]$ associated to each entry $Z_{ij}$ as a measure of confidence on the similarity estimation. For the purpose of this paper, we will mostly be focusing on the case when all the weights are equal, $w_{ij}=1, \forall (i,j) \in E$.

The synchronization of clocks in a distributed network from noisy measurements of their time offsets is another example of synchronization, where the underlying group is the real line  $\mathbb{R}$.
The eigenvector and semidefinite programming methods for solving an instance of the synchronization problem were originally introduced by Singer in \cite{sync} in the context of angular synchronization, 
where one is asked to estimate $n$ unknown angles $\theta_1,\ldots,\theta_n \in [0,2\pi)$ given $m$ noisy measurements $\delta_{ij}$ of their offsets $\theta_i - \theta_j \mod 2\pi$.  The difficulty of the problem is amplified on one hand by the amount of noise in the offset measurements, and on the other hand by the fact that $m \ll {n \choose 2}$, i.e., only a very small subset of all possible pairwise offsets are measured. In general, one may consider other groups $\mathcal{G}$ (such as SO($d$), O($d$)) for which there are available noisy measurements $g_{ij}$ of ratios
between the group elements
\begin{equation}
g_{ij} = g_i g_j^{-1}, g_i, g_j \in \mathcal{G}.
\end{equation}
We remind the reader that O($d$) denotes the  group of $d \times d$ orthogonal matrices, and its subgroup SO($d$) 
(the special orthogonal group) denotes the group of $d \times d$ orthogonal matrices of determinant 1. 
The set $E$ of pairs $(i,j)$ for which a noisy measurement of ratio of group elements is available
can be realized as the edge set of a graph $G=(V,E)$, $|V|=n, |E|=m$,
with vertices corresponding to the group elements $g_1,\ldots,g_n$ and edges
corresponding to the available pairwise measurements $ g_{ij} = g_i g_j^{-1}$.
As long as the group $\mathcal{G}$ is compact and has a real or complex
representation, one may construct a real or Hermitian matrix (which may also be a
matrix of matrices) where the element in the position $(i,j)$ is the matrix
representation of the measurement $g_{ij}$ (possibly a matrix of size $1 \times
1$, as it is the case for $\mathbb{Z}_2$), or the zero matrix if there is no direct measurement for the ratio of $g_i$
and $g_j$. For example, the rotation group SO(3) has a real representation using
$3 \times 3$ rotation matrices, and the group SO(2) of planar rotations has a complex
representation as points on the unit circle
$e^{\imath \theta_i} = \cos \theta_i + \imath \sin \theta_i$.
Given the above matrix of pairwise group measurements, the approach initiated  in \cite{sync} proposed to compute the top eigenvectors(s) of the pairwise measurement matrix and use them to estimate the unknown group elements. Alternatively, one may cast the synchronization problem as a semidefinite programming (SDP) problem \cite{Vandenberghe94sdp}, and extract the
unknown group elements from a low-rank approximation of the noisy incomplete matrix of pairwise group measurements.

We have successfully used the above synchronization methods (first introduced by Singer in \cite{sync}) to solve the graph realization problem in the context of sensor network localization \cite{asap2d}, and the \textit{molecule problem} in structural biology \cite{asap3d}. 
Another recent application of synchronization is to the \textit{Structure from Motion} problem \cite{AmitVisionSync}, a fundamental task in computer vision where one is asked to recover three-dimensional structure from a collection of images. In general, the synchronization problem can be applied in such settings where the underlying problem exhibits a group structure, and one has readily available (possibly noisy) pairwise  measurements of ratios of the group elements.

In this paper, we focus exclusively on the group synchronization problem over the group $\mathbb{Z}_2$, denoted from now on as SYNC($\mathbb{Z}_2$). We apply the eigenvector synchronization method to a multiplex network, representing voting patterns for the U.S.  Congresses during the years 1927 to 2009, with the goal of robustly detecting the two political parties across time. We motivate the robustness to noise of the eigenvector synchronization method when the underlying graph of pairwise measurements is the Erd\H{o}s-R\'{e}nyi random graph, using recent results form the random matrix theory community, following a similar perturbation analysis introduced in \cite{sync} for the case of the group $SO(2)$. We also consider a message passing formulation of the synchronization problem, and compare its performance with the other spectral and SDP-based methods. Finally, we consider a variant of SYNC($\mathbb{Z}_2$) when additional constraints are available on the polarity of certain subgroups of nodes, and compare the performance of our proposed algorithms for this particular instance of SYNC($\mathbb{Z}_2$). We remark that the SDP formulation, though computationally expensive to solve for large networks, provides the most robust solution for the case when additional constraints are available.

The structure of this paper is as follows: 
Section \ref{sec:sync} describes the synchronization problem over $\mathbb{Z}_2$ and existing methods for solving it. 
Section \ref{sec:syncCongress} is an application of synchronization to a U.S. Congress roll call voting multiplex network.  
Section \ref{sec:rmtx} is a noise sensitivity analysis of the eigenvector method using tools from random matrix theory (detailed in Appendix A).
In Section \ref{sec:MPD}, we propose a message passing algorithm for solving the synchronization problem, and compare its performance with that of the spectral and SDP relaxations under different graph and noise models, also when anchor information is available (we review in Appendix B the quadratically constrained quadratic formulations for incorporating anchor information, considered in \cite{asap3d}).  
In Section \ref{sec:sync_partition}, we consider the synchronization problem with partition constraints, propose several algorithms for computing an approximate solution, and detail the numerical results when applying our methods to two synthetically generated networks and the U.S. Congress voting data set. 
Finally, Section \ref{sec:summary} is a summary and a discussion of possible future research directions. 


\section{Synchronization over $\mathbb{Z}_2$} \label{sec:sync}

In this section, we describe several different methods for solving SYNC($\mathbb{Z}_2$), introduced in \cite{asap2d,asap3d} as one of the building blocks of the ASAP localization algorithm.
Mathematically, the problem of synchronization over the group $\mathbb{Z}_2$, whose elements we will denote by $\{\pm 1 \}$, can be stated as follows. Given a graph $G=(V,E)$, with the node set $V$ of size $|V|=n$ corresponding to a set of $n$ group elements $z_1,\ldots,z_n \in
\mathbb{Z}_2$, and the edge set $E$ of size $m \ll {n\choose 2} $ corresponding to an incomplete
set of $m$ (possibly noisy) pairwise group measurements of $z_i z_j^{-1}$, for
$(i,j) \in E$, the goal is to provide accurate estimates
$\hat{z}_1,\ldots,\hat{z}_n \in \mathbb{Z}_2$ for the unknown group elements
$z_1,\ldots,z_n$. Recall that in $\mathbb{Z}_2$ an element is its own inverse $z_i^{-1}=z_i$, thus from now on we shall write the group elements ratio $z_i z_j^{-1}$ as $z_i z_j$. 
We denote by $Z=(Z_{ij})_{1 \leq i,j \leq n}$ the symmetric matrix of the available pairwise group measurements. We remark here that one may associate a non-negative weight $w_{ij}$ to each measurement $Z_{ij}$ that reflects the precision of the pairwise measurement, if such information is available to the user, in which case the matrix $Z$ used throughout the paper would be replaced by $Z^{(W)}$, with $Z^{(W)}_{ij} = Z_{ij} W_{ij}$. Note that if all measurements have equal precision, then one can set all weights equal $w_{ij}=1$, for $(i,j) \in E$. For simplicity of the exposition we assume that all measurements have weight equal to 1, and choose to work with matrix $Z$ instead of $Z^{(W)}$. We refer the reader to \cite{asap2d,asap2d,AmitVisionSync} for other applications of the group synchronization problem (over groups other than $\mathbb{Z}_2$) that discuss and benefit from the incorporation of weights into the pairwise measurements, if such additional information is available.


\begin{figure}[h]
\begin{center}
\includegraphics[width=0.4\columnwidth]{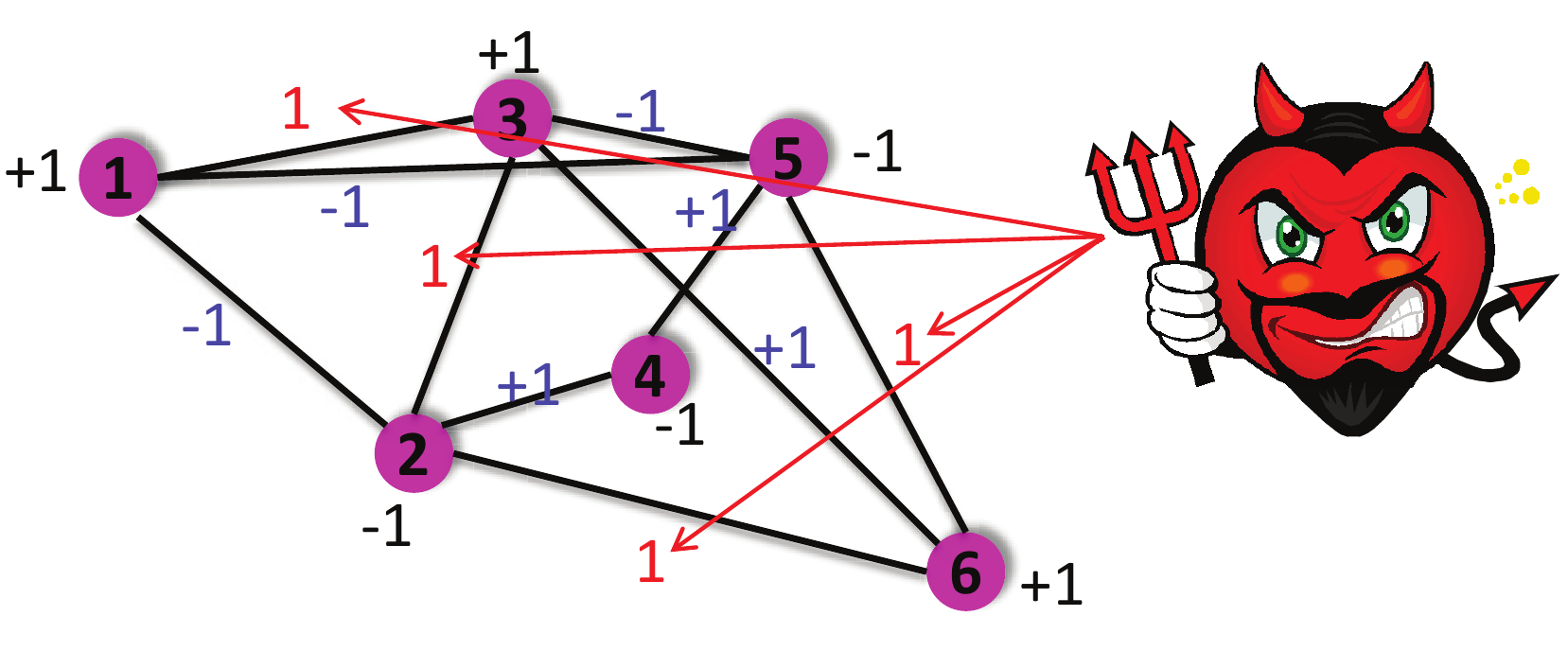}
\end{center}
\caption{An instance of the noisy synchronization problem. The goal is to recover the $\pm 1 $ values at each node of the graph, based on a sparse subset of pairwise measurements, i.e., using only the information on the edges. The value of each edge $Z_{ij}$ is a (perhaps noisy) measurement of the ratio of its endpoints $z_i z_j^{-1} = z_i z_j$. The blue, respectively red, edges denote correct, respectively incorrect, pairwise measurements.}
\label{fig:examplesyncgraph}
\end{figure}

If the data is noise free, i.e., the pairwise group measurements perfectly satisfy $ Z_{ij} = z_i z_j$, for all $(i,j) \in E$, one can simply recover the $n$
group elements by choosing a spanning tree of the graph $G$, fixing the root
(for example $z_1 = 1$), and propagating the information across the spanning tree to compute the values at the remaining nodes. However, in the noisy case, this approach is no longer robust as a single noisy edge induces errors for the remaining nodes of the spanning tree, and thus the problem becomes significantly harder.


In the context of the sensor network localization problem \cite{asap2d}, the measurements
$Z_{ij} \in \mathbb{Z}_2$ denote the relative reflection of pairs of
overlapping subgraph embeddings, and the task is to recover the global orientation of each such embedding. In the context of the U.S. Congress roll call voting network that we are about the investigate, where Democrats and Republicans have conflicting interests and their voting patterns are overall opposite, the measurement $Z_{ij} \in \mathbb{Z}_2$ reflects the similarity of the voting pattern of two senators, and represents a noisy proxy for whether they belong to the same party or no.


If the pairwise measurements are noiseless, the maximum  of the following quadratic form
\begin{equation}
\max_{x_i \in \mathbb{Z}_2} \sum_{i,j=1}^{n} x_i {Z}_{ij} x_j = \max_{x \in
\mathbb{Z}_2^n} x^{T} Z x
\label{maxZ2}
\end{equation}
is attained when $ \boldsymbol{x} = \boldsymbol{z}$ (where $\boldsymbol{z}$ denotes the vector of ground truth elements with components $z_1, \ldots,z_n$), which yields $ \boldsymbol{z}^{T} Z \boldsymbol{z} = 2m$, where $m$ is the number of edges in the graph. This holds true since each edge which correctly estimates
the similarity of two senators contributes with 
        $$ z_i {Z}_{ij} z_j = + 1 $$
to the sum in Equation (\ref{maxZ2}).  However, when the pairwise group measurements contain errors the problem becomes significantly more challenging. In the noisy case, we write $Z_{ij}=z_i z_j \delta$, where $\delta$ is a Bernoulli random variable
taking values $-1$, respectively $+1$, with probability $\eta$, respectively $p=1-\eta$. Thus, we denote by $\eta$ the likelihood of each edge being corrupted with noise, i.e., having its sign flipped. Note that if the underlying graph $G$ is a complete graph and the measurements are noiseless, i.e., $G=K_n$ and $\eta=0$, then $Z = \boldsymbol{z} \boldsymbol{z}^T$ is a rank-one matrix. However, in most practical applications, $G$ is a sparse graph and $\eta>0$, and one can interpret the matrix $Z$ available to the user as a sparsified and perturbed version of the rank-one matrix $\boldsymbol{z} \boldsymbol{z}^T$.

Since NP-hard problems, such as the maximum clique problem \cite{Bomze99themaximum,karp}, can be formulated as a quadratically constrained quadratic program (QCQP) such as the one above, the problem in (\ref{maxZ2}) is itself NP-hard.
In the spirit of the approach introduced in \cite{sync} for the group SO(2), in \cite{asap2d} we considered the following relaxation for the group $\mathbb{Z}_2$,
\begin{equation}
\max_{x_1, \ldots, x_n \in \mathbb{Z}_2 ; \sum_{i=1}^{n}|x_i|^2=n}  \sum_{i,j=1}^{n}
x_i {Z}_{ij} x_j = \max_{ \parallel \boldsymbol{x} \parallel^2 = n} \boldsymbol{x}^T Z \boldsymbol{x},
\label{relmaxZ2}
\end{equation}
whose maximum is achieved when $ \boldsymbol{x} = \boldsymbol{v_1}$, where $\boldsymbol{v}_1$ is the normalized top
eigenvector of $Z$, satisfying $Z \boldsymbol{v_1} = \lambda_1 \boldsymbol{v_1}$ and $\parallel \boldsymbol{v_1} 
\parallel^2=n$, with $\lambda_1$ being the largest eigenvalue. Thus, an
approximate solution to the maximization problem in (\ref{maxZ2}) is given by the
top eigenvector of the symmetric  matrix $Z$, as illustrated in equation
(\ref{solstep1}).
Prior to computing the top eigenvector of the matrix $Z$, as initially proposed in
\cite{sync}, we choose to normalize $Z$ as follows. Let $D$ be an $n \times n$
diagonal matrix, whose entries are given by $D_{ii} = \sum_{j=1}^n |Z_{ij}|$. In
other words,
\begin{equation}
D_{ii} = deg(i),
\end{equation}
where $deg(i)$ is the degree of node $i$ in $G$. We define the matrix $\mathcal{Z}$ as
\begin{equation}
\label{Z-norm}
\mathcal{Z} = D^{-1}Z,
\end{equation}
and note that, although not necessarily symmetric, it is similar to the
symmetric matrix $D^{-1/2} Z D^{-1/2}$ through $$\mathcal{Z} = D^{-1/2}
(D^{-1/2} Z D^{-1/2}) D^{1/2}.$$ Therefore, the matrix $\mathcal{Z}$ has $n$
real eigenvalues $ \lambda_1^\mathcal{Z} > \lambda_2^\mathcal{Z} \geq \cdots
\geq  \lambda_n^\mathcal{Z}$ and $n$ orthonormal eigenvectors
$\boldsymbol{v_1^\mathcal{Z}},...,\boldsymbol{v_n^\mathcal{Z}}$, satisfying $\mathcal{Z} \boldsymbol{v_i^{\mathcal{Z}}}
= \lambda_i^{\mathcal{Z}}   \boldsymbol{v_i^{\mathcal{Z}}}$.
In the eigenvector method, we compute the top eigenvector $ \boldsymbol{v_1^\mathcal{Z}} \in
\mathbb{R}^n$ of $\mathcal{Z}$, which satisfies
\begin{equation}
\mathcal{Z} \boldsymbol{v_1^\mathcal{Z}} = \lambda_1^\mathcal{Z} \boldsymbol{v_1^\mathcal{Z}},
\end{equation}
and use it to obtain estimators $\hat{z}_1,\ldots,\hat{z}_n$ for the unknown group elements, in the
following way:
\begin{equation}
   \hat{z}_i = \operatorname{sign}(   \boldsymbol{v_1^\mathcal{Z}}(i)) =
\frac{  \boldsymbol{v_1^\mathcal{Z}}(i)}{|  \boldsymbol{v_1^\mathcal{Z}}(i)|}, \;\;\;\; i=1,2,\ldots,n.
\label{solstep1}
\end{equation}
The top eigenvector recovers the initial group elements up to a global sign,
since if $ \boldsymbol{v_1^\mathcal{Z}} $ is the top eigenvector of $\mathcal{Z}$ then so is
$-\boldsymbol{v_1^\mathcal{Z}}$. We refer the reader to Section 4 of \cite{asap2d} for a detailed explanation of the increased noise robustness the above normalization brings, and its connection to the normalized discrete graph Laplacian.

Note that one may use a different objective function as an alternative to (\ref{maxZ2}), that allows  to formulate the group  synchronization problem as a least squares problem \cite{asap3d}, by minimizing the following quadratic form
\begin{eqnarray}
\min_{ \boldsymbol{x} \in \mathbb{Z}_2^n }  \sum_{(i,j) \in E} (x_i - Z_{ij}x_j)^2  \nonumber
& 
= & \min_{ \boldsymbol{x}  \in \mathbb{Z}_2^n } \sum_{(i,j) \in E}   \left( x_i^2 + Z_{ij}^2 x_j^2  - 2 Z_{ij} x_i x_j \right) 
=   \min_{ \boldsymbol{x}  \in \mathbb{Z}_2^n }  \sum_{(i,j) \in E}  \left( x_i^2 + x_j^2  - 2 Z_{ij} x_i x_j \right)\nonumber \\
& 
= & \min_{ \boldsymbol{x}  \in \mathbb{Z}_2^n } \left(  \sum_{i=1}^n d_i x_i^2  -  \sum_{(i,j) \in E} 2 Z_{ij} x_i x_j \right) 
= \min_{ \boldsymbol{x}  \in \mathbb{Z}_2^n }  \left( \boldsymbol{x}^T D \boldsymbol{x} - \boldsymbol{x}^T Z \boldsymbol{x} \right) = \min_{ \boldsymbol{x}  \in \mathbb{Z}_2^n }  \boldsymbol{x}^T( D - Z) \boldsymbol{x}
\label{LS_DZ}
\end{eqnarray}
where $d_i$ denotes the degree of node $i$ in $G$. Since such a QCQP program is NP-hard to solve, one may replace the $n$ individual constraints for each of the variables $x_i = 1$ by a single and much weaker constraint which requires that the sum of squared magnitudes is $|| \boldsymbol{x} ||^2=n$. The solution to this relaxation of the above minimization problem, is given by the  eigenvector (normalized such that $|| \boldsymbol{x} ||^2=n$ ) corresponding to the smallest eigenvalue of the more popular (symmetric) combinatorial Laplacian $D-Z$.

As previously mentioned, an alternative approach for solving the synchronization problem was initially introduced in \cite{sync} for the group SO(2) of planar rotations, and relied  on casting the synchronization problem as a semidefinite programming (SDP) optimization problem. The objective function in (\ref{maxZ2}) can be written as
\begin{equation}
  \sum_{i,j=1}^{n} x_i {Z}_{ij} x_j = Trace(Z \Upsilon)
\end{equation}
where $\Upsilon$ is the $n \times n$ symmetric rank-one matrix with $\pm 1$ entries
denoting the correct pairwise measurements of ratios of group elements $\Upsilon_{ij}=x_ix_j^{-1}$. Note that $\Upsilon$ is a rank-one matrix, with ones on its diagonal $ \Upsilon_{ii} =1, \forall i=1,\ldots,n$. Enforcing the low-rank constraint renders the problem no longer convex, and we write the SDP relaxation of (\ref{maxZ2}) as
\begin{equation}
	\begin{aligned}
	& \underset{\Upsilon \in \mathbb{R}^{n \times n}}{\text{maximize}}
	& & Trace(Z \Upsilon) \\
	& \text{subject to}
	& & \Upsilon_{ii} =1, i=1,\ldots,n \\
		& & &   \Upsilon \succeq 0
	\end{aligned}
\label{SDP_max}
\end{equation}
where the maximization is taken over all semidefinite positive real-valued matrices  $\Upsilon \succeq 0$ \cite{Vandenberghe94sdp}. Note that the solution of the SDP is not necessarily of rank-one as desired. We compute the top eigenvector of $\Upsilon$, and estimate $\bar{z}_1,\ldots,\bar{z}_n$ based on the sign of its entries.

In Sections \ref{sec:syncCongress} and \ref{sec:sync_partition} we apply the eigenvector and SDP-based synchronization methods (and their variations described in later sections) to the analysis of a Congress data set of roll call voting patterns in the U.S. Senate across time, with the goal of identifying the two existing communities, i.e., the Democratic and Republican parties. For any two senators $i$ and $j$ that participate in the same Congress, $Z_{ij}=1$ if they cast similar votes, $Z_{ij}=-1$ if they cast opposite votes, and $Z_{ij} = 0$ if they belong to two different Congresses and no pairwise measurement is available.

\section{The synchronization of a voting network} \label{sec:syncCongress}

In this section, we explore the application of the synchronization method to the
Congress data set \cite{Voteview} of roll call voting patterns in the U.S. Senate
across time~\cite{PR97,WPFMP09_TR,multiplex_Mucha}. 
We considered Senates in the $70^{th}$ Congress through the $110^{th}$ Congress, covering the years
$1927$ to $2009$. During this time, United States went from $48$ to $50$ states, each of which delegates two senators (with a few exceptions) and thus the number of senators $S$ in each of the $C=41$ Congresses was roughly the same, i.e., $S=100$. The number of unique senators throughout the $C=41$ Congresses was $k=735$, and the total number of senators (counting repetitions) was $n=4196$. Note that, for simplicity, we assume that $n=C \cdot S$, although there are states that delegate more than two senators. We denote by $m_i$, $i=1,\ldots,k$, the number of Congresses on which each unique senator $i$ has served, during the interval $1927$ to $2009$, and note that $m_1+m_2 + \ldots + m_k=n$.

The available data set is in the following form. For each Congress, we have available a complete weighted matrix $W^{(t)}$, $t = 1,\ldots,C$, with $W^{(t)}_{ij} \in [0,1]$ denoting the fraction of bills on which two senators voted in the same manner. In other words, the element $W^{(t)}_{ij}$ is equal to the number of times that senators $i$ and $j$ voted in the same way, divided by the total number of bills for which both $i$ and $j$ cast a vote, during the $t^{th}$ Congress. Such networks are often encountered in the literature, and are referred to as ``similarity networks" since the weights on the edges represent a measure of similarity between adjacent nodes \cite{mason11,Zhang2008c,Traud2011, masonUnited, masonHouseRep}. 


Next, we integrate all the available information into one large block-diagonal square matrix $H$ of size $n=C \cdot S$, where the diagonal blocks are given by the (complete) weighted matrices $W^{(1)},\ldots,W^{(C)}$
\begin{equation}
H = \operatorname{Diagonal}(W^{(1)},\ldots,W^{(C)}).
\label{defH}
\end{equation}
Furthermore, we also want to take into account the fact that a senator may participate in several (not necessarily consecutive) Congresses. To this end, we introduce an inter-Congress connection between the entries corresponding to the same senator in Congresses $u$ and $v$. In other words, if the $i^{th}$ senator from Congress $u$ is the same as the $j^{th}$ senator from Congress $v$, then we set
\begin{equation}
\Omega_{(u-1)S + i,(v-1)S + j} = 1,
\label{defOmega}
\end{equation}
and set $\Omega_{(v-1)S + j, (u-1)S + i} = \Omega_{(u-1)S + i,(v-1)S + j}$ to preserve the symmetry. One may think of $\Omega$ as the adjacency matrix of a graph which has $k=735$ non-overlapping complete subgraphs (one for each unique senator), and each node contained in such a complete subgraph denotes a distinct Congress on which the corresponding senator has served. Finally, we build the $n \times n$ nonnegative matrix $W$ which couples together multiple adjacency matrices $W^{(1)},\ldots,W^{(C)}$ via an interslice coupling parameter $\epsilon$, and thus captures the similarity of the voting patterns across all $C$ Congresses, taking into account the information that senators may serve for multiple terms
\begin{equation}
W_{ij} = \left\{
     \begin{array}{rl}
    \in [0,1] & \;\; \text{ if $i$ and $j$ belong to the same Congress}  \\
   \epsilon         & \;\; \text{ if $i$ and $j$ denote the same senator in different Congresses}\\
    0  & \;\; \text{ otherwise.} \\
     \end{array}
   \right.
\label{entries_W}
\end{equation}
Note that we choose to use \textit{categorical} coupling and add a pairwise similarity link between nodes from different Congresses that correspond to the same senator, in order to add more information to the problem and increase the likelihood of having a consistent party affiliation for different occurrences of the same senator across time. For example, if a senator appears in $q$ Congresses, using an \textit{ordinal} coupling that connects only adjacent Congresses (as used in \cite{multiplex_Mucha}) yields only $q-1$ self-similarity measurements as opposed to ${q \choose 2}$ in the case of categorical coupling.

\begin{figure}[t]
\begin{center}
\subfigure[Illustration of the Congress network as a ``spy'' plot.]{\includegraphics[width=0.23\columnwidth]{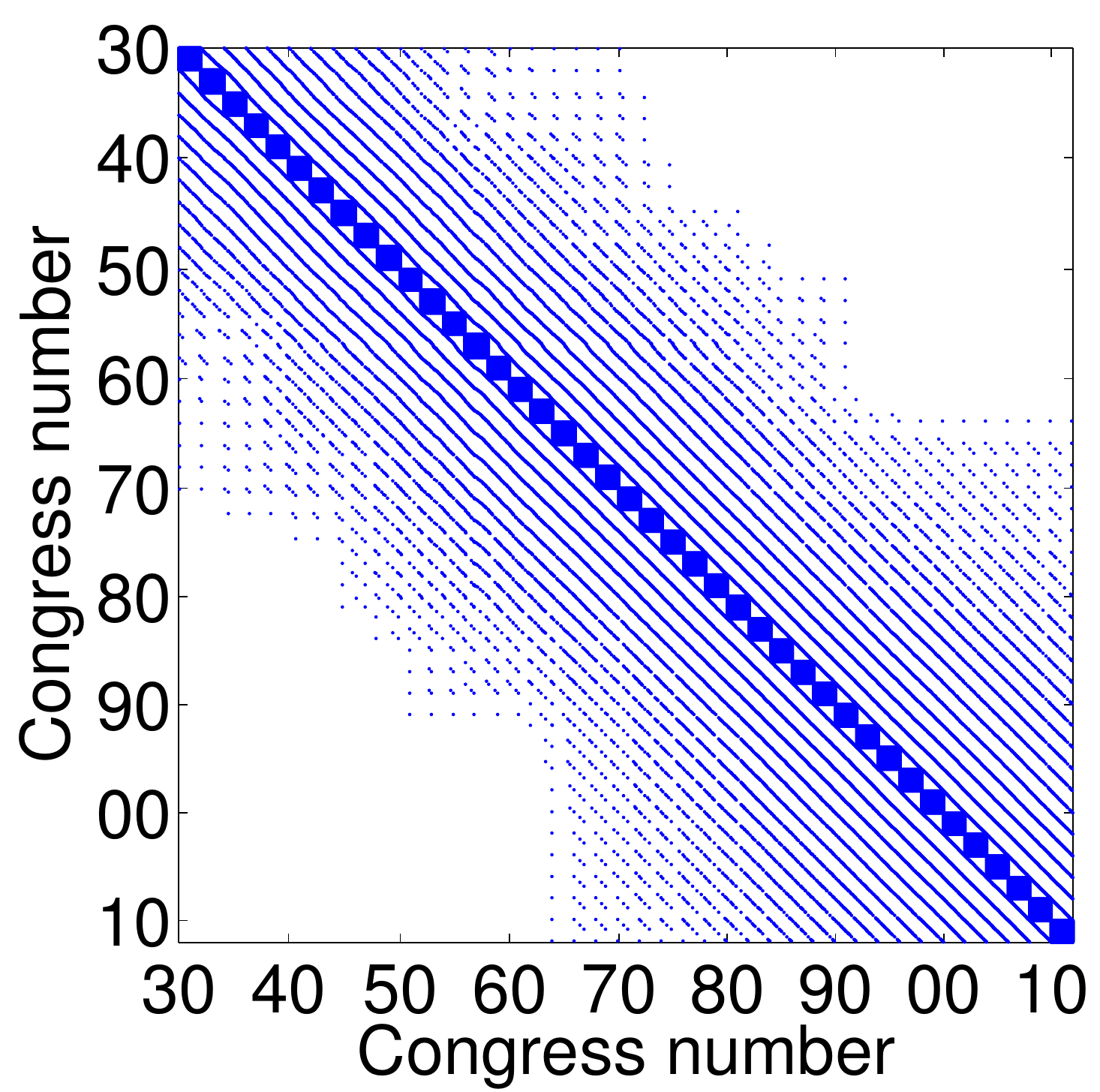}}
\subfigure[Normalized square spectrum $\epsilon=1$.]
{\includegraphics[width=0.24\columnwidth]{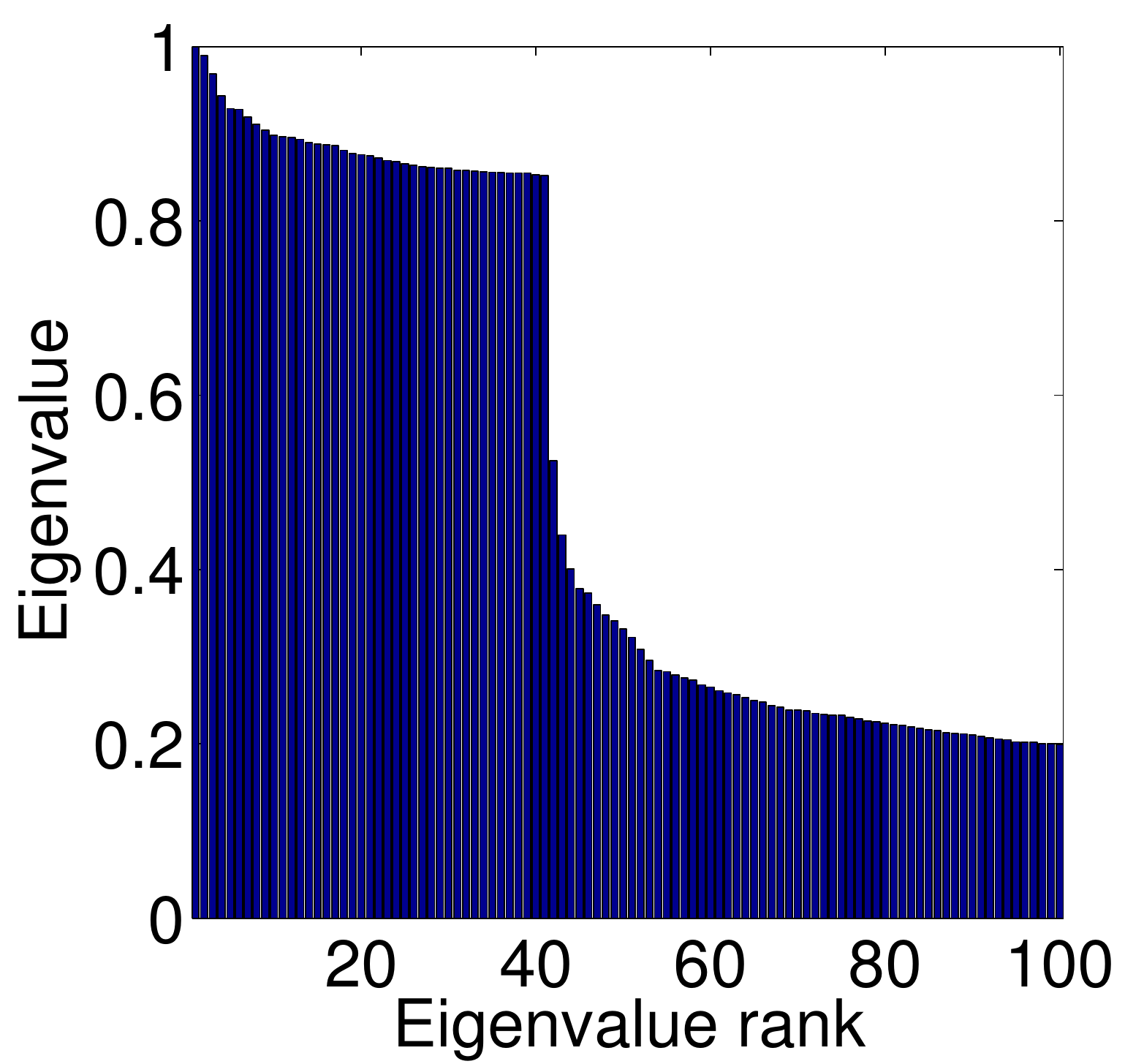}}
\subfigure[Normalized square spectrum $\epsilon=0.1$.]
{\includegraphics[width=0.24\columnwidth]{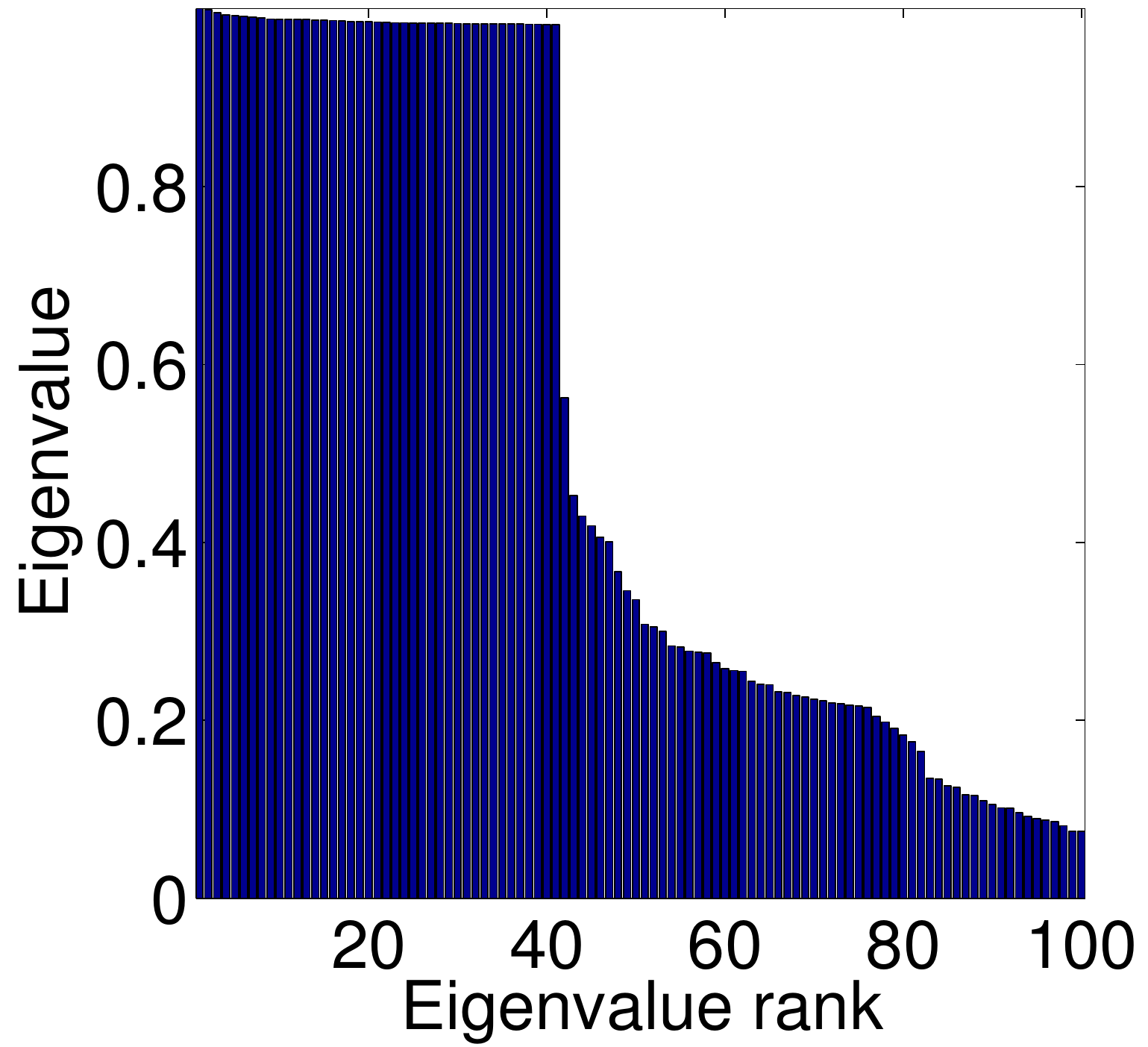}}
\subfigure[Partitioning based on $v^{(1)}_{41}$.]
{\includegraphics[width=0.25\columnwidth]{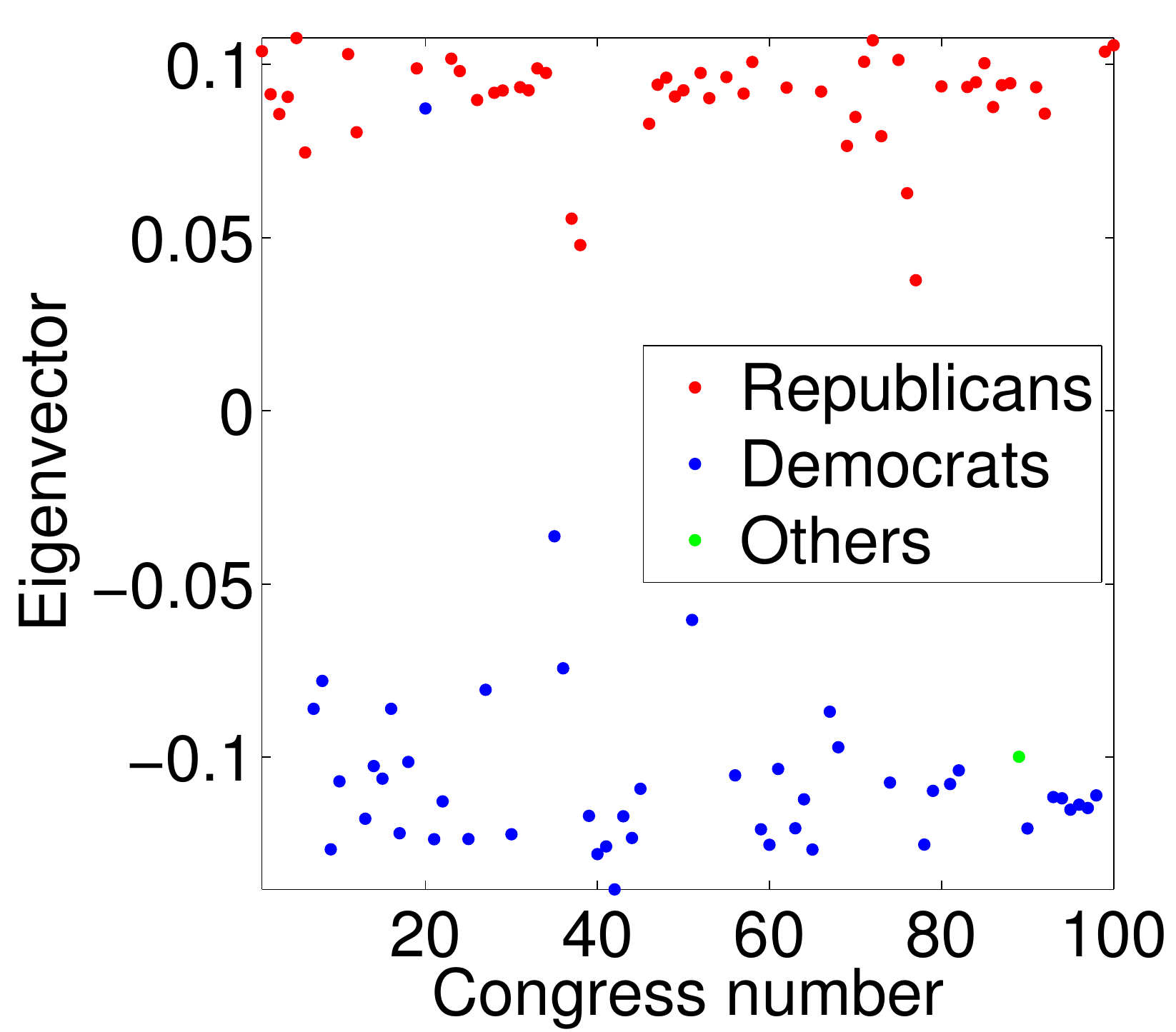}}  
\end{center}
\caption{
(a) A visualization of matrix W described  in equation (\ref{entries_W}).  The blocks on the diagonal correspond to the complete matrices of voting patterns in each of the $C+41$ Congresses, while the off-diagonal entries take the value $\epsilon$ denoting a single individual that served in two  different Congresses. (b) (respectively (c)) Barplot of the normalized square spectrum of the Congress matrix, i.e., $\frac{\lambda_i^2}{\sum_{j=1}^{n} \lambda_j^2}$, for $i=1,\ldots,100$, for $\epsilon=1$ (respectively, $\epsilon=0.1$) showing that the large eigenvalues account for a large fraction of the variance of the data.
(c) Plot of the first nontrivial eigenvector
$ v^{(1)}_{41} $ of the Laplacian associated to matrix $W^{(41)}$, which corresponds to senators from the $110^{th}$ Congress, showing a good separation between Republicans and Democrats.
}
\label{fig:cong-details}
\end{figure}

Throughout our computations, we keep constant the interslice coupling $\epsilon=1$ that controls the strength of the connection between nodes across different Congresses. We refer the reader to Mucha et al. \cite{multiplex_Mucha} for a thorough discussion on the role of the interslice coupling, and various structural results obtained by varying this parameter. A high-level view of the network defined by matrix $W$ is as a set of inter-coupled layered subnetworks, and hence the name of a multiplex network. We refer the reader to the extensive review of multilayer networks (of which multiplex networks are a subset) surveyed recently in \cite{reviewMultiplex}. While each subnetwork (i.e., Congress in our case) may have its own particular features, the ensemble of subnetworks together with their interactions across time define a richer structure that can provide further information. Note that, in light of Equations (\ref{defH}), (\ref{defOmega}), and (\ref{entries_W}), one may decompose the matrix $W$ as
\begin{equation}
 W = H + \epsilon \Omega,
 \label{W_decomp}
\end{equation}
a structure which may also be observed in Figure \ref{fig:cong-details}(a). Furthermore, the band around the main diagonal and the lack of entries in the top-right and bottom-left corners are due to the fact that usually senators serve on consecutive Congresses (although there are many exceptions), and it is less likely for a senator to serve on Congresses which are very far apart in time, mostly due to natural causes. The above decomposition (\ref{W_decomp}) of $W$ highlights the inter-layer and intra-layer connections present within the network, and motivates the terminology of \textit{supra-adjacency matrices} used recently in the literature to denotes matrices with such particular structure  \cite{reviewMultiplex}. 
Note that a different decomposition of matrix $W$ is given by
\begin{equation}
 	W =  \epsilon B +  \Theta,
\label{W_decomp2}
\end{equation}
for the same value of the $\epsilon$ parameter as in \ref{W_decomp}, where $B=\operatorname{Diagonal}(\mb{1}_{m_1},\ldots,\mb{1}_{m_k})$ is a block-diagonal matrix whose blocks are all ones matrices of size $m_i$ denoting the multiplicity of senator $i$, and $\Theta_{ij}$ denotes the similarity vote between two distinct senators for a given Congress. In reconciling the two decompositions (\ref{W_decomp}) and (\ref{W_decomp2}), we note that $\Omega$ and $B$ are both unweighted graphs (with 0/1  entries), where $ \Omega$ corresponds to the positions of the nonzero entries in the diagonal blocks of the adjacency matrix show in Figure \ref{fig:cong-details} (a), while the $B$ matrix corresponds to the positions of the nonzero entries outside of the main diagonal blocks of the same Figure.

\begin{figure}[t]
\begin{center}
\subfigure[$\epsilon = 0.1$]{
\includegraphics[width=0.19\columnwidth]{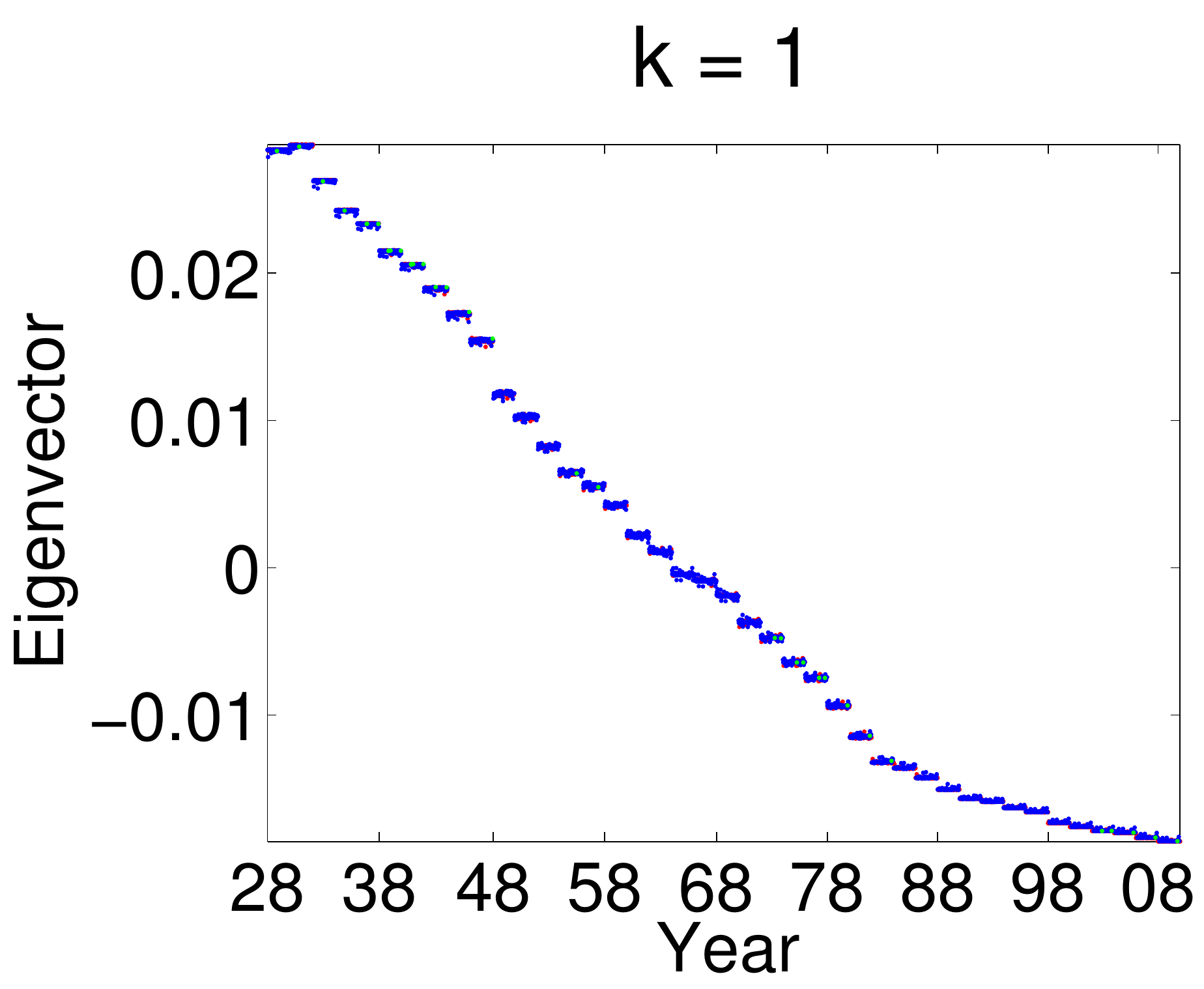}
\includegraphics[width=0.19\columnwidth]{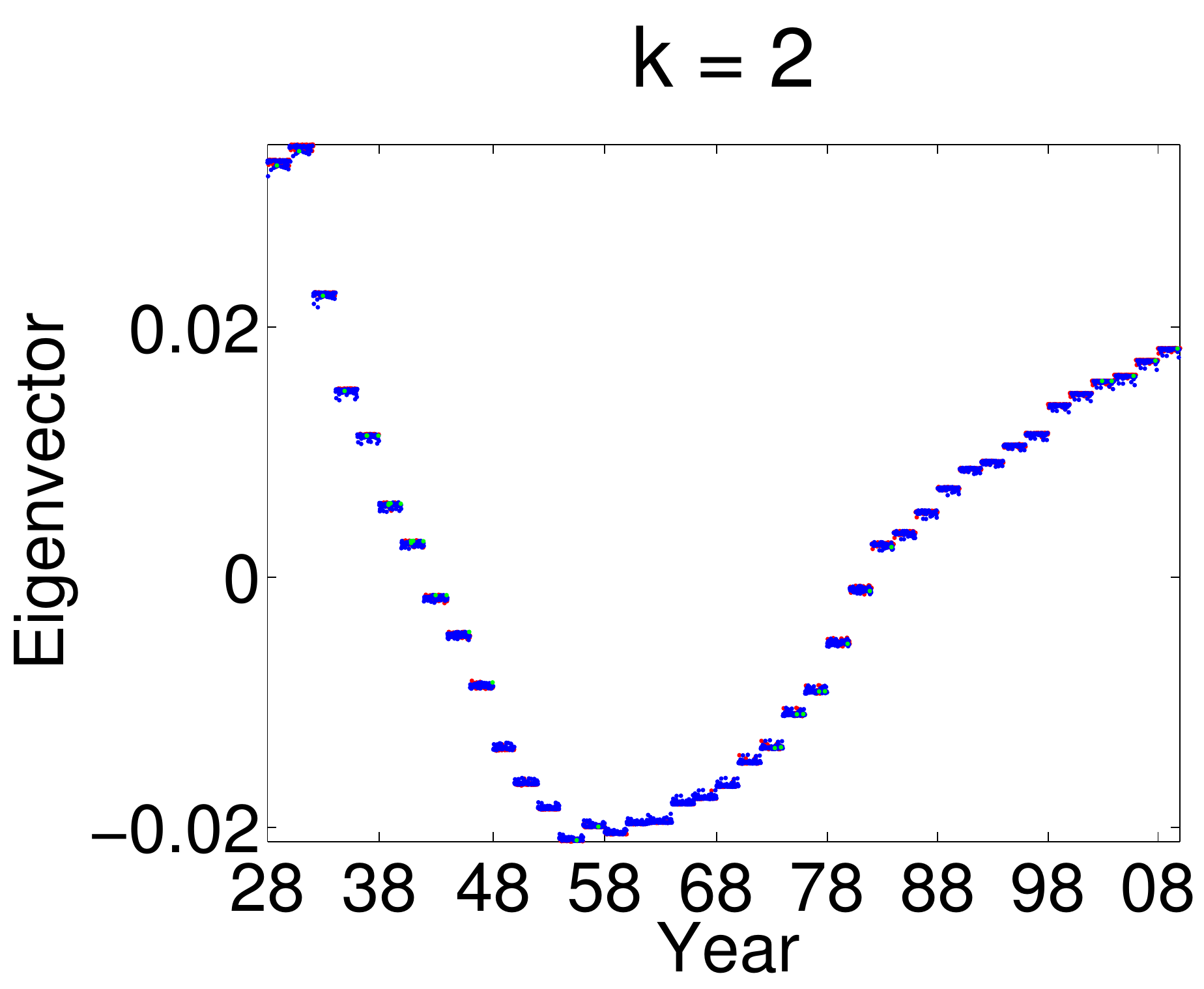}
\includegraphics[width=0.19\columnwidth]{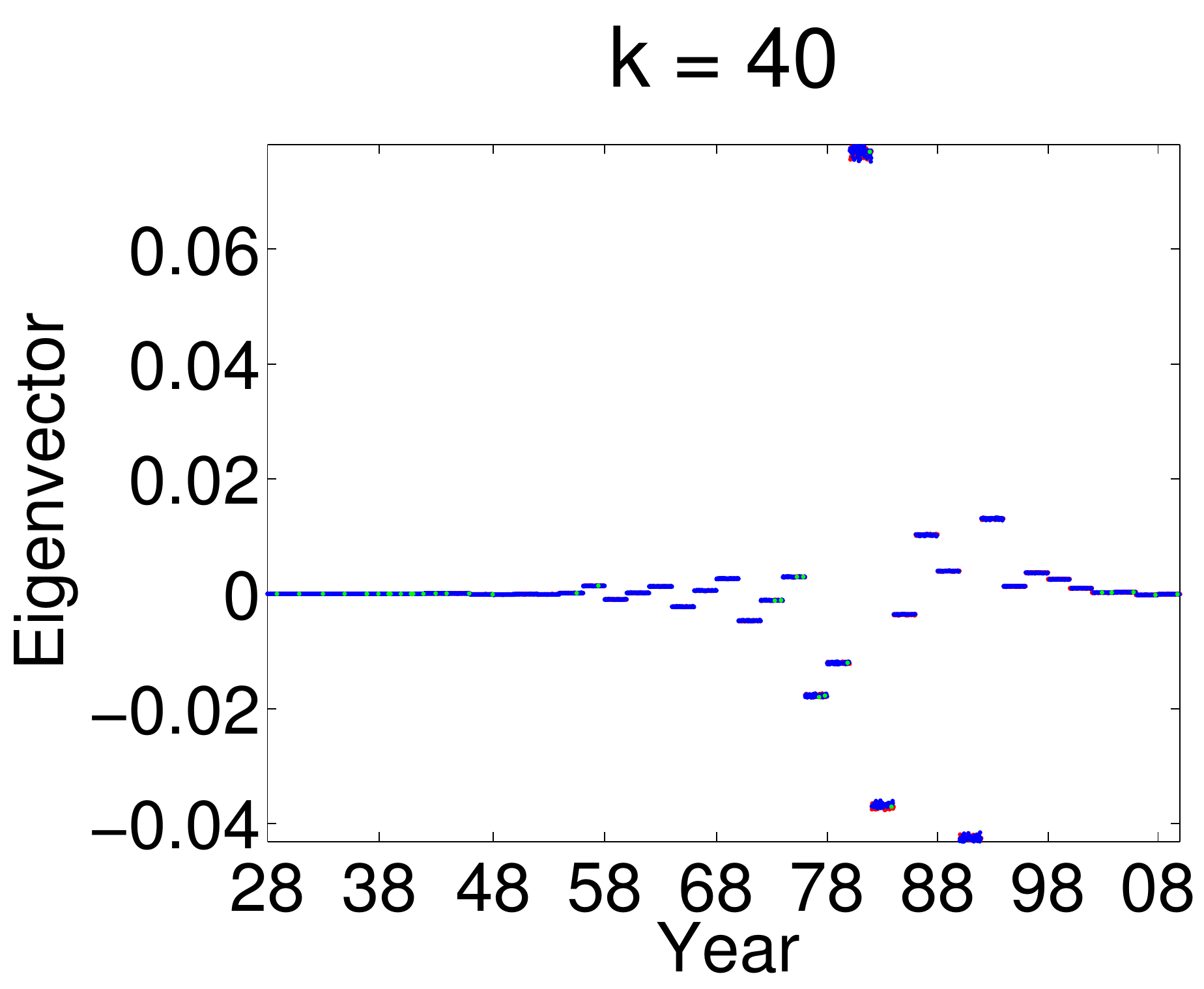}
\includegraphics[width=0.19\columnwidth]{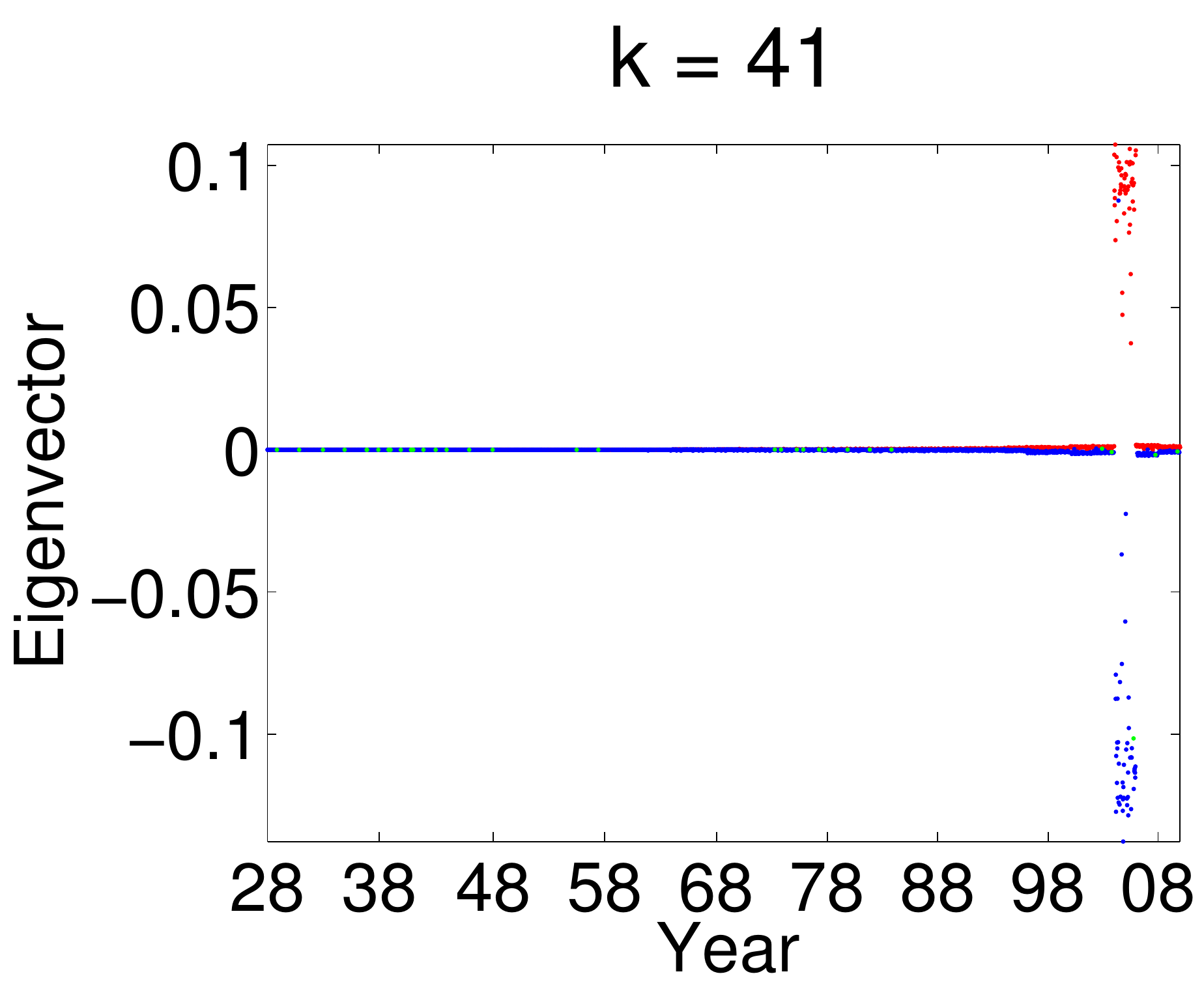}
\includegraphics[width=0.19\columnwidth]{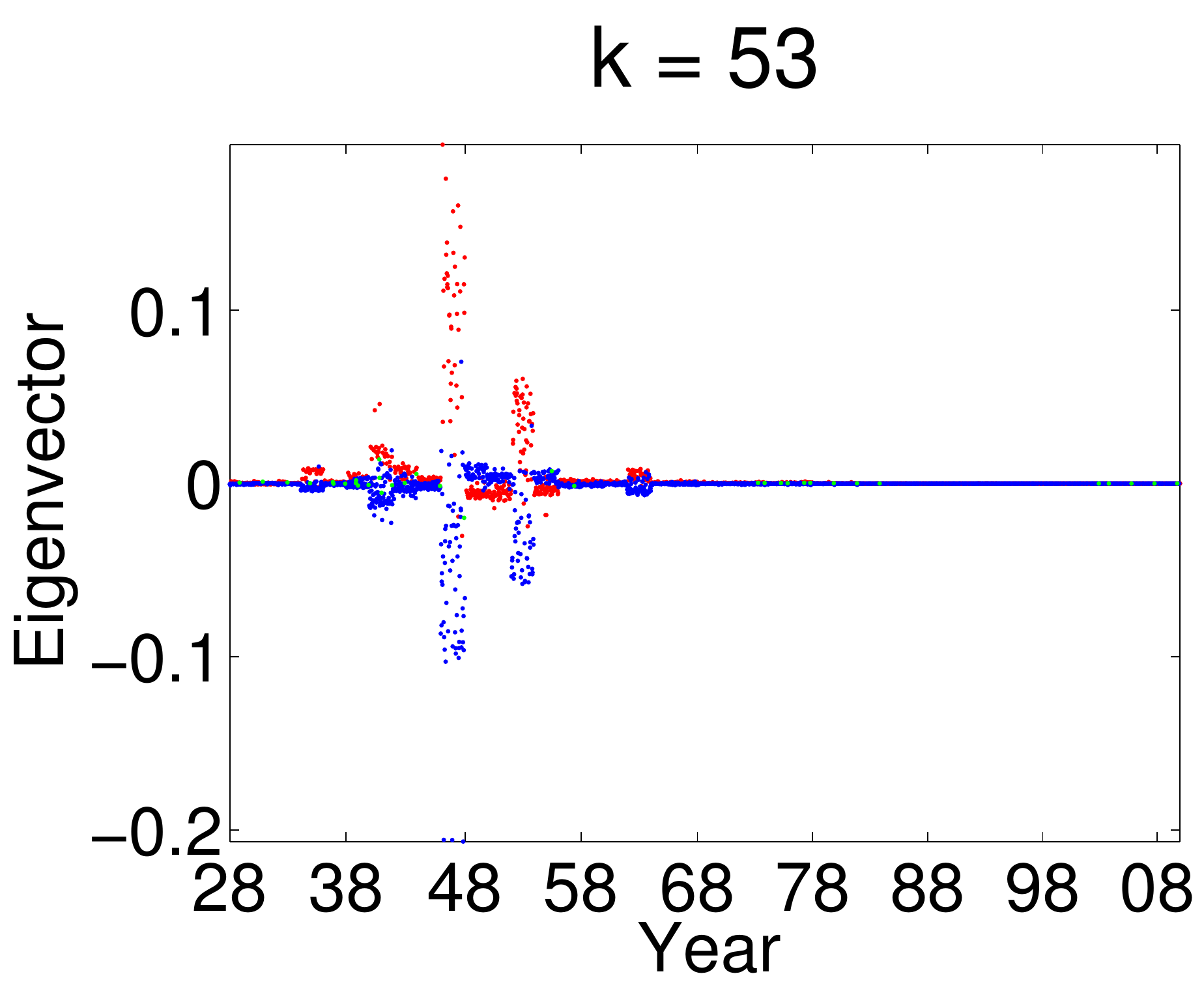}}
\subfigure[$\epsilon = 1$]{
\includegraphics[width=0.19\columnwidth]{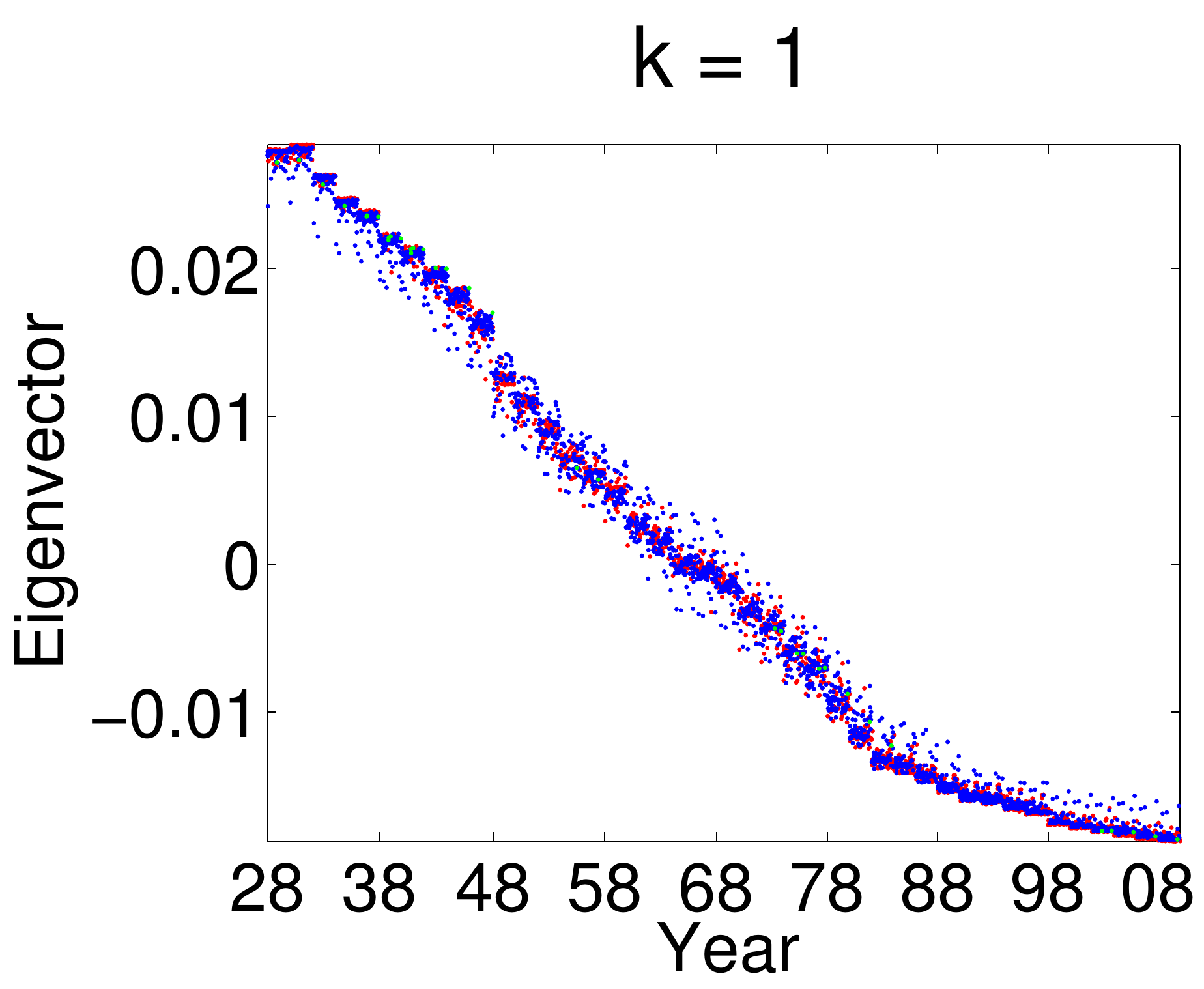}
\includegraphics[width=0.19\columnwidth]{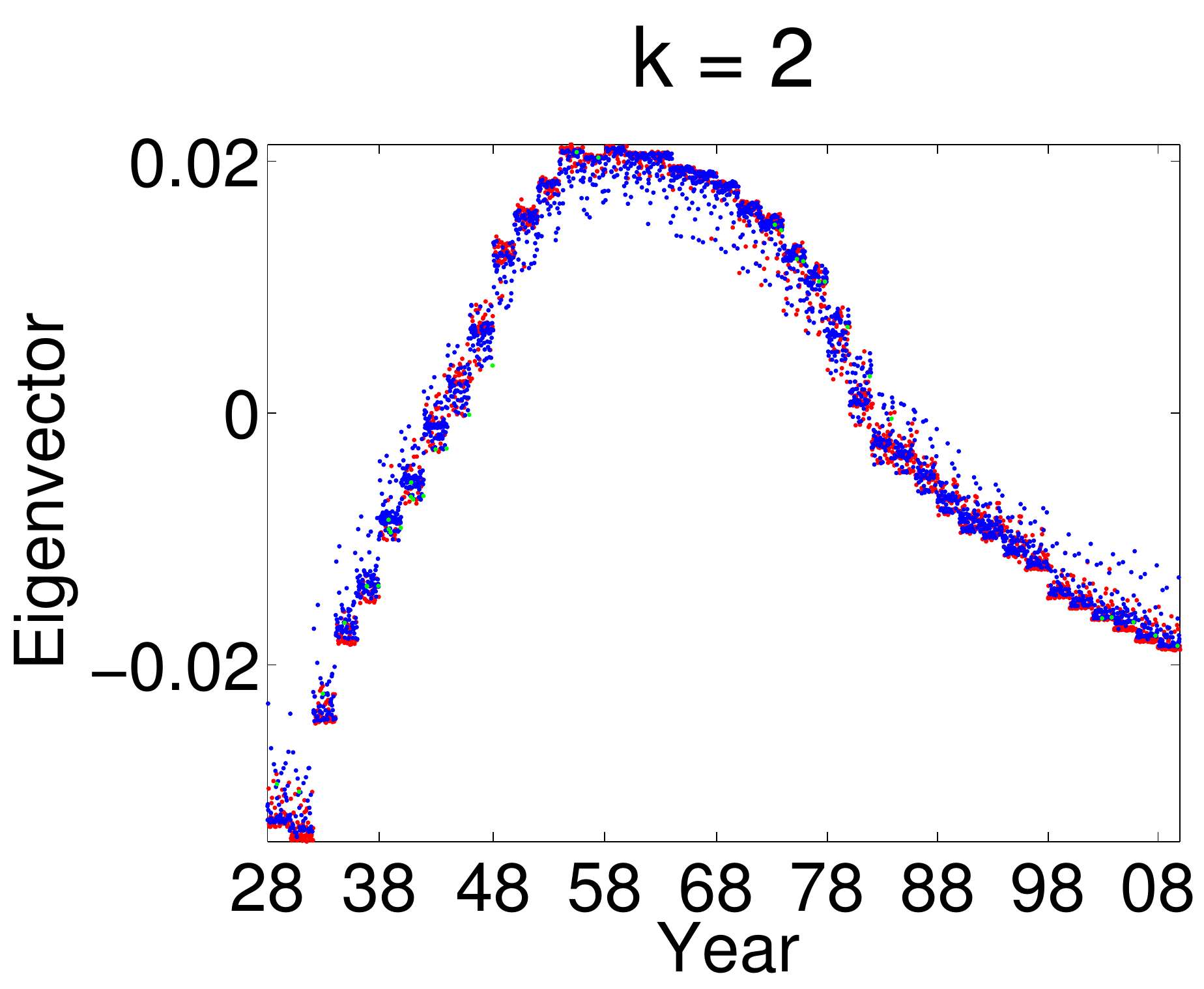}
\includegraphics[width=0.19\columnwidth]{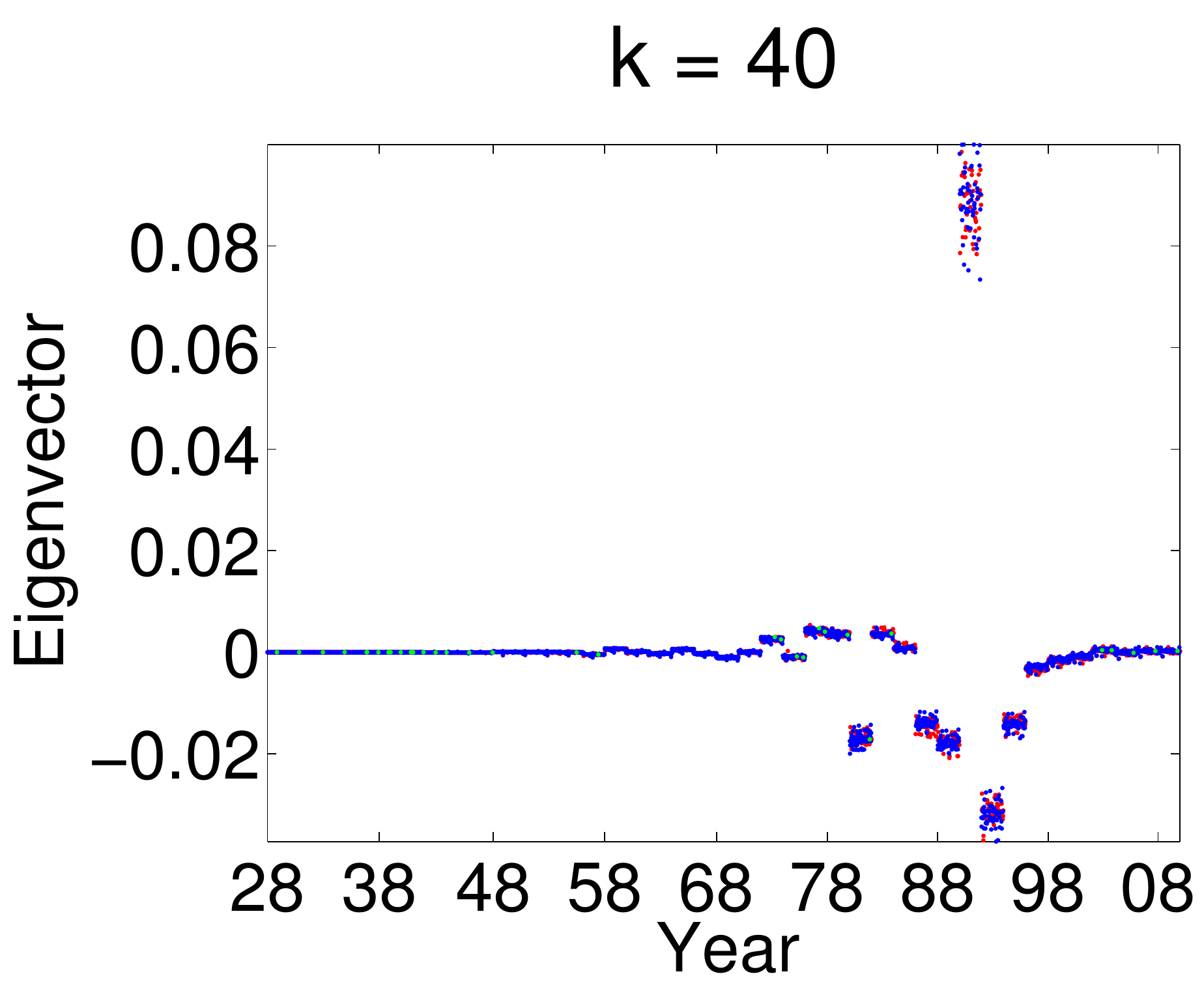}
\includegraphics[width=0.19\columnwidth]{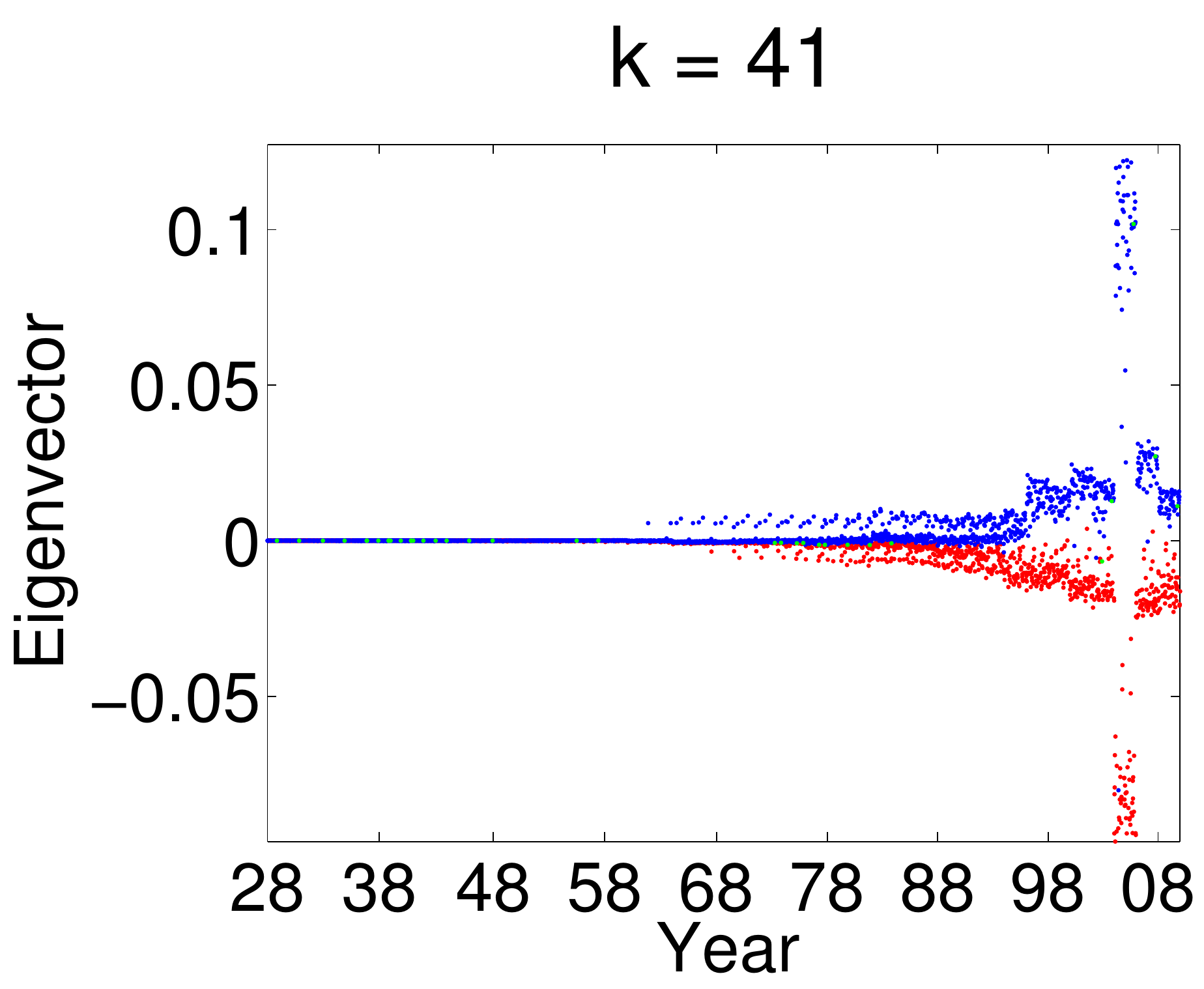}
\includegraphics[width=0.19\columnwidth]{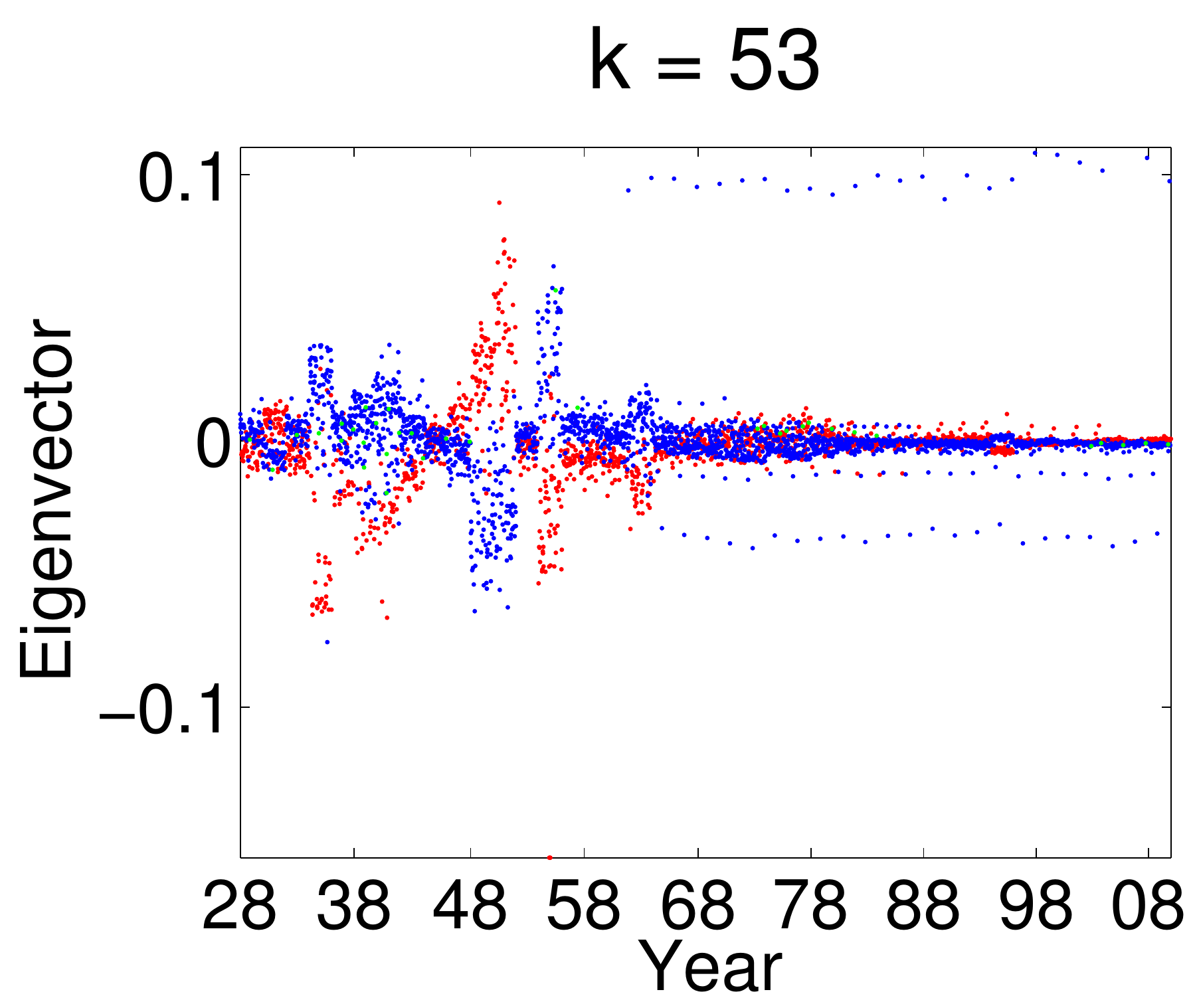}}
\end{center}
\caption{A selection of top eigenvectors of the Laplacian associated to the Congress data set, exhibiting varying degrees of localization. $\epsilon$ denotes the values of the inter-Congress coupling constant. The x-axis denotes time, the y-axis the eigenvector entries, while the coloring highlights the Republican and Democratic parties. Note that in the numbering of the eigenvectors, $k=1$ denotes the first nontrivial eigenvector, and that the x-axis corresponds to the year of the Congress associated to each subset of senators.}
\label{fig:cong_vect2012}
\end{figure}

Given the above matrix of voting patterns $W$, the goal is to recover the two parties, i.e., to classify each of the (non-unique) $n=4196$ senators as either a Republican or a Democrat. It should not be surprising that the top eigenvector of matrix $W$ (or of its associated Laplacian) no longer captures the bipolarity of the Congress, as the problem does not fit the hat of synchronization. As illustrated in Figure \ref{fig:cong_vect2012} \cite{MCLocalization}, none of the top eigenvectors capture the bipolarity of the network, and only eigenvectors buried deeper in the spectrum do so ($l \geq 41$), but only for individual Congresses or for sets of temporally adjacent Congresses. The normalized spectrum of matrix $W$ shown in Figure \ref{fig:cong-details} (b) shows that a large fraction of the variance of the data is captured by the first largest eigenvalues.  We refer the reader to our recent work in \cite{MCLocalization} for a detailed investigation of the above eigenvectors, and of the eigenvector localization phenomenon encountered in numerous other real data sets. As detailed in  \cite{MCLocalization}, eigenvector $l=1$ (the first nontrivial eigenvector) and $l=2$ define the coarsest modes of variation in the data, are delocalized at a global level but localized on individual congresses, and exhibit global oscillatory behavior characteristic of sinusoids, which is expected for such one-dimensional data. Eigenvector $l=40$ is the last one to exhibit similar behavior, although the localization is more prominent, and $l=41$ is the first one to identify the two parties, but only for a subset of adjacent Congresses. Lower eigenvectors, such as $l=53$, exhibit a surprising degree of localization, very often on a single Congress or a few temporally-adjacent Congresses, and reveal the existence of the two parties. However, they do so only for single Congresses or a few adjacent ones, which still does not resolve of task of finding the party affiliation of all the senators, across all the Congresses we have considered.

Alternatively, one may consider only a single Congress at a time, $W^{(t)}, t=1,\ldots,C$, and use the top eigenvector of each $W^{(t)}$ to extract the two parties, as exemplified in Figure \ref{fig:cong-details} (c). In this case, the top eigenvector will separate the Democrats and the Republicans, as this is the main course of variation within an individual Congress. However, this approach has two main obvious disadvantages. The bipolar structure present within each Congress is detected up to a global sign (since the eigenvector is computed up to a global sign), and aggregating this information across all Congresses requires further work. Second, we no longer make use of the information that some senators participate in several Congresses, and usually maintain the same party affiliation across time, although as we shall see, some senators are often inconsistent in their voting, and vote as if they belong to the opposite party.

A simple solution to obtaining a separation between the two parties at a global level is given by the following procedure, which lays the ground for using the synchronization methods over $\mathbb{Z}_2$. By mapping the entries $w_{ij} \in [0,1]$  within the diagonal blocks $W^{(1)},\ldots,W^{(C)}$ of the congress matrix $W$ to $\{-1,1\}$, we are able to highlight the group structure of the problem. One simple option for such a transformation is given by
\begin{equation}
\overline{W}^{(t)}_{ij} = \operatorname{sign}(2 \cdot W^{(t)}_{ij} - 1), t=1,\ldots,C,
\label{transformation}
\end{equation}
which is enough to make the problem fit the framework of synchronization. If two senators agreed on more than $50\%$ of their votes, then we consider that they belong to the same party, and set $\overline{W}^{(t)}_{ij}=+1$. Otherwise, we decide they belong to
opposite parties and set $\overline{W}^{(t)}_{ij} = -1$. Note that an alternative approach to (\ref{transformation}) would be to weight the correlation between nodes and use a mapping such as $\overline{W}^{(t)}_{ij} = 2 \cdot W^{(t)}_{ij} - 1$. For simplicity, we choose to work with (\ref{transformation}), and refer the reader to (\ref{H_ij_def_b}) for an instance of the synchronization problem with weights. Next, we let $\overline{H}$ denote the block diagonal matrix
$$ \overline{H} = \operatorname{Diagonal}(\overline{W}^{(1)},\ldots,\overline{W}^{(C)}),$$
and note that we may decompose the matrix $Z$ of pairwise measurements in a manner similar to the decomposition shown in equation (\ref{W_decomp})
\begin{equation}
Z = \overline{H} + \epsilon \Omega.
\end{equation}
Note that if the interslice coupling constant $\epsilon$ is set to $0.1$ or a different transformation is used in (\ref{transformation}) whose range is other than $\pm 1$, the entries of the matrix $Z$ are no longer elements of $Z_2$, and we interpret the fractional entries as confidence weights associated to the pairwise measurements. We point out that the formulation in (\ref{H_ij_def_b}) is an instance of such a synchronization problem with weights, since each edge measurement has an associated weight that quantifies how similar or dissimilar the two nodes (i.e., partitions) are.



\begin{figure}[h]
\begin{center}
\includegraphics[width=0.45\columnwidth]{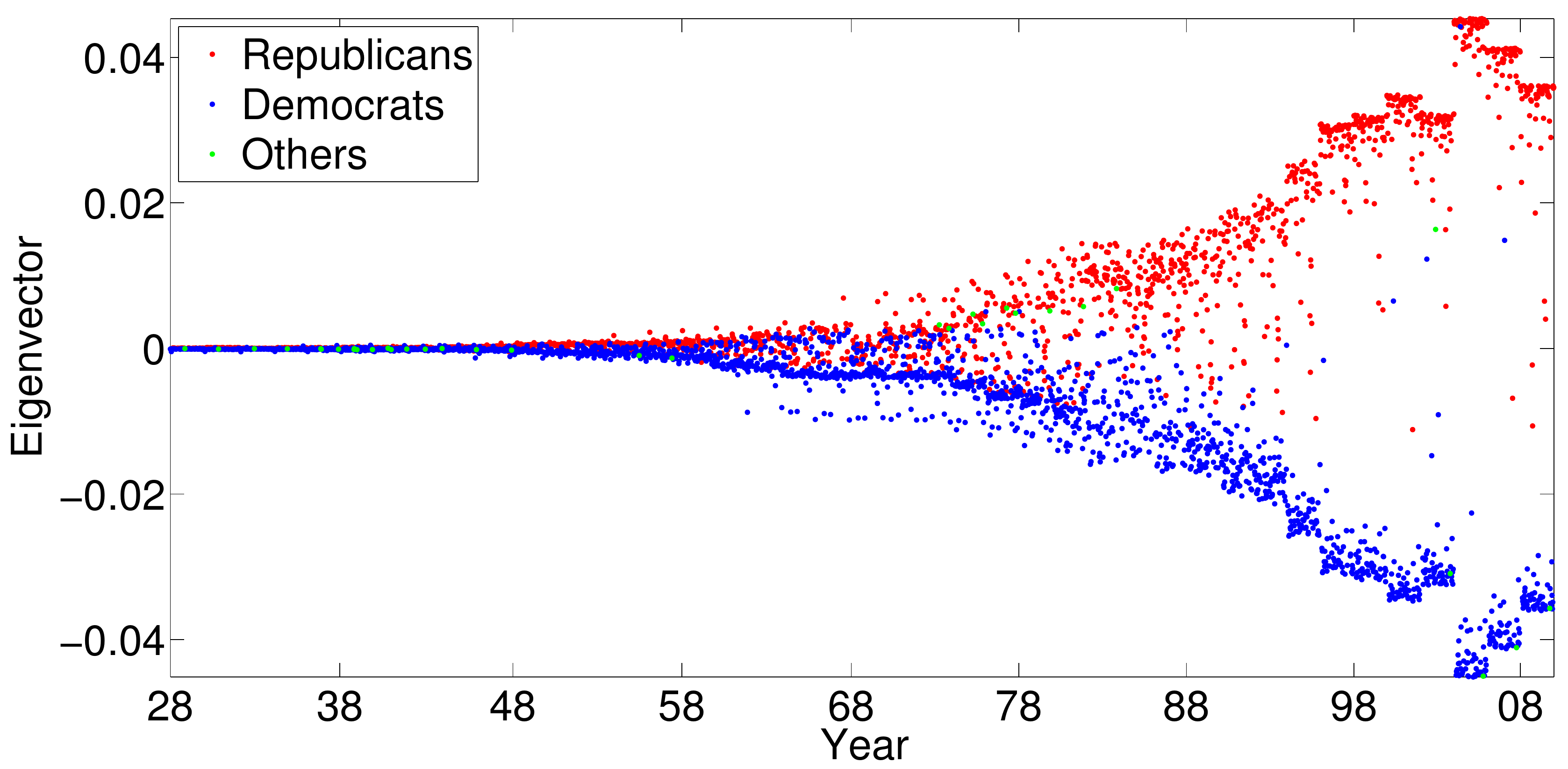}
\includegraphics[width=0.45\columnwidth]{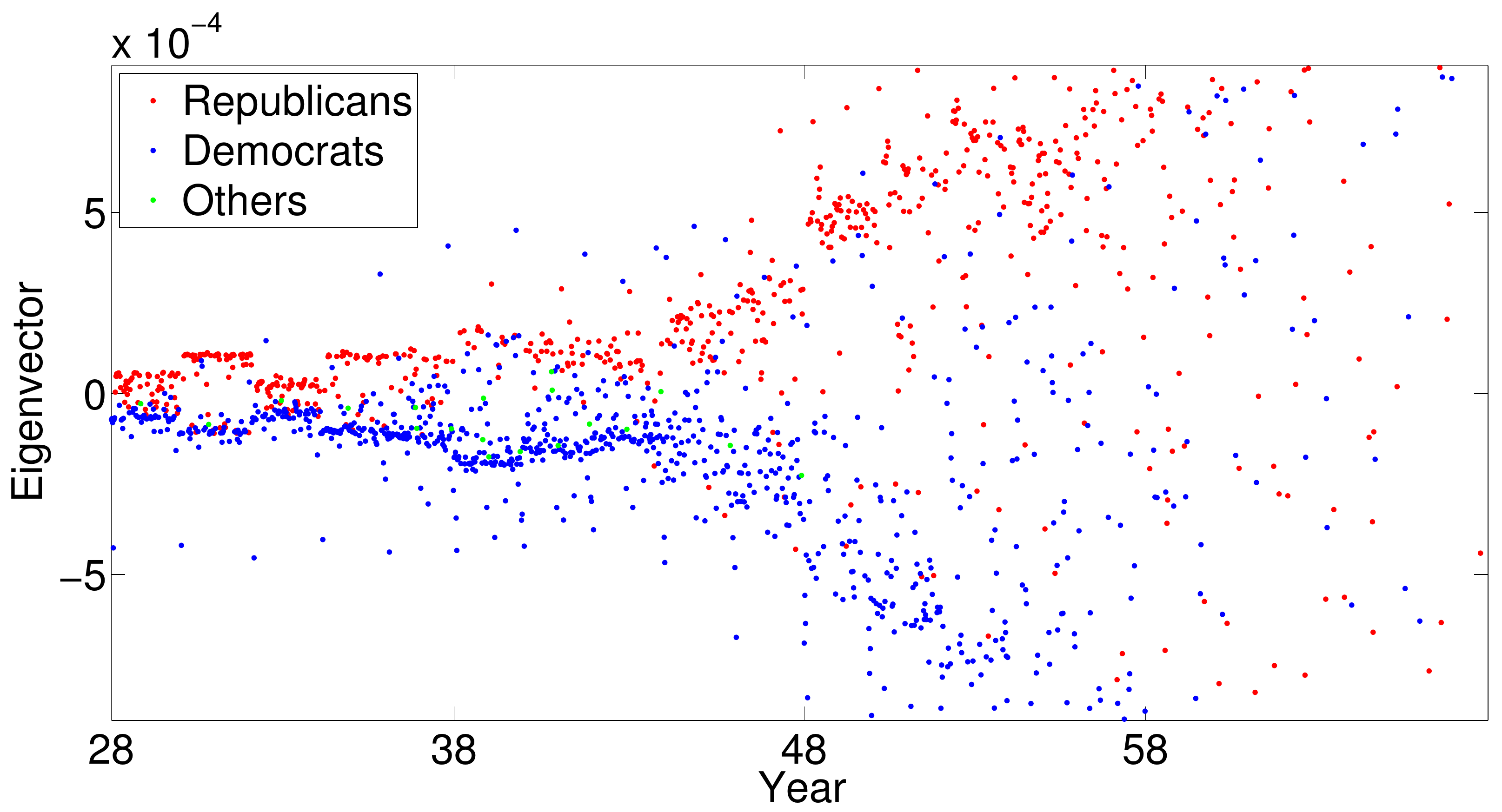}
\end{center}
\caption{ Left: plot of $v_1^{\mathcal{Z}}, $ the top eigenvector of matrix $\mathcal{Z}$ of size $n=4196$, that corresponds to 41 Congresses, starting with the $70^{th}$ Congress. Right: zoom in on the entries of the same eigenvector $v_1^{\mathcal{Z}}$, corresponding to the first 20 Congresses. We label the x-axis with the number of the Congress corresponding to a subset of senators (for example the first $\approx 100$ entries correspond to the $70^{th}$ Congress, while the last $\approx 100$ entries correspond to the $110^{th}$ Congress). 
Note that within each Congress, the ordering of the nodes corresponds to the alphabetical ordering of the states that senators represent.}
\label{fig:cong_sep}
\end{figure}

\begin{figure}[h]
\begin{center}
\includegraphics[width=0.4\columnwidth]{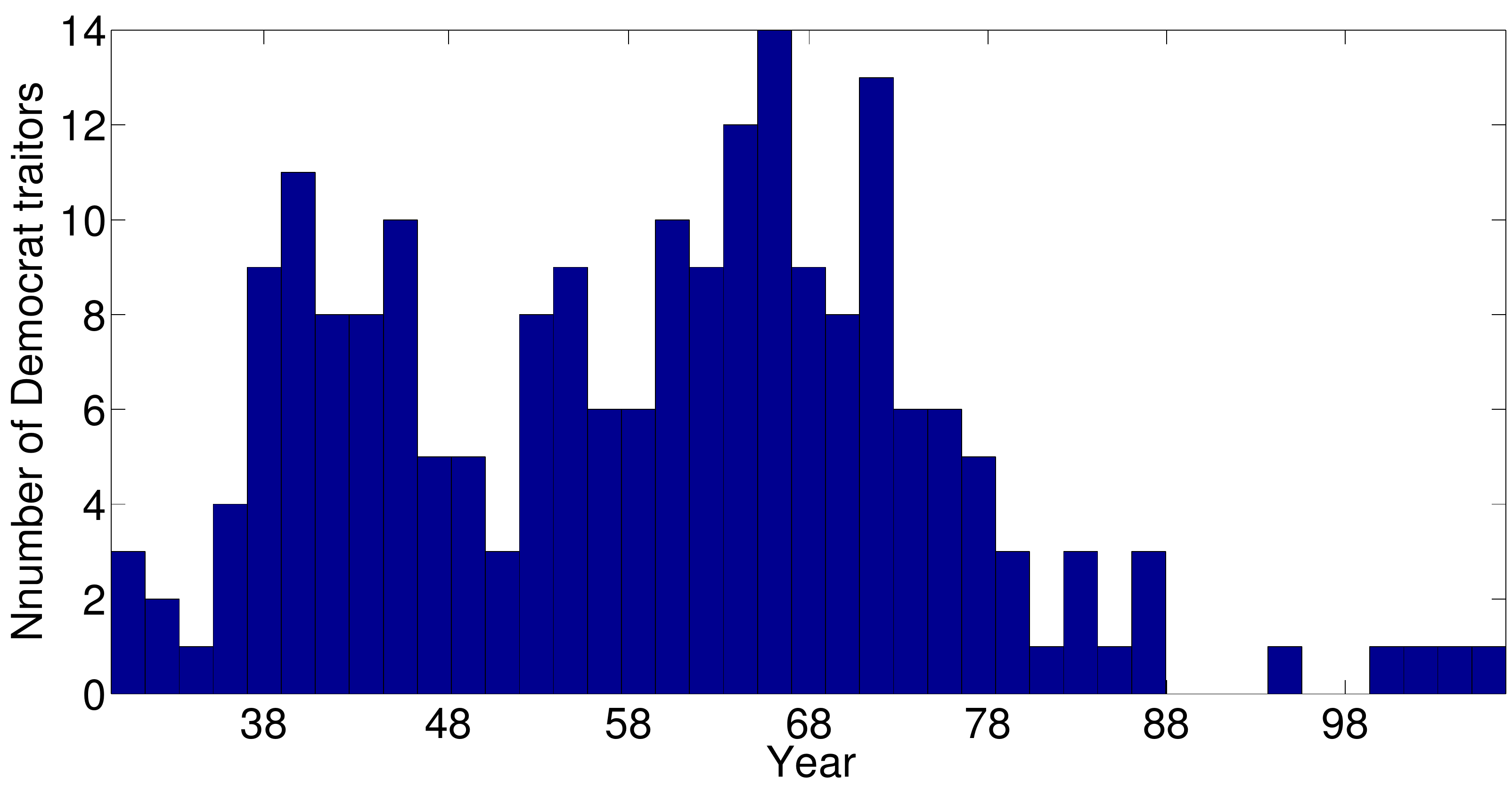}
\includegraphics[width=0.4\columnwidth]{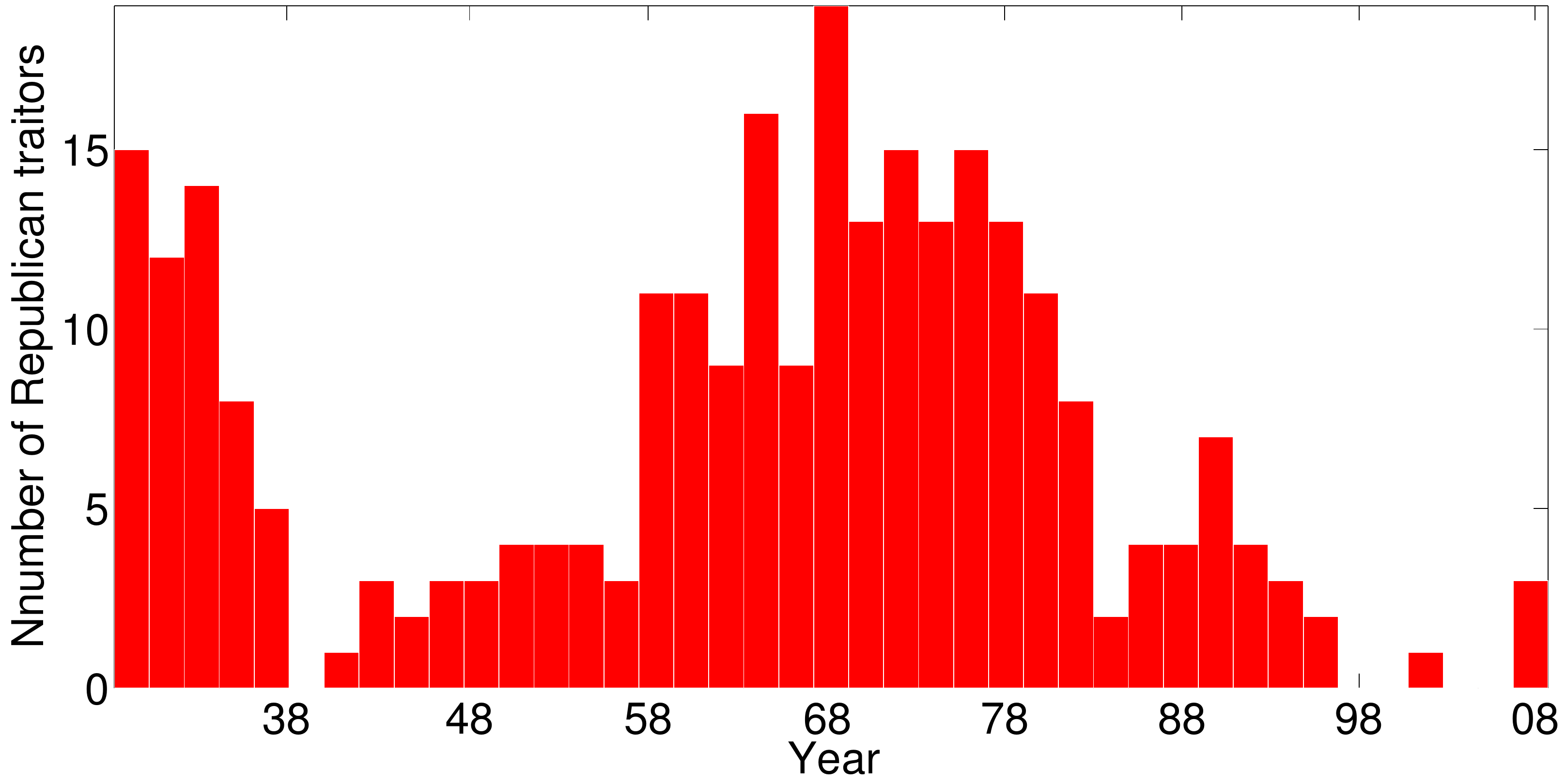}
\end{center}
\caption{ Barplot of the number of Democrat (left) and Republican (right) senators that voted as if they served in the opposite party, for each of the 41 Congresses under investigation. The  x-axis denotes time in years, while the y-axis denotes the number of misclassifications in each Congress.}
\label{fig:hist_traitors}
\end{figure}

In practice, for increased robustness to noise, we use the normalization $\mathcal{Z} = D^{-1}Z$ described in Section \ref{sec:sync}, and introduced for the first time in our previous work \cite{asap2d}. We refer the reader to Section 4 of \cite{asap2d} for a detailed explanation of the additional noise robustness the above normalization brings, and its connection to the normalized discrete graph Laplacian. As shown in Figure \ref{fig:cong_sep}, the top eigenvector of the matrix $\mathcal{Z}$ shows a clear separation of the two political parties. Due to the existing noise in the data, as often senators actually vote as if they were in the opposition, there are of course errors in classifying each senator as a Republican or Democrat, but the overall accuracy is still high. Of the $n=4196$ non-unique senators, we were able to correctly identify $84\%$ of the Republicans, and $91\%$ of the Democrats (we ignore the 35 senators which are neither Republicans or Democrats). Note that for the first half of the Congresses, the entries of the top eigenvector take values very close to zero, misleading one into thinking that perhaps a large number of misclassifications are being made during that interval. However, in the right plot of Figure \ref{fig:cong_sep} we zoom in on the first half of the Congresses, and the separation becomes more visible, although there are clearly many more misclassifications during  this time interval. To this end, we plot in Figure \ref{fig:hist_traitors} the histogram of misclassifications, separately for the Republicans and the Democrats, which shows indeed that the most ``treasons" occurred between 1960s and 1980s, and the least occurred after the year 2000.

\begin{figure}[h]
\begin{center}
\includegraphics[width=0.19\columnwidth]{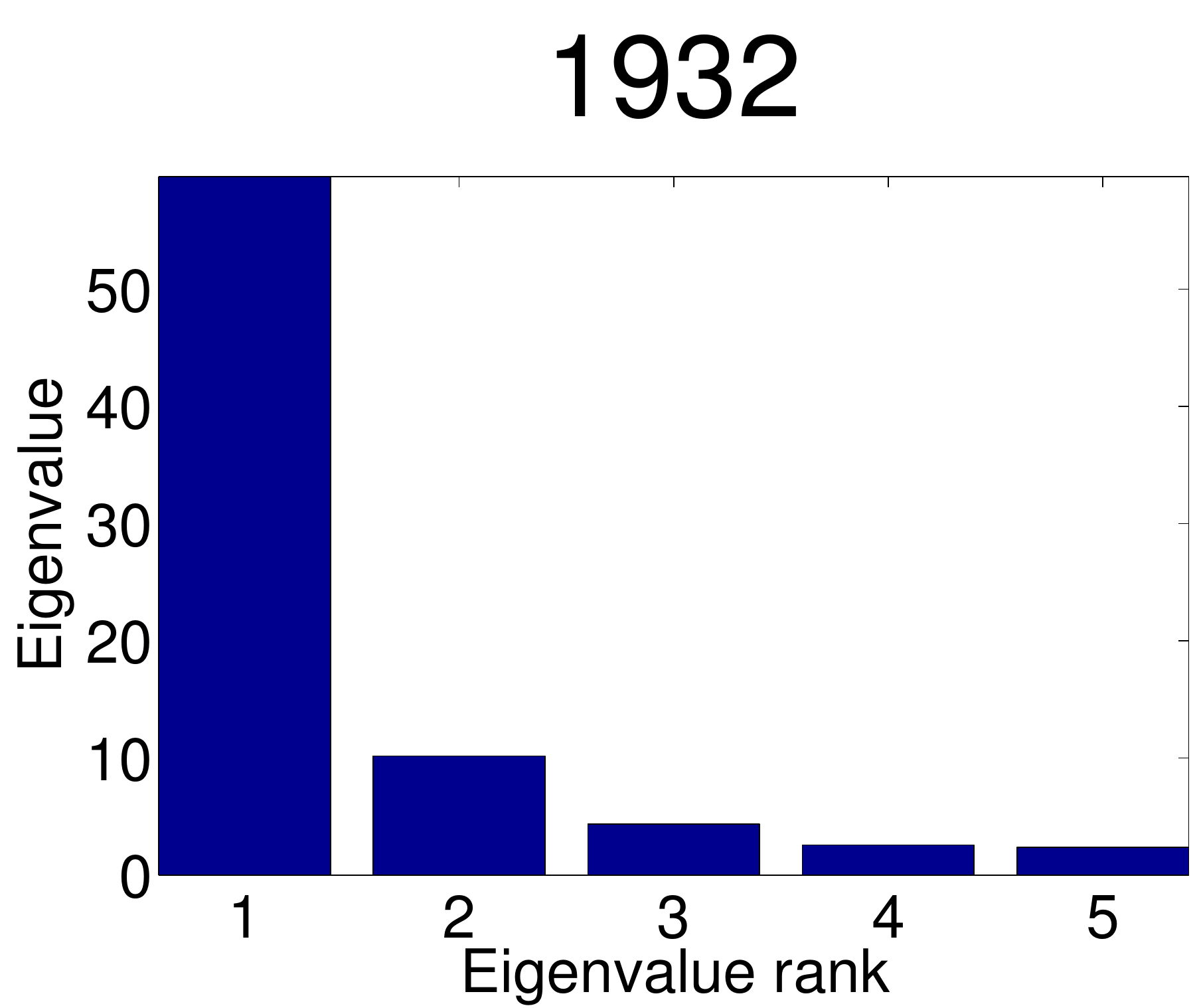}
\includegraphics[width=0.19\columnwidth]{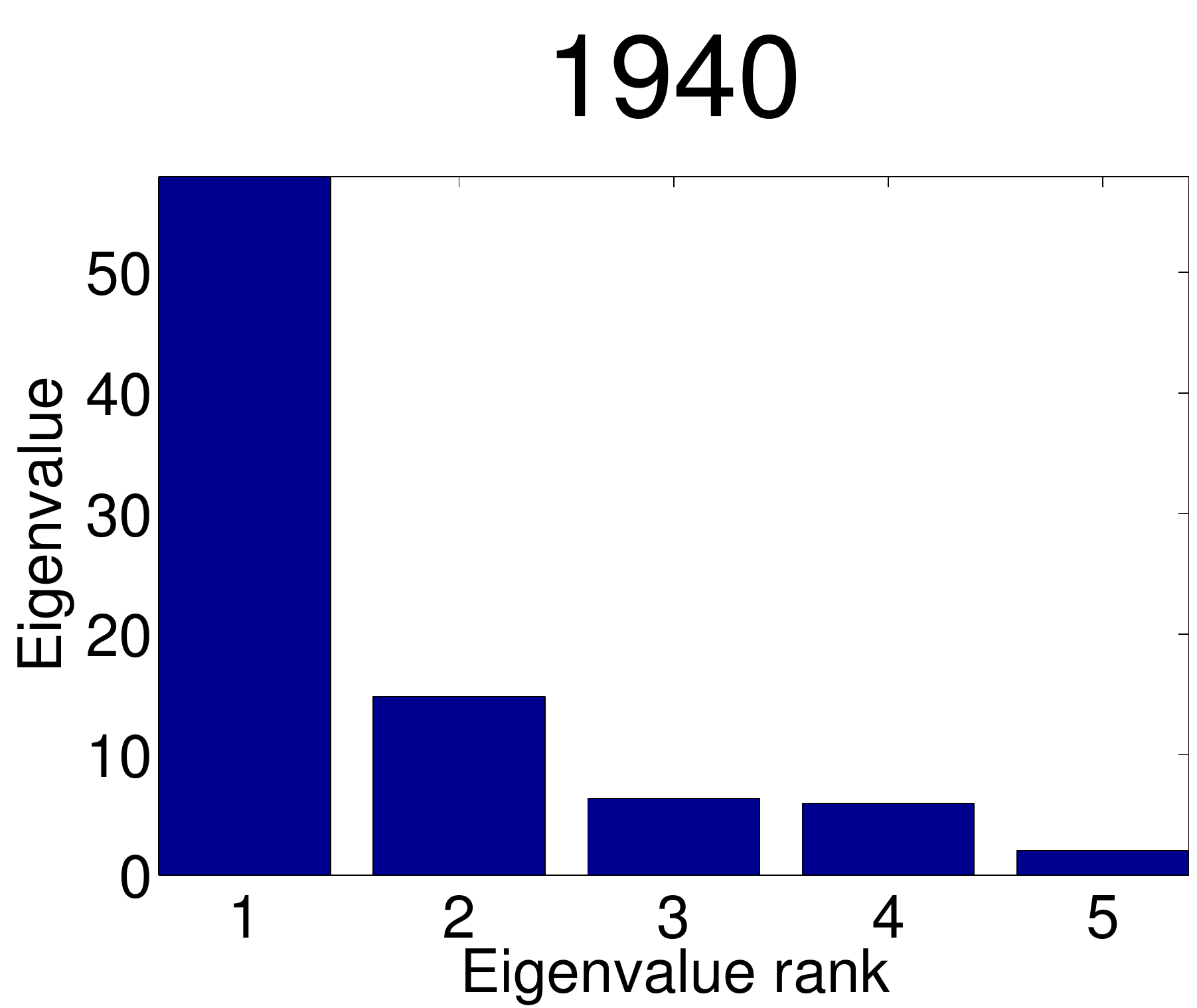}
\includegraphics[width=0.19\columnwidth]{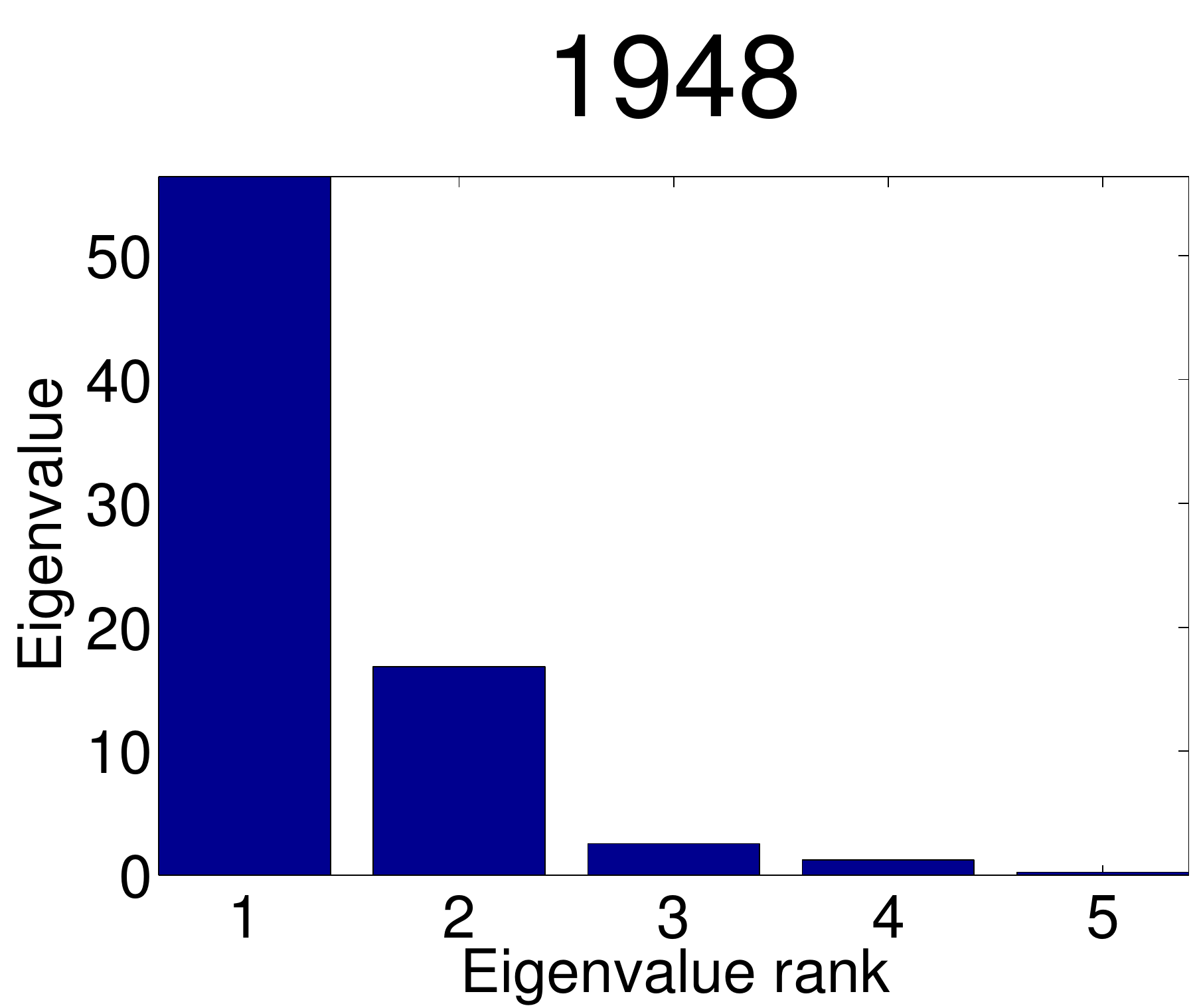}
\includegraphics[width=0.19\columnwidth]{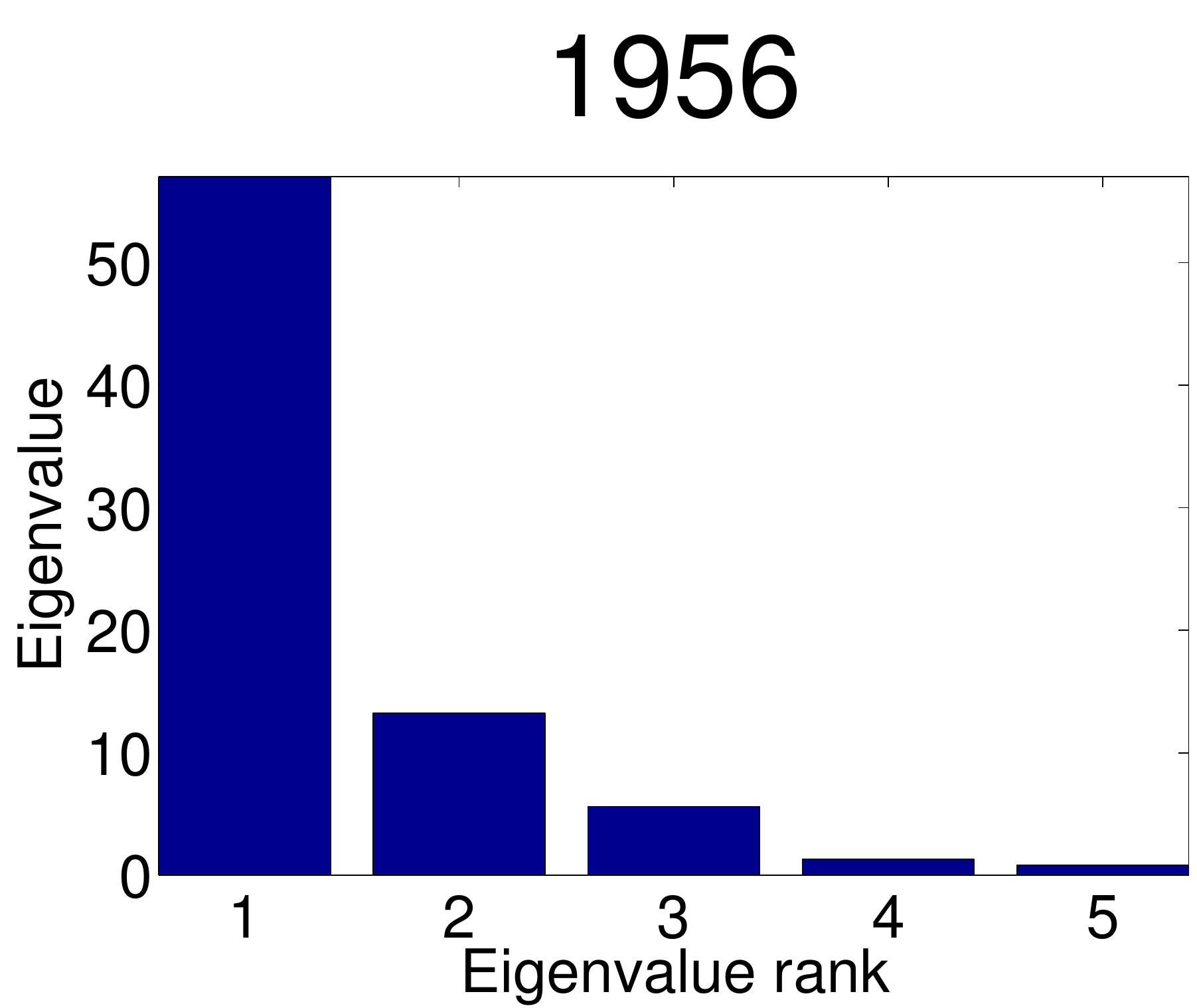}
\includegraphics[width=0.19\columnwidth]{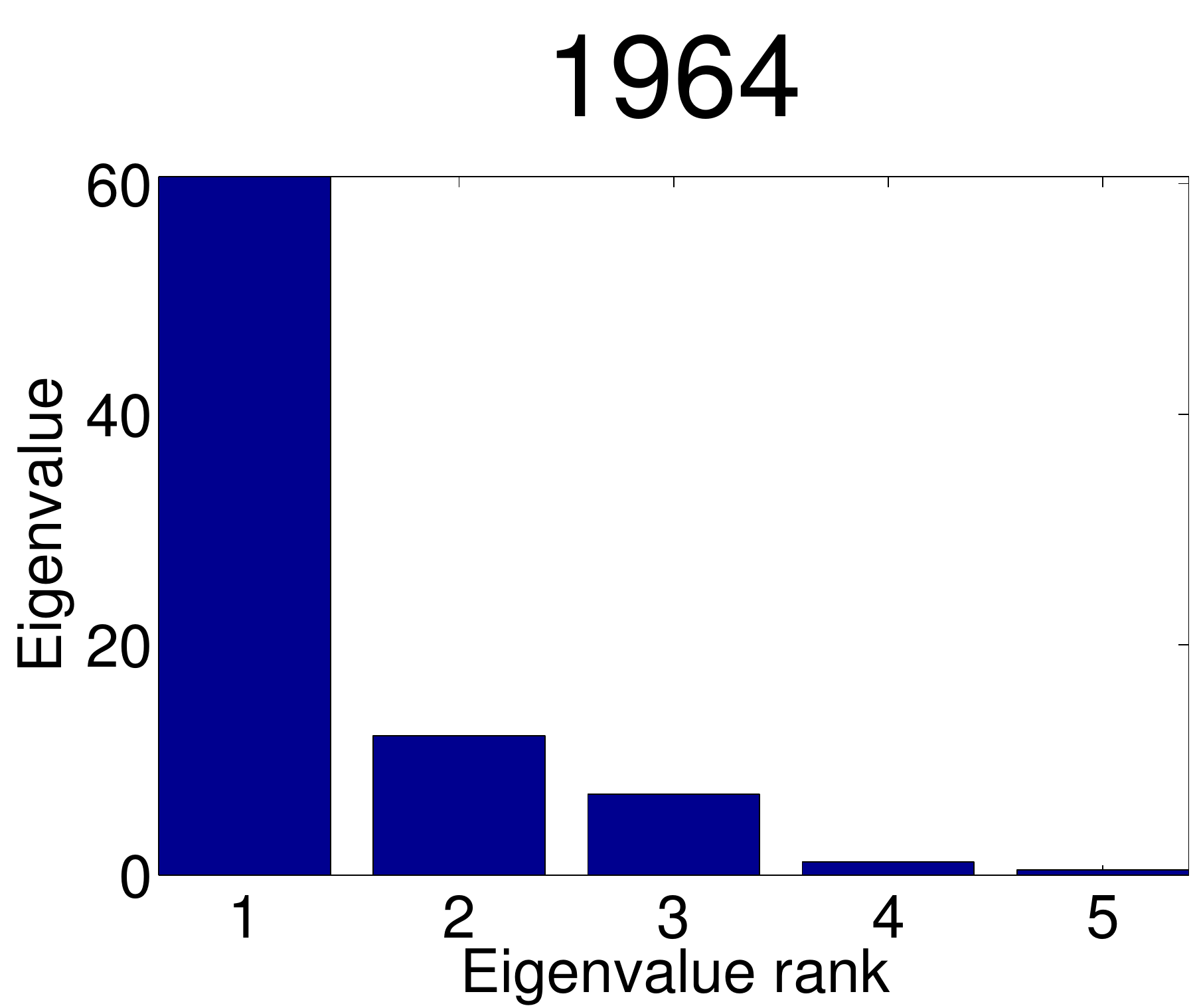}
\includegraphics[width=0.19\columnwidth]{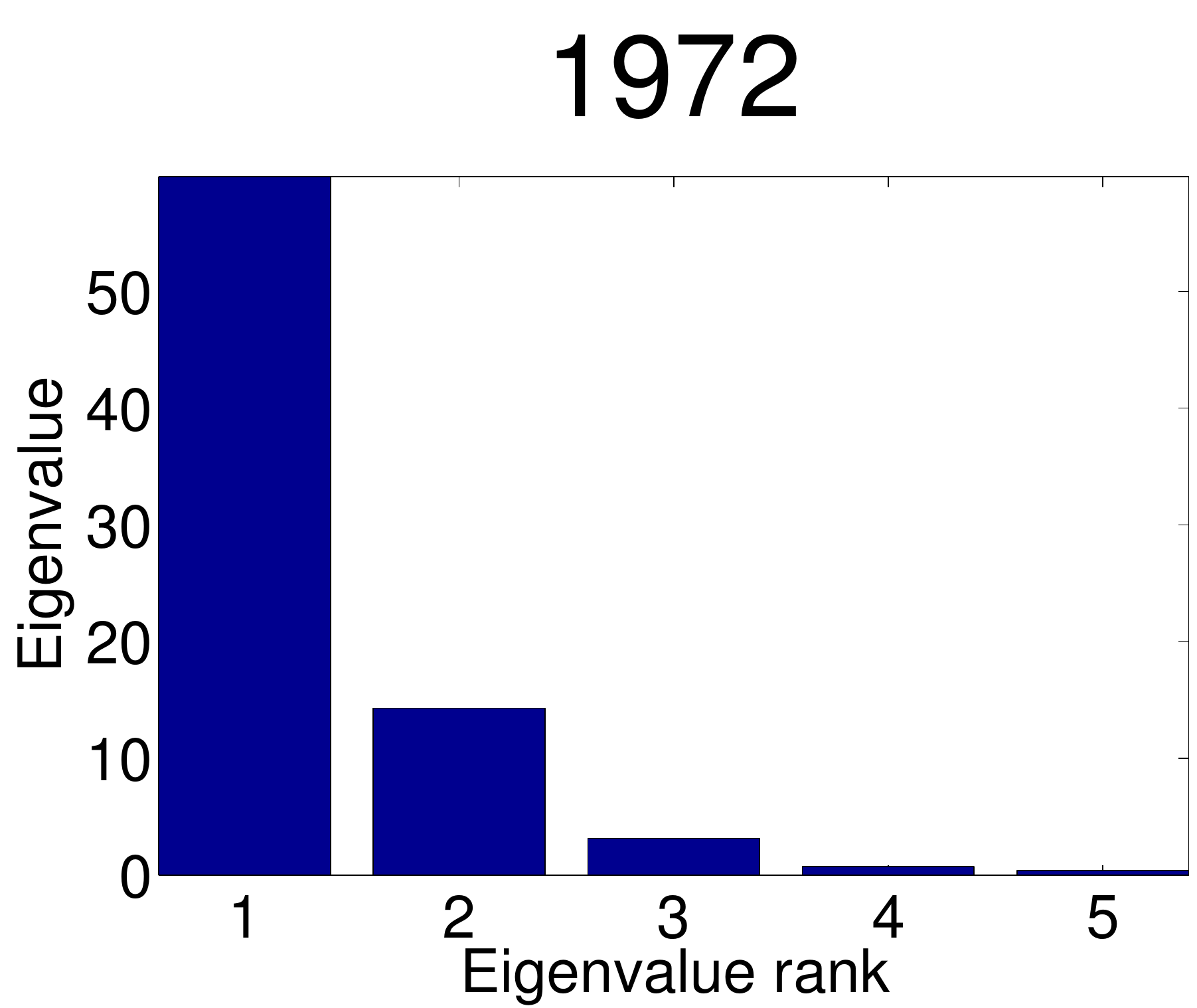}
\includegraphics[width=0.19\columnwidth]{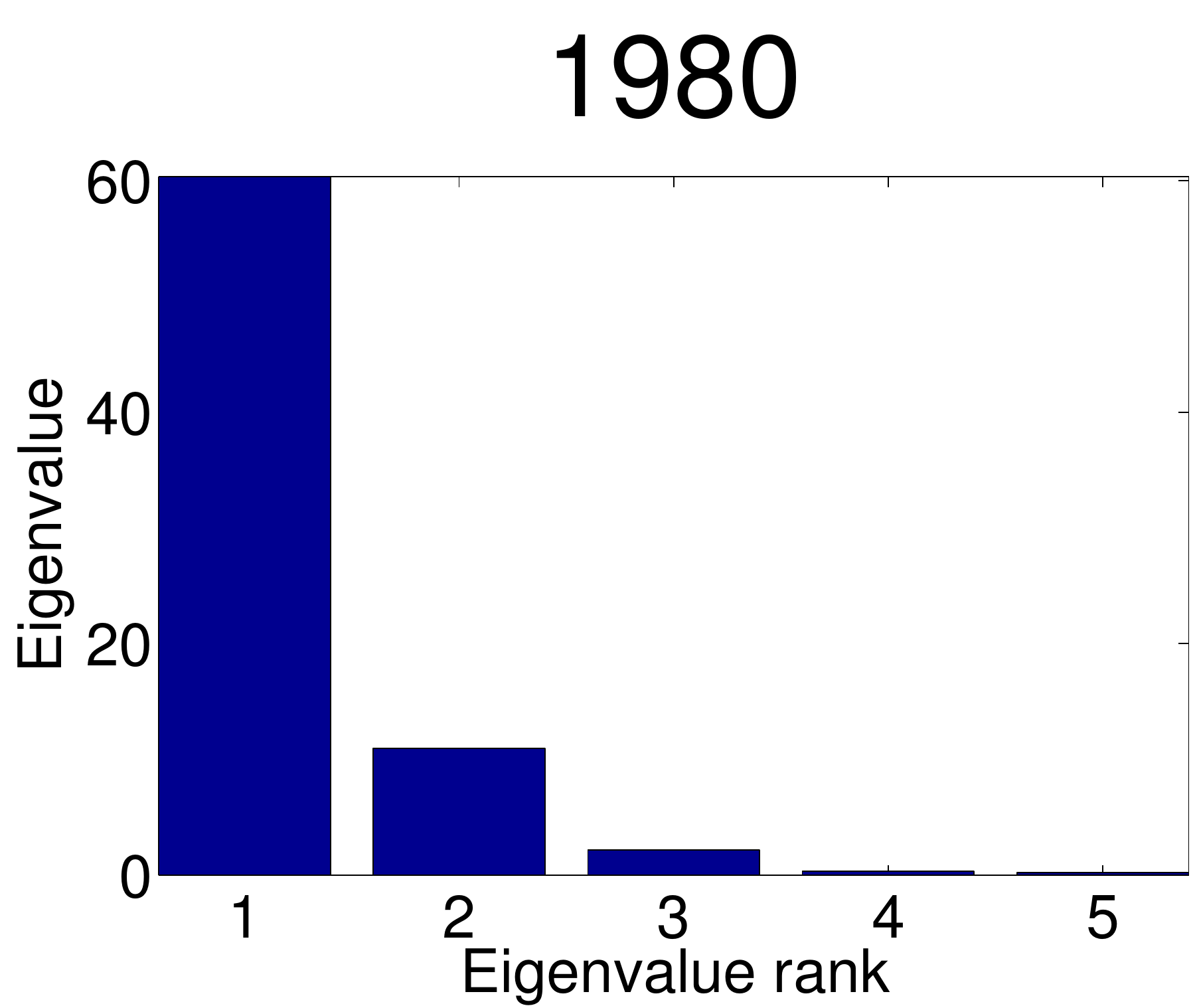}
\includegraphics[width=0.19\columnwidth]{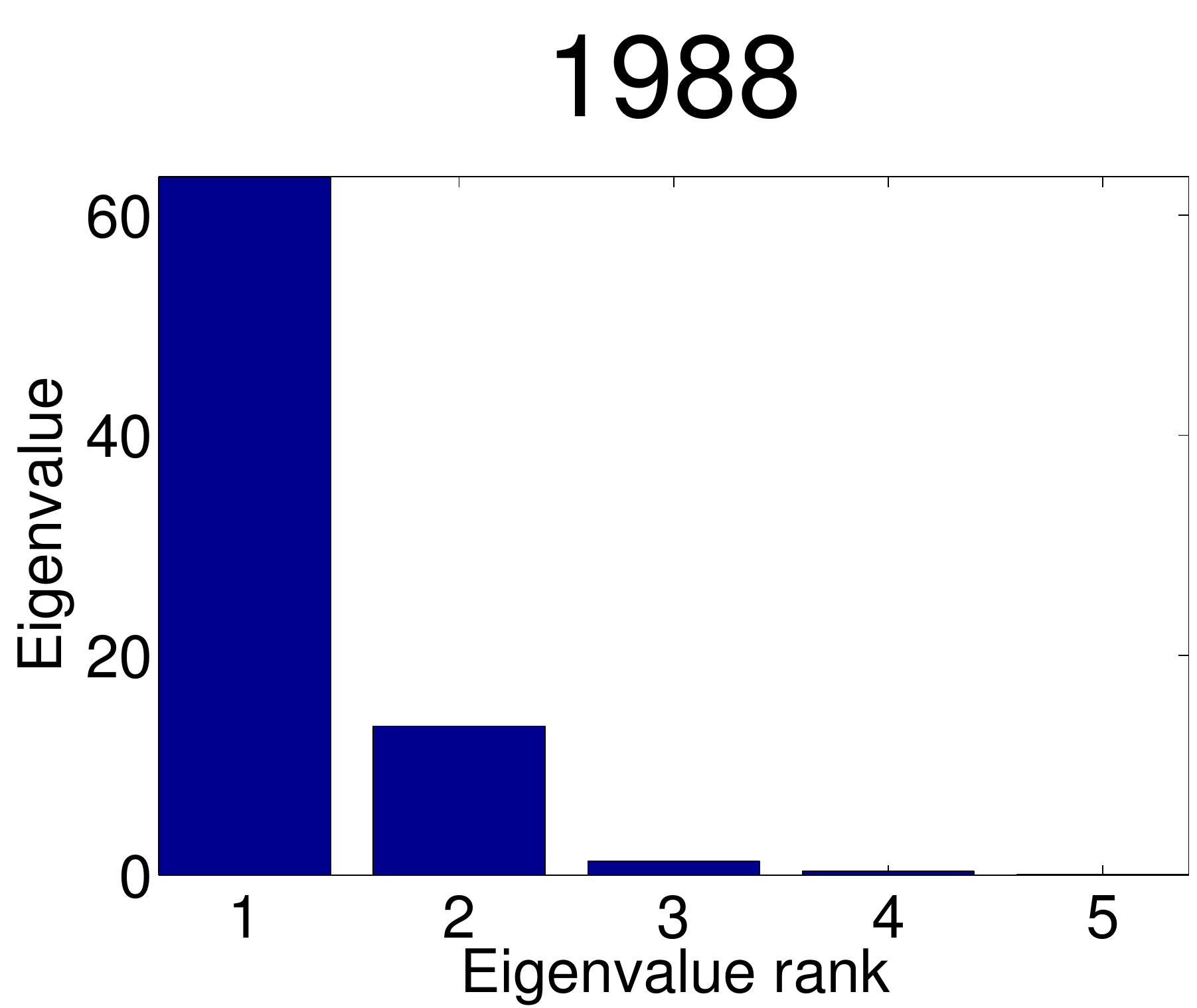}
\includegraphics[width=0.19\columnwidth]{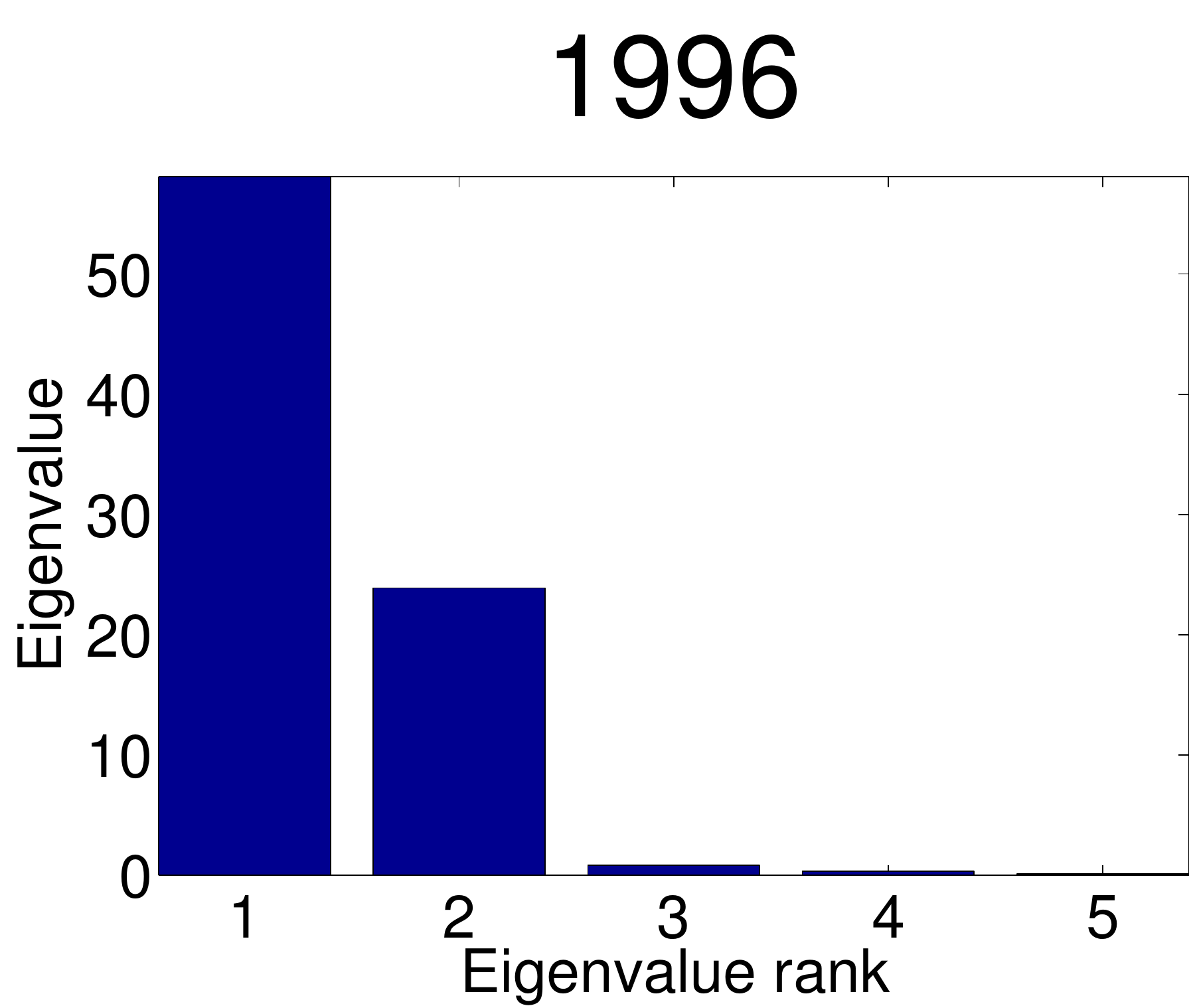}
\includegraphics[width=0.19\columnwidth]{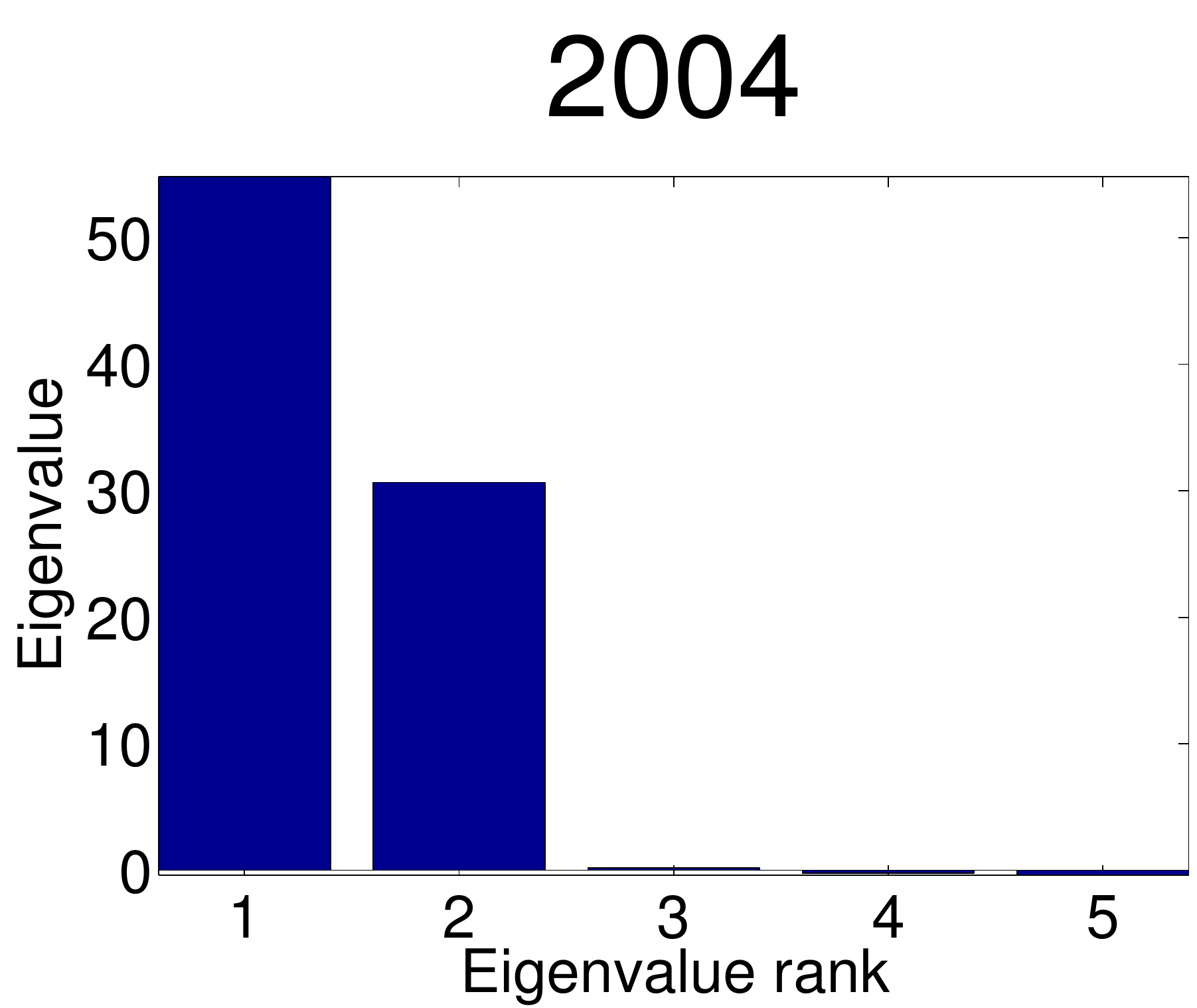}
\end{center}
\caption{ Barplot of the top five largest eigenvalues for several of the voting similarity matrices $W^{(i)}$, corresponding to the years $\{1932,1946,1960,1974,1988,2002\}$.}
\label{fig:spect_each_congress}
\end{figure}

\begin{figure}[h]
\begin{center}
\includegraphics[width=0.4\columnwidth]{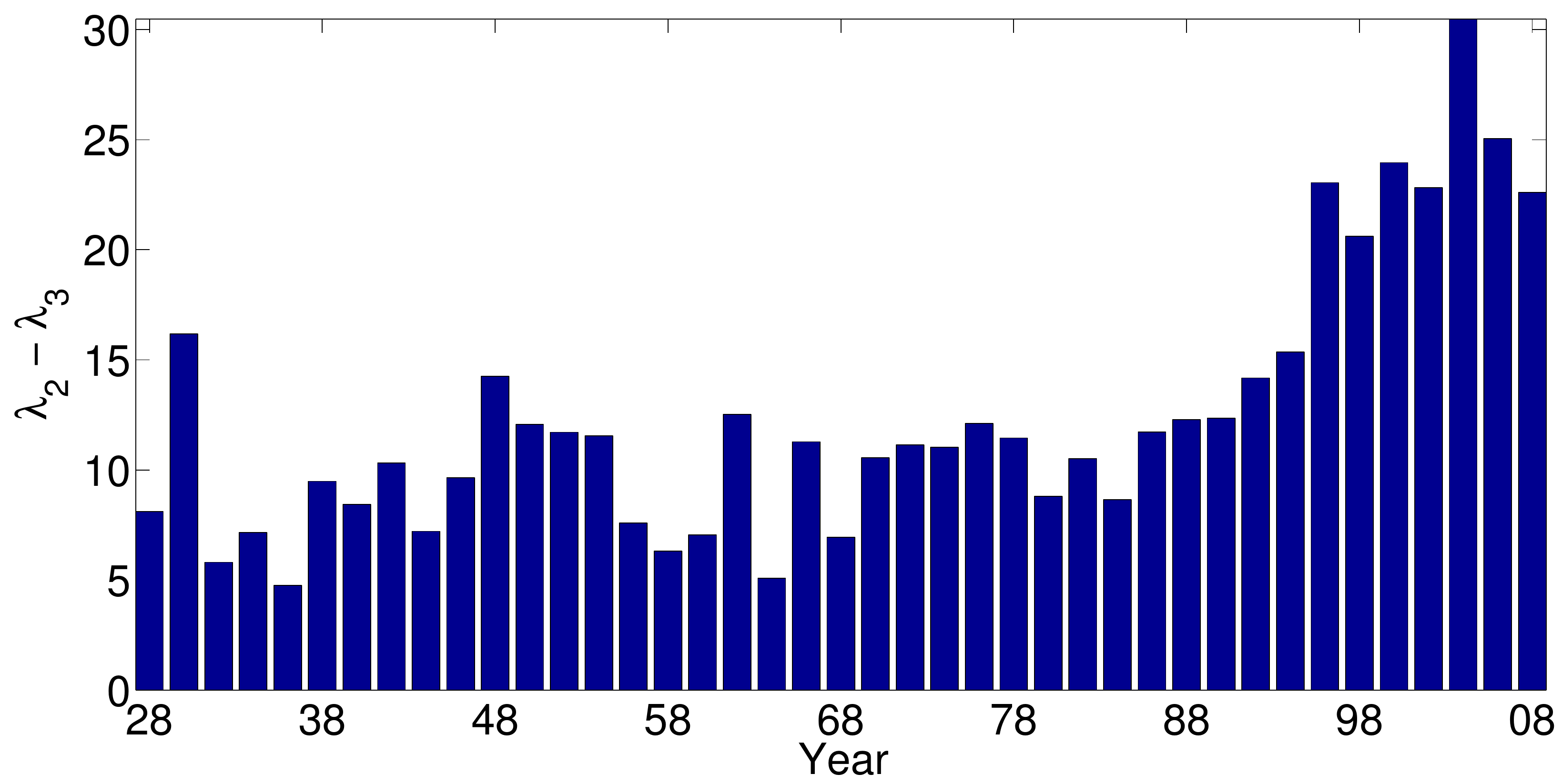}
\includegraphics[width=0.4\columnwidth]{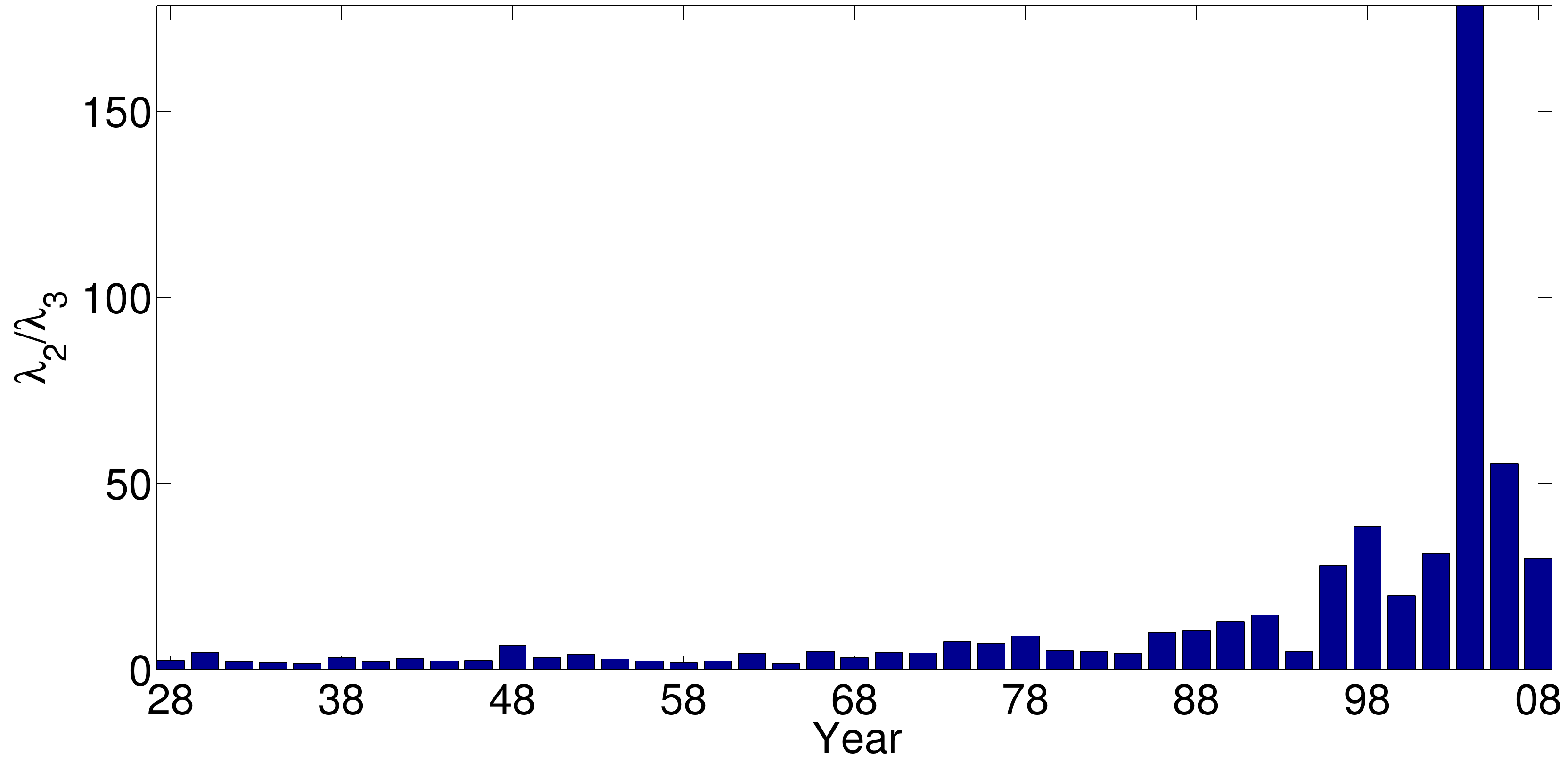}
\end{center}
\caption{Barplot of the gap $\lambda_2 - \lambda_3$ (left) and barplot of the ratio
$\frac{\lambda_2}{\lambda_3}$ (right), between the second and third eigenvalue of matrices $W^{(1)},\ldots,W^{(41)}$ corresponding to the $41$ Congresses between the years 1927-2009.}
\label{fig:gap_dif_rat}
\end{figure}

We confirm our findings above by an analysis of the spectrum of each of the $C=41$ Congresses.  To this end, we plot in Figure \ref{fig:spect_each_congress}  the top 5 eigenvalues for a subset of the Congresses, illustrating the fact that in very recent years, the third, fourth and fifth eigenvalues are very small (close to zero) when compared to their counterparts for the Congresses up until the 1960s. In addition, we record in the left plot of Figure \ref{fig:gap_dif_rat} the spectral gap $\lambda_3 - \lambda_2$ between the second the third eigenvalue, a proxy for measuring the bipolarity  of each Congress. Similarly, the right plot of the same Figure \ref{fig:gap_dif_rat} shows the ratio $\frac{\lambda_3}{\lambda_2}$, for each of the 41 Congresses in the interval 1927-2009. Both plots clearly show that the spectral gaps for the last 10 Congresses are significantly larger than the spectral gaps of all previous Congresses, thus validating our earlier conclusion from Figure \ref{fig:hist_traitors} that the bipolar structure of the network was more prominent  during the last 10 Congresses. In addition, in Table \ref{tab:asdasd2wf} we show a list of the top 10 senators that we have classified correctly most often, and a list of the top 10 senators that have been misclassified most often. In other words, the latter 10 senators were the ones whose voting patterns resemble the least the voting pattern of their respective parties, and voted as if they were in the opposition.

Another explanation that accounts for the different behavior of the top eigenvector of matrices $W$ and $Z$ is the following. One may choose to interpret that the zero entries denoting the temporal separation in the $W$ matrix somewhat cause ''confusion`` in the eigenvector computation. This ambiguity stems from the fact that a small $W_{ij}$ entry, for example $W_{ij}=0.01$, means that senators $i$ and $j$ are
close in time (actually they belong to the same Congress) and their voting patterns completely disagree. On the other hand, a slightly smaller entry $W_{ij}=0$ denotes the fact that the two senators do not belong to the same Congress and can actually be very far apart in time, thus a totally different interpretation. In other words, the eigenvector cannot distinguish whether the very small entries (zero or almost zero) denote completely opposite parties or temporal separation, and it is precisely this ambiguity that we remove by using the transformation in (\ref{transformation}) and conducting our analysis based on the resulting matrix $Z$.


\begin{table}[tpb]
\begin{minipage}[b]{0.46\linewidth}
\begin{center}
\begin{tabular}{|c|c|c|c|l|l|}
\hline
Party & $\eta$ & T & M & State & Name \\
\hline
D &  1.00 &  19 &  19 &   MS & Eastland \\
D &  1.00 &  18 &  18 &   AR    &  McClellan\\
D &  1.00 &  17 &  17 &   VA    &  Byrd\\
R &  0.94 &  16 &  17 &   VT     &  Aiken \\
D &  0.75 &  15 &  20 &   GA     &  Russel  \\
R &  0.87 &  13 &  15 &   OR      &  Hatfield  \\
R &  1.00 &  12 &  12 &   NJ  &  Case  \\
R &  1.00 &  12 &  12 &   NY    &  Javits  \\
D &  0.52 &  11 &  21 &   MS &  Stennis  \\
R &  1.00 &  10 &  10 &   ND &  Langer  \\
\hline
\end{tabular}
\end{center}
\end{minipage} \;\;\;\;\;\;\;\;\;
\begin{minipage}{0.46\linewidth}
\begin{center}
\begin{tabular}{|c|c|c|c|l|l|}
\hline
Party & $\eta$ & T & M & State & Name \\
\hline
D &  0.50  & 1 &  2  &  SC & Blease  \\
D &  0.33  & 1 &  3  &  LA      & Broussard  \\
R &  1.00  & 1 &  1  &  ND   & Brunsdale  \\
R &  0.09  & 1 &  11 &  KS         & Capper  \\
D &  0.33  & 1 &  3  &  OH           & Donahey  \\
D &  0.17  & 1 &  6  &  DE       & Frear  \\
R &  1.00  & 1 &  1  &  ID          & Gooding  \\
D &  0.33  & 1 &  3  &  OK       & Gore  \\
R &  0.14  & 1 &  7  &  MI       & Griffin  \\
D &  0.20  & 1 &  5  &  NC & Hoey  \\
\hline
\end{tabular}
\end{center}
\end{minipage}
\caption{List the senators who were misclassified most often (i.e., ``traitors'', on the left) and the least (i.e., those most ``faithful" to their party, on the right). D and R denote a Democrat, respectively a Republican. $\eta = \frac{T}{M}$, where $M$ denotes the total number of terms a senator served on, and $T$ denotes the number of misclassifications. Results are sorted in decreasing order as a function of $T$.}
\label{tab:asdasd2wf}
\end{table}

\section{Robustness to noise of eigenvector synchronization over $\mathbb{Z}_2$} \label{sec:rmtx}


In this section, we discuss the robustness to noise of the eigenvector method  for synchronization over $\mathbb{Z}_2$. We defer to Appendix A the associated analysis of the eigenvector method when the underlying graph $G$ of pairwise measurements is an Erd\H{o}s-R\'{e}nyi random graph (or possibly a complete graph $K_n$), and the subgraph of noisy edges is a random subgraph of $G$.  We follow closely the analysis detailed in previous work by Singer \cite{sync}, pertaining to the group of planar rotations SO(2), and note that crucial to the analysis are recent random matrix theory results of  F\'eral and P\'ech\'e \cite{FeralPeche} on the largest eigenvalue of rank-one deformation of (real, symmetric) large random matrices. 

Under the Erd\H{o}s-R\'{e}nyi random graph model assumption, where the graph of corrupted measurements with a random subgraph of the existing graph, the initial measurement matrix $Z=(Z_{ij})$ is given by the following  model
\begin{equation}
Z_{ij} = \left\{
     \begin{array}{rll}
   z_i  z_j^{-1}  & \;\; \text{ for a correct edge}  		& \text{with probability } p \alpha 	\\
 - z_i  z_j^{-1}  & \;\; \text{ for a incorrect edge} 	& \text{with probability } (1-p) \alpha	\\
		     0   & \;\; \text{ for a missing edge}, 	& \text{with probability } 1-\alpha	\\
     \end{array}
   \right.
\label{SHORT_noiseModelZ2}
\end{equation}
whose expected value can be written as a rank-one matrix. We refer to $\eta = 1-p$  as the \textbf{noise level} in the data. Following the approach used in \cite{sync} and using recent results on the largest eigenvalue of rank-one deformation matrices, one can give a similar analysis for the robustness to noise of the eigenvalue method for synchronization over  $\mathbb{Z}_2$ (see Appendix A), by decomposing the given matrix $Z$ into a sum of a rank-one matrix and a random matrix R corresponding to the noise in the measurement graph.

\begin{figure}[h]
\begin{center}
\subfigure[$p= 55\%, \tau=12\%$]{\includegraphics[width=0.24\columnwidth]{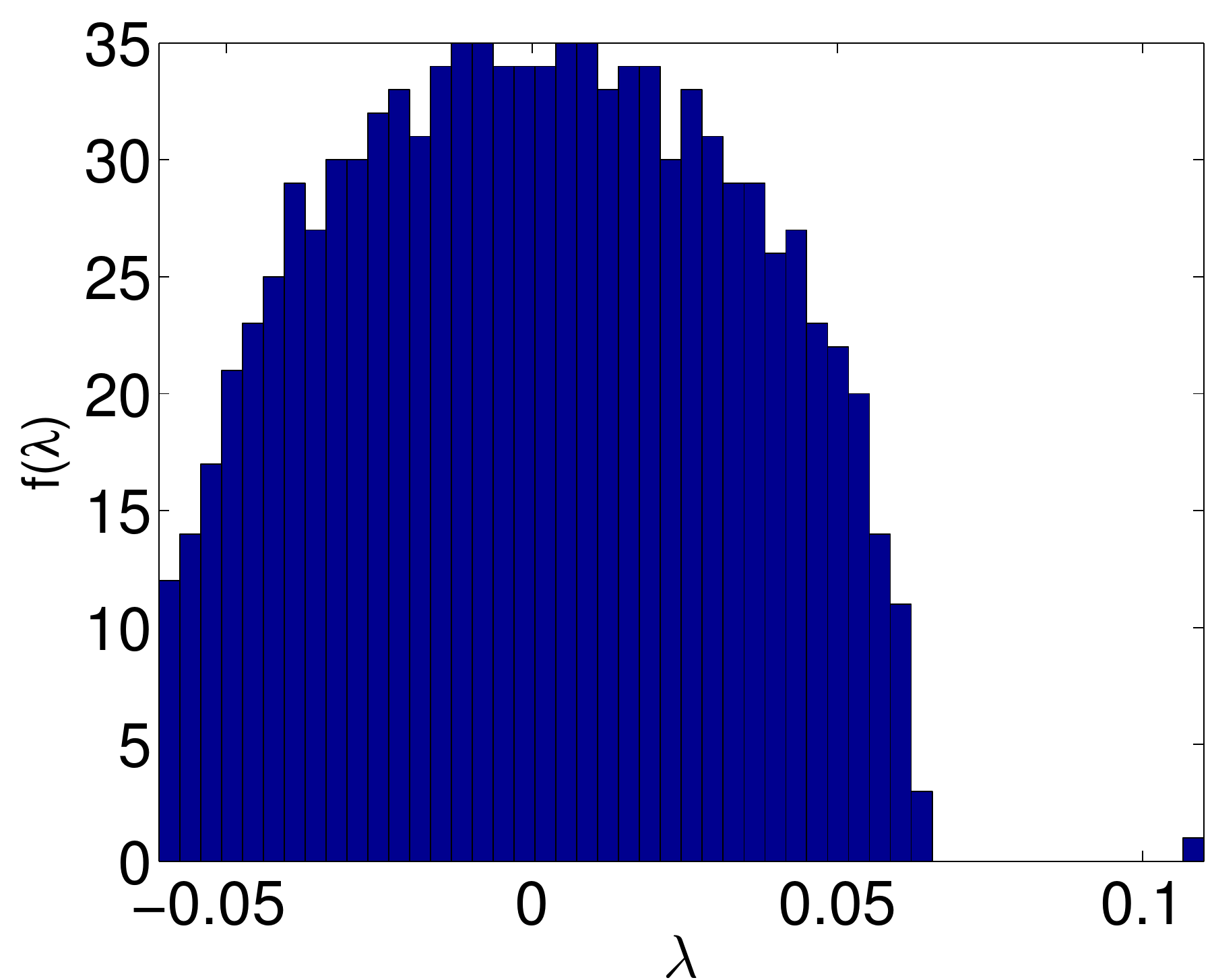}}
\subfigure[$p= 52.5\%, \tau= 44\%$]{\includegraphics[width=0.24\columnwidth]{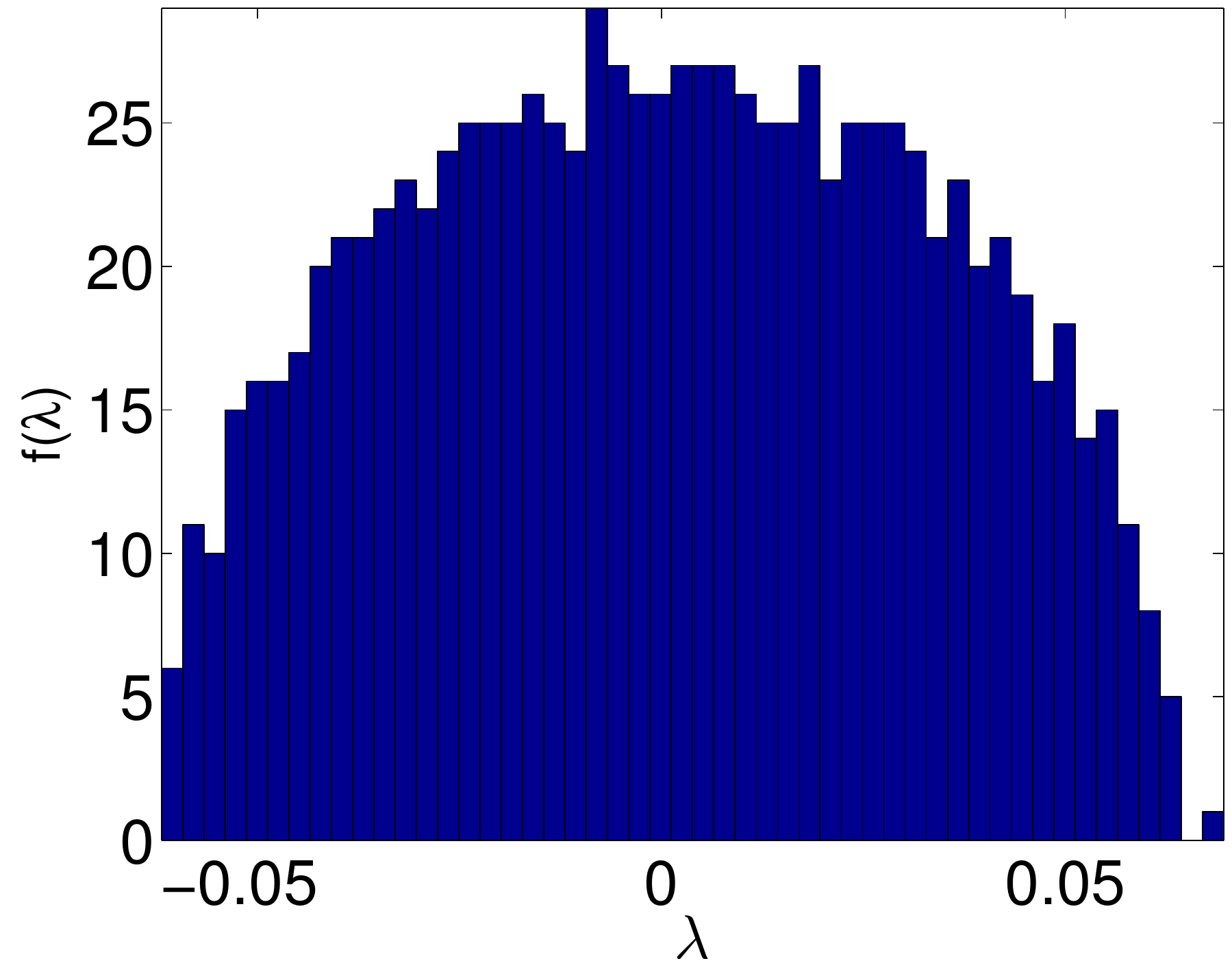}}
\subfigure[$p= 51.4\%, \tau= 44\%$]{\includegraphics[width=0.24\columnwidth]{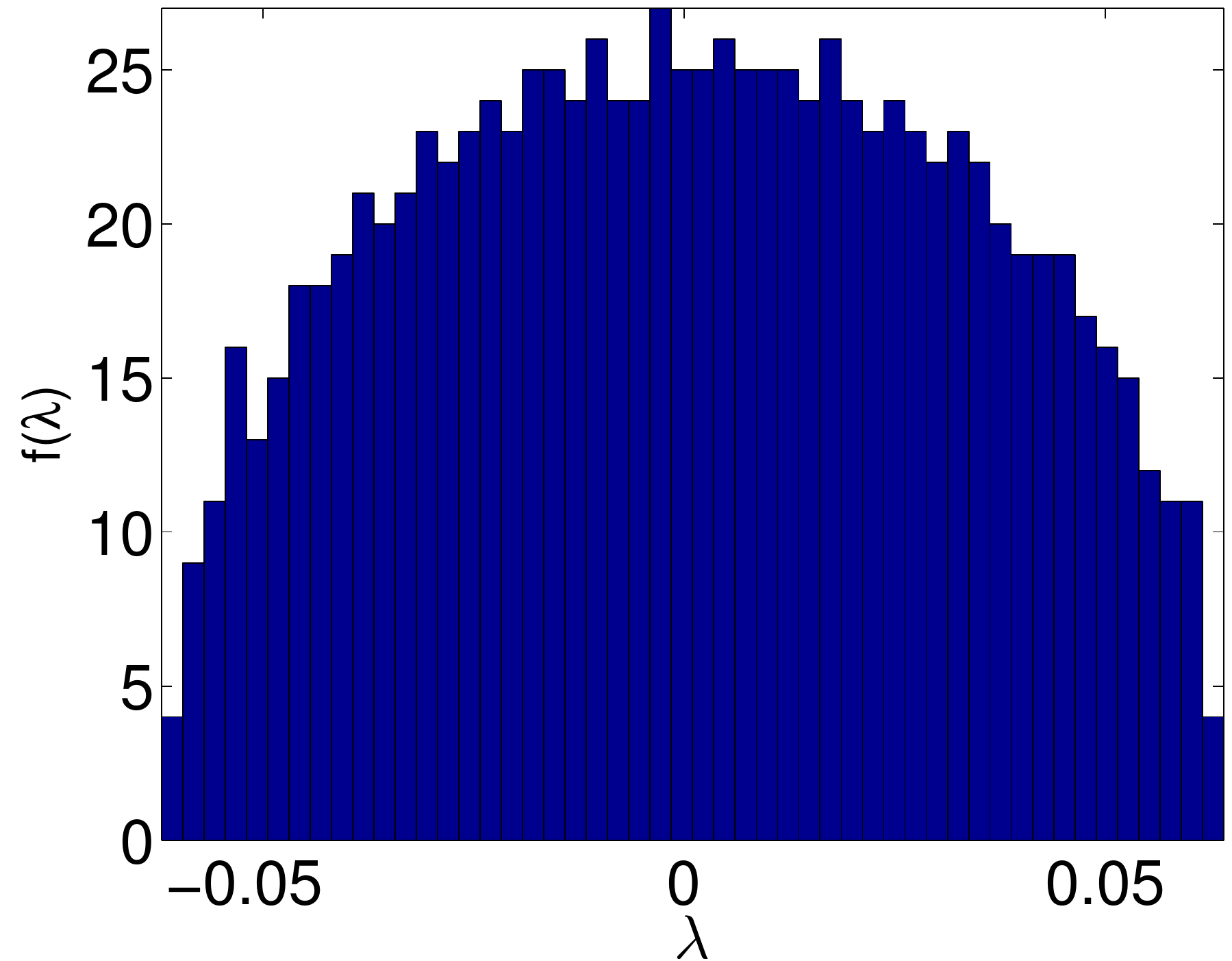}}
\subfigure[$p= 50\%, \tau= 49 \% $]{\includegraphics[width=0.24\columnwidth]{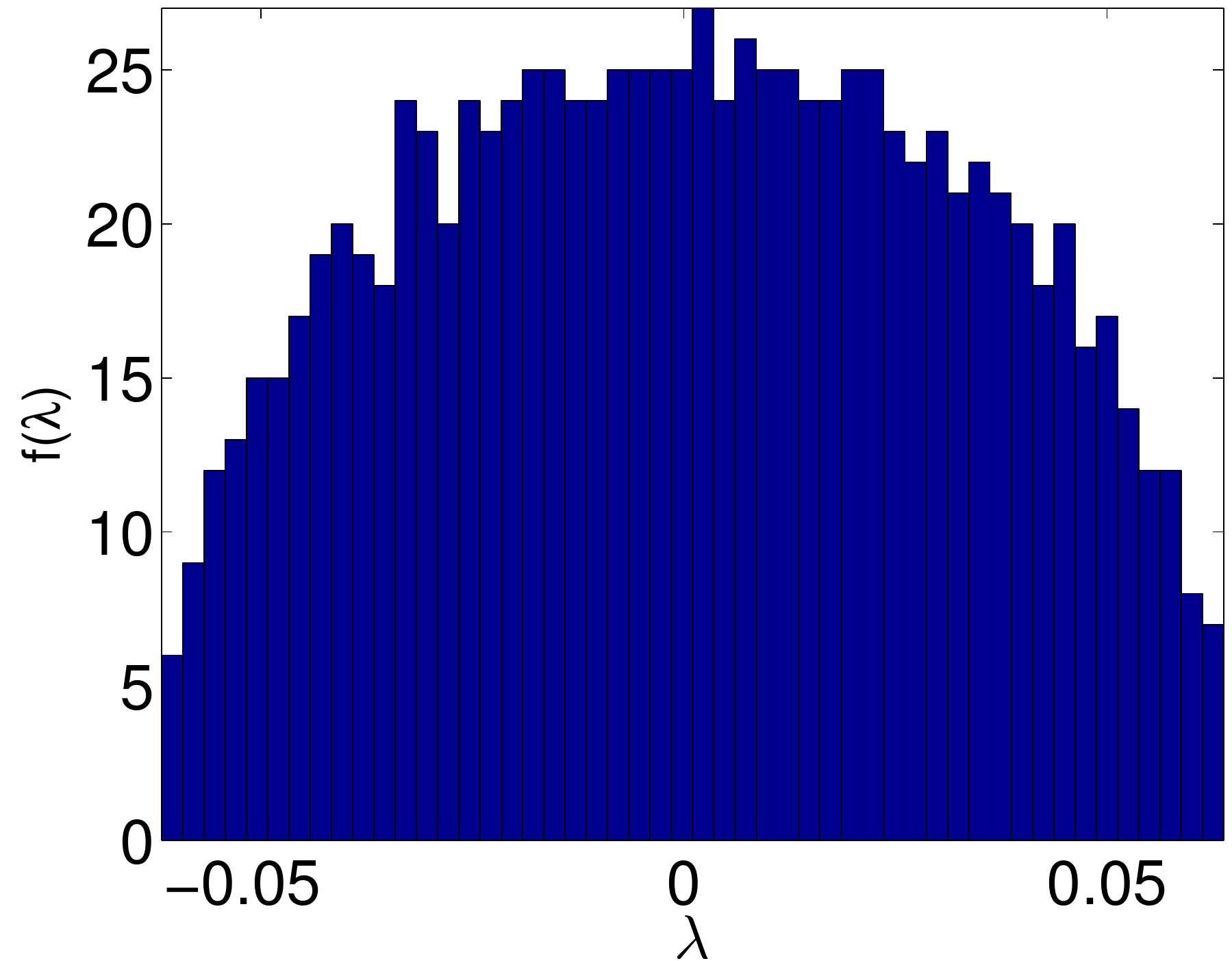}}
\end{center}
\caption{Histograms of the eigenvalues of the complete matrix $Z$ of size $n=1000$ of all pairwise measurements, for different values of $p$, and the associated error levels $\tau$.}
\label{fig:spectra_Compl_ErdosR_short}
\end{figure}

\begin{figure}[h]
\begin{center}
\includegraphics[width=0.48\columnwidth]{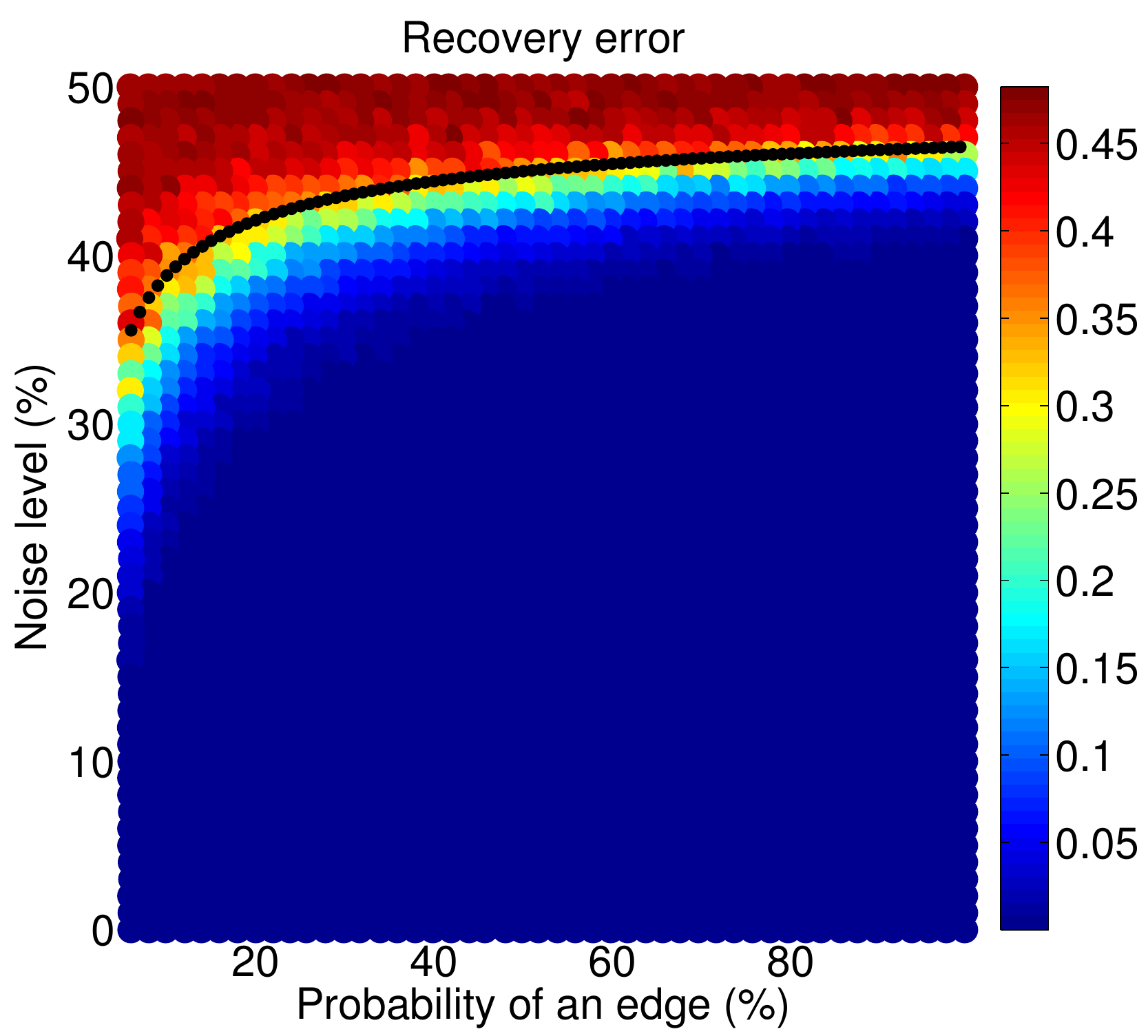}
\includegraphics[width=0.48\columnwidth]{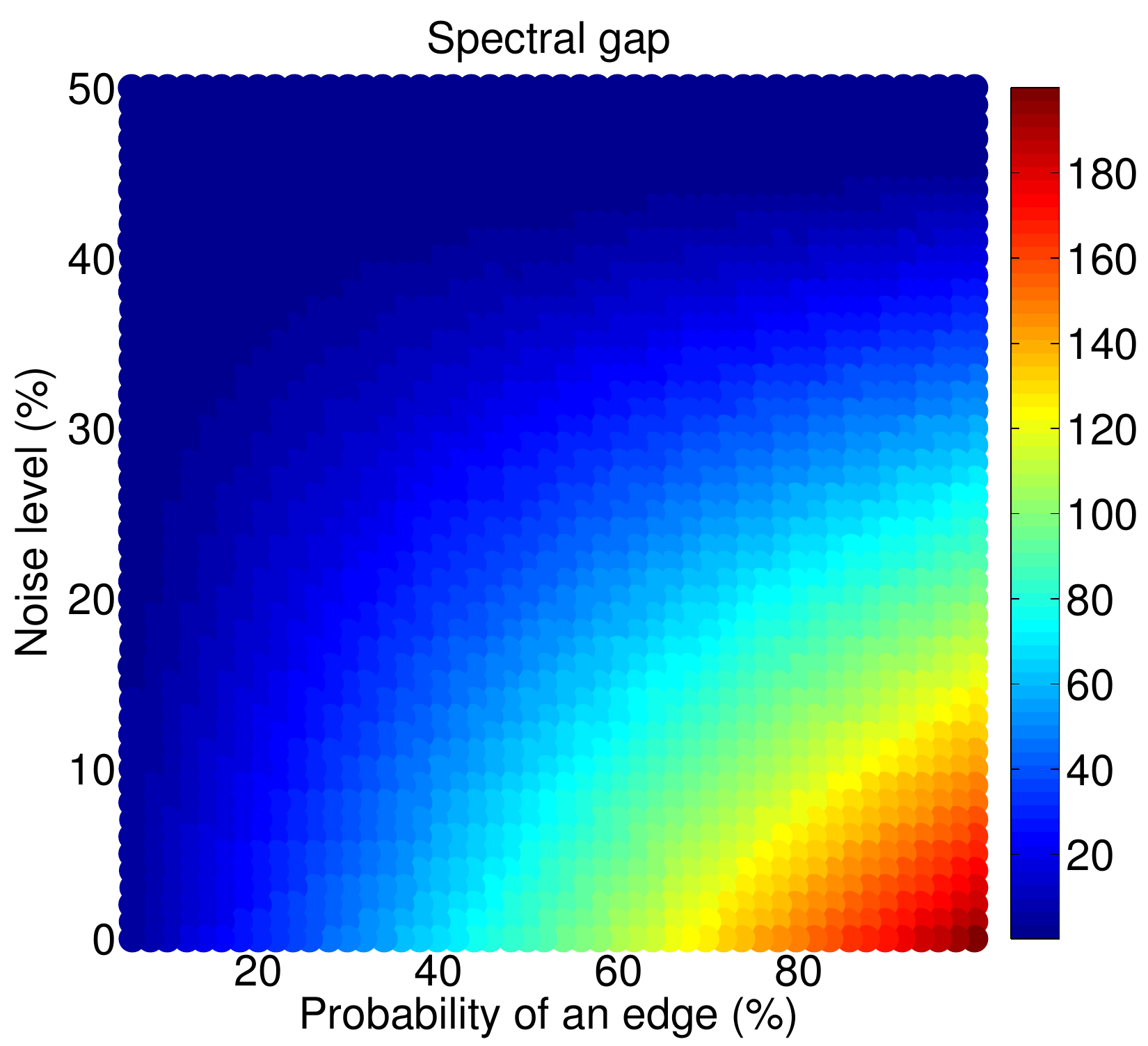}
\end{center}
\caption{Left: heat map showing the error level $\tau$ as we vary the noise level $\eta$
and the edge probability $\alpha$, based on the original matrix $Z$ (without any normalization). The black curve corresponds to the threshold values given by equation (\ref{condpTh}). Right: heat map showing the spectral gap $\lambda_1^{(Z)} - \lambda_2^{(Z)}$, as we vary $\eta$ and $\alpha$.
}
\label{fig:Zspectrumanalysis}
\end{figure}

\begin{figure}[h]
\begin{center}
\includegraphics[width=0.48\columnwidth]{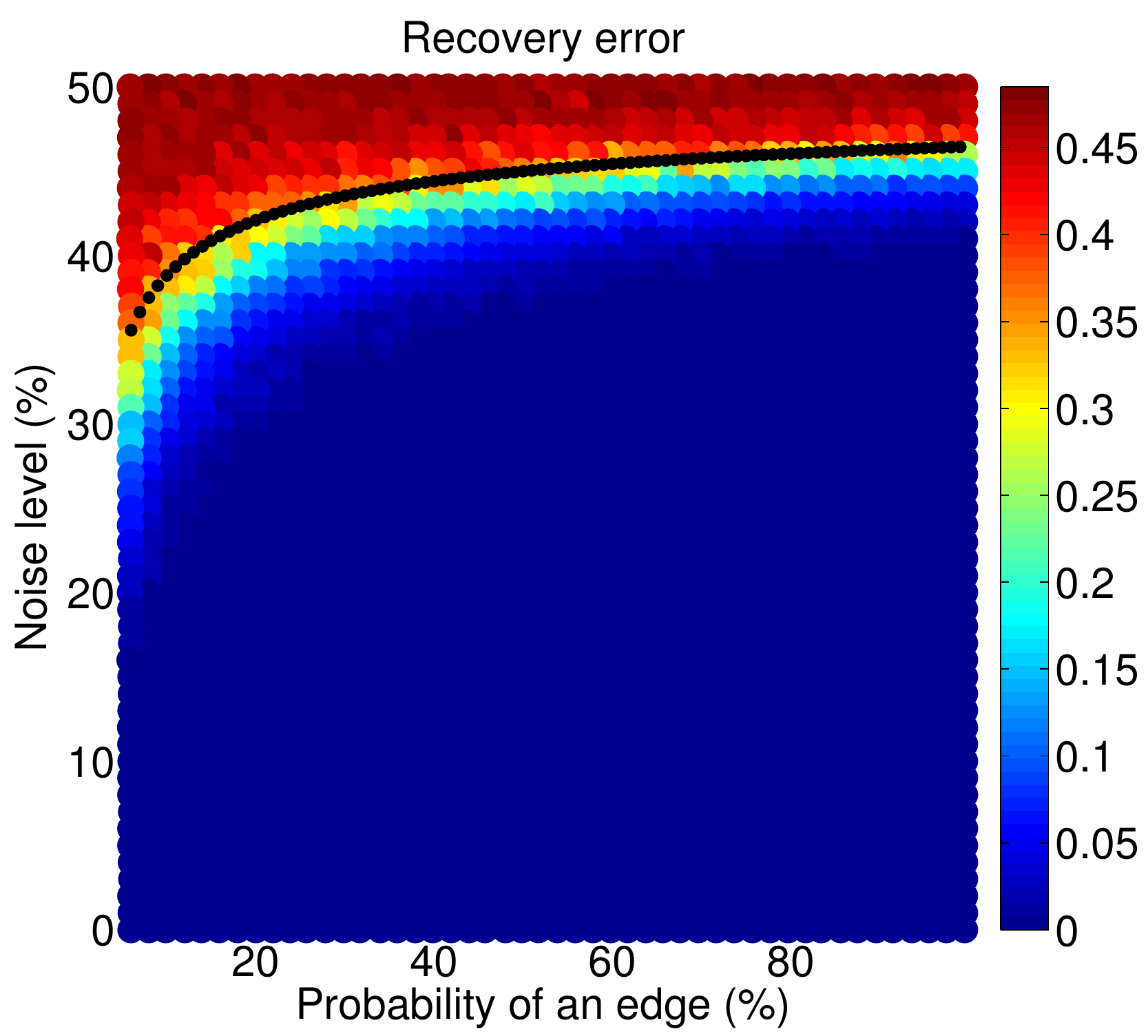}
\includegraphics[width=0.48\columnwidth]{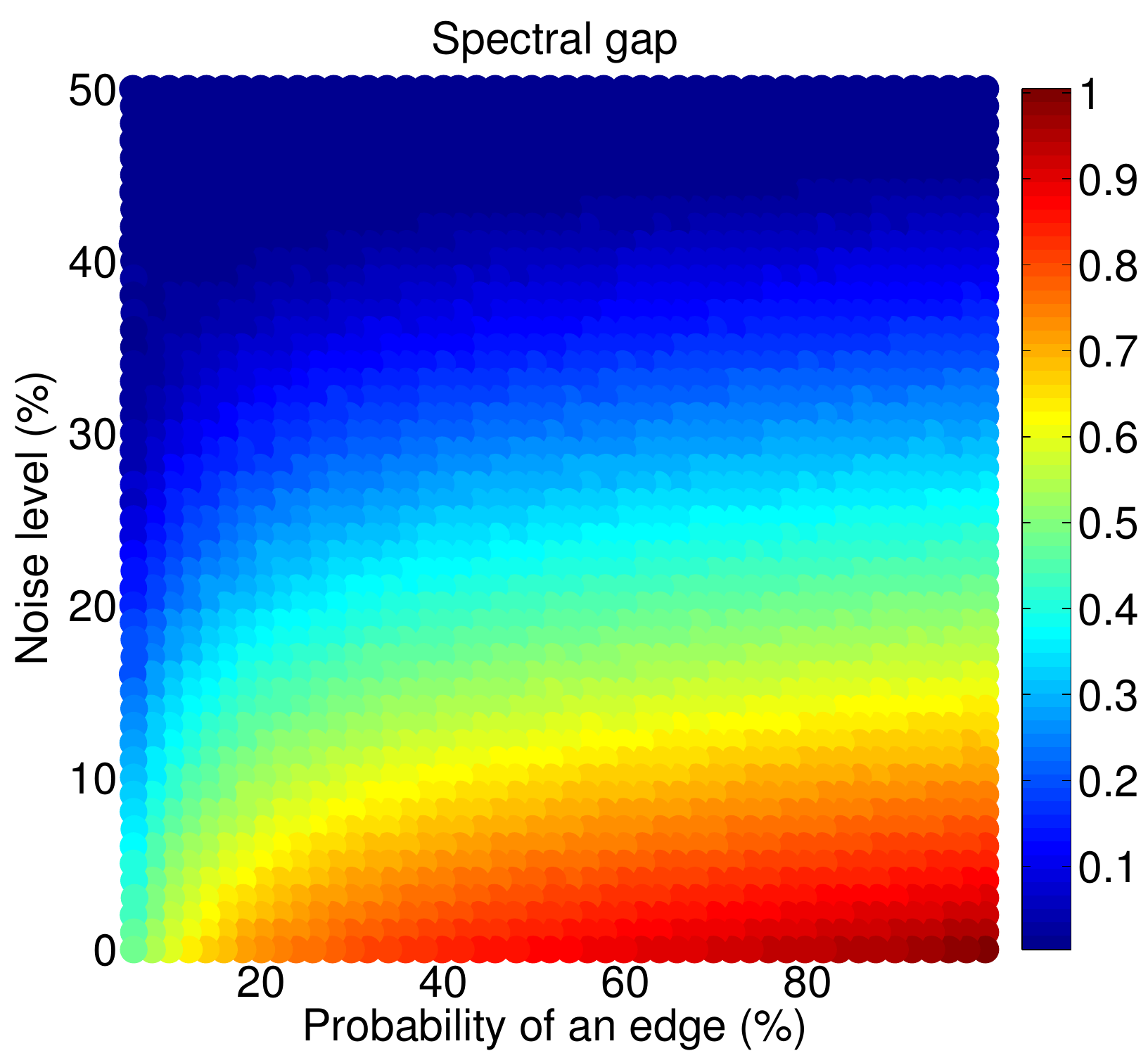}
\end{center}
\caption{
Left: heat map showing the error level $\tau$ as we vary the noise level $\eta$ and the edge probability $\alpha$, using the normalized matrix $\mathcal{Z}$. Right: heat map showing the spectral gap $\lambda_1^{(\mathcal{Z})} - \lambda_2^{(\mathcal{Z})}$, as we vary $\eta$ and $\alpha$.
}
\label{fig:tildaZ_spectrumanalysis}
\end{figure}

In Figure \ref{fig:spectra_Compl_ErdosR_short}, we show histograms of the eigenvalues of the measurement matrix $Z$  of size $n = 1000$ for the complete graph case (thus $\alpha = 1$), as well as the resulting error levels $\tau$, for different values of the noise level $p$ in the data. 
The additional numerical simulations shown in Figures \ref{fig:Zspectrumanalysis} and \ref{fig:tildaZ_spectrumanalysis} further confirm the results from Appendix A. We denote the error level by $\tau$, i.e., the percentage of nodes whose sign has been estimated incorrectly. The heat map in the left plot of Figure \ref{fig:Zspectrumanalysis} shows the error rates for the eigenvector synchronization algorithm based on the original matrix $Z$ of pairwise measurements, as we vary the edge probability $\alpha$ on the $x$-axis, and the noise level $p$ on the $y$-axis.  The intensity of the color denotes the error level $\tau$, with dark blue indicating a perfect recovery, and dark red indicating an erroneous solution that is very close to random. We measure accuracy by the percentage of nodes correctly classified. On the right plot of the same Figure \ref{fig:Zspectrumanalysis}, we plot the spectral gap $\lambda_1^{(Z)} - \lambda_2^{(Z)}$ as we vary the same two parameters, $\alpha$ and $p$. 
We show similar plots in Figure \ref{fig:tildaZ_spectrumanalysis}, but this time we use the normalized matrix $\mathcal{Z}=D^{-1}Z$. Note that while the recovery rates are very similar, the heat maps showing the corresponding spectral gaps differ significantly. Therefore, for the random Erd\H{o}s-R\'{e}nyi graph model, the lack of a large spectral gap in the former case does not lead to an erroneous solution, and we are still able to recover the rank-one structure. However, we refer the reader to Section 3.2 in \cite{asap2d}, for a detailed explanation of the additional noise robustness the above normalization brings for the case of large degree nodes in the graph, and its connection to the normalized discrete graph Laplacian.

\section{Synchronization by Message Passing (MPS)}  \label{sec:MPD}
This section introduces a 
message passing algorithm that solves the synchronization problem over $\mathbb{Z}_2$.
The algorithm we propose is reminiscent of the message passing algorithms responsible for some of the recent breakthroughs in the field of information theory, in particular error correcting codes \cite{RothCodingTh,GallagerThesis}. Long after Shannon's theorems from the 1950s, computationally efficient codes have been found which approach Shannon's theoretical limit, with turbo codes and low density parity check (LDPC) codes achieving unprecedented error-correcting performance, and at the same time providing a theoretical understanding of their good performance. One of the main ingredients of these decoding schemes are message passing algorithms, such as the popular belief propagation \cite{montanari,Pearl1988}.

The algorithm we propose in this section to solve the group synchronization problem is also inspired by the standard belief propagation algorithm for approximating marginal distributions for unobserved nodes conditioned on the observed nodes \cite{Pearl1988,Yedidia}. Our iterative algorithm relies on messages that are real numbers, which, for each node $x_i$, aggregate the beliefs of its neighbors to estimate the probability distribution of node $x_i$ taking value $-1$ or $1$. An analysis of the proposed algorithm is beyond the scope of this paper, and we merely aim to present an algorithm that is able to provide a robust solution to the synchronization problem even when additional 
constraints are imposed.

The synchronization problem SYNC($\mathbb{Z}_2$) can also be regarded as a decoding problem. The information $z_i, i=1,\ldots,n$ at the nodes of the graph is encoded in $m$ edges, in the form of pairwise measurements $e_{t} = z_i z_j^{-1}, (i,j) \in E$, $t = 1,\ldots,m$. The coded message, i.e., the set of edge measurements $e_1, \ldots, e_m$, has been corrupted by a noisy channel, and the problem is to infer the message that was initially transmitted, i.e., the (group) elements $z_i \in \mathbb{Z}_2, \forall i=1,\ldots,n$. In other words, we are trying to send $n$ bits of information encoded in a block message of $m$ bits. In the absence of noise, any $m=n-1$ edges forming a spanning tree of $G$ will provide an accurate solution. Once we start adding noise to the edge measurements, more and more edges (i.e., redundant information) will be required in order to recover a meaningful solution. In other words, $m$ will be greater than $n$ so as to provide redundancy that can be used to recover the original message from the errors induced by the noisy channel.

The rest of this section summarizes the message passing synchronization (MPS) algorithm we propose to solve SYNC($\mathbb{Z}_2$). At each iteration of the algorithm, we update the belief at each node on what its value should be, i.e., $+1$ or $-1$. We denote by $p_i^{+, r}$ (respectively, $p_i^{-, r}$) the probability that, at iteration $r$, the value of node $x_i$ is $+1$ (respectively, $-1$), and note that $p_i^{+, r} + p_i^{-, r}=1$.
At the same time, at each iteration we update the weight of each existing edge, to reflect the current level of confidence on the associated measurement  being correct. We denote by $w_{ij}^{+,r}$ (respectively, $w_{ij}^{-,r}$) the probability that, at iteration $r$, the edge $(i,j) \in E$ is correct, given that the value of the available measurement $Z_{ij}$ equals 1 (respectively, $-1$).

Let us first have a short digression on elementary probability, that will give some intuition for the messages passed along the graph at each iteration. Let $X_i$ and $X_j$ denote binomial random variables $\{\pm1\}$ associated to nodes $i$ and $j$, with success probabilities $p_i^{+}$ and $p_j^{+}$, i.e., $P[X_i=1] = p_i^{+}$ and $P[X_i=-1] = p_i^{-}  = 1 -  p_i^{+}$ and similarly for $X_j$. We let $\pi_{ij}=X_i X_j$, denote its associated success probability $P[\pi_{ij}=1]$ by  $\pi_{ij}^{+}$, and note that $\pi_{ij}^{+} = p_i^{+} p_j^{+} + p_i^{-} p_j^{-}$. To model the noise in the channel, we associate to every edge $(i,j) \in E$ a random variable $F_{ij}$
\begin{equation}
F_{ij} = \left\{
    \begin{array}{rl}
   1    & \;\; \text{ with probability  } p \\
  -1    & \;\; \text{ with probability  } \eta=1-p. \\
    \end{array}
  \right.
\end{equation}
In other words, a measurement $Z_{ij}$ is correct if and only if $F_{ij}=1$, which happens with probability $p$. We can now think of the observed measurement on edge $(i,j) \in E$ as a random variable $Z_{ij}$ given by $Z_{ij} = X_{i} X_{j} F_{ij} = \pi_{ij} F_{ij} $.
We use Bayes' Law to estimate the edge weights as follows
\begin{eqnarray}
w_{ij}^{+} \overset{def} = P[F_{ij} = 1 | Z_{ij}=1] &=& \frac{P[Z_{ij}=1 | F_{ij}=1] P[ F_{ij}=1] }{P[ Z_{ij}=1 ]}    \nonumber \\
 		         & = & \frac{ P[X_{i} Y_{j}=1] P[ F_{ij}=1] }{P[ Z_{ij}=1 ]}
\end{eqnarray}
where
\begin{eqnarray}
P[Z_{ij}=1]  &=& P[ Z_{ij}=1 | F_{ij}=1] P[F_{ij}=1] + P[Z_{ij}=1 | F_{ij}=-1] P[F_{ij}=-1]  \nonumber \\
		     &=& P[X_{i} X_{j}=1] P(F_{ij}=1) + P[ X_{i} X_{j}=-1] P(F_{ij}=-1) \nonumber \\
		     &=& (p_i^{+} p_j^{+} + p_i^{-} p_j^{-})p + (1-p_i^{+} p_j^{+} - p_i^{-} p_j^{-}) (1-p) =  
		     (p_i^{+} p_j^{+} + p_i^{-} p_j^{-})(2p-1) + 1-p.
\end{eqnarray}
Since we will be updating the edge weights at each iteration, we denote by  $w_{ij}^{+,r}$ and $w_{ij}^{-,r}$ the edge weights at iteration $r$. The update rules at round $r+1$ are given by
\begin{eqnarray}
w_{ij}^{+,r+1}  &=&
\frac { (p_i^{+,r} p_j^{+,r} + p_i^{-,r} p_j^{-,r}) p } { (p_i^{+} p_j^{+} + p_i^{-} p_j^{-})(2p-1) + 1-p } = \frac { \pi_{ij}^{+} p } {  \pi_{ij}^{+} (2p-1) + 1-p }, 
\label{defWij_plus}
\end{eqnarray}
for the positive edges (i.e., for which $Z_{ij}=1$), and similarly for the negative edges (i.e., for which $Z_{ij} = -1$ )
\begin{eqnarray}
w_{ij}^{-,r+1} = P[F_{ij} = 1 | Z_{ij}=-1] & = & 
\frac {(1-p_i^{+,r} p_j^{+,r} - p_i^{-,r} p_j^{-,r}) p} {1-      (p_i^{+} p_j^{+} + p_i^{-} p_j^{-})(2p-1) - 1 + p } = \frac{ (1 - \pi_{ij}^{+}) p}{ p - \pi_{ij}^{+} (2p-1) }.  \nonumber
\end{eqnarray}

Within the same iteration $r+1$, after updating the edge weights as shown above, we update the local beliefs at each node using the following rule
\begin{equation}
p_i^{+,r+1} \mapsto \sum_{j \sim i}  p_j^{+,r} w_{ij}^{+,r+1}  \mathbf{1}_{\{Z_{ij}=1\}} + \sum_{j \sim i}  p_j^{-,r} w_{ij}^{-,r+1} \mathbf{1}_{\{Z_{ij}=-1\}},
\label{nodeupdate}
\end{equation}
and similarly for $p_i^{-,r+1}$. In order to interpret the beliefs (i.e., weights) at each node as probabilities, we normalize them such that $p_i^{+,r+1} + p_i^{-,r+1} = 1$.  Note that the beliefs updated at each iteration of our proposed procedure are the edge beliefs $w_{ij}^{+,r} \in [0,1]$ in (\ref{defWij_plus}) (respectively $ w_{ij}^{-,r} \in [0,1] $) for very edge $(i,j)$ in the graph $G$, and the vertex beliefs  $p_i^{+,r} \in [0,1]$ in (\ref{nodeupdate}) (respectively, $p_i^{-,r} \in [0,1]$) corresponding only to the non-anchor nodes in the graph, since the beliefs of the anchor nodes remain unchanged throughout the iterative process.

The intuition for (\ref{nodeupdate}) is given by the following. If we observe that edge $(i,j)$ has value $Z_{ij} = +1$, then the contribution to the belief that node $i$ has value $+1$ is given by the probability that node $j$ has value $+1$ (i.e., $p_j^{+,r}$) weighted by the confidence that edge $(i,j)$ is correct. Similarly, if we observe a measurement $(i,j)$ with value $ Z_{ij} = -1$, then the contribution to the belief that node $i$ has value $+1$ is the probability that node $j$ has value $-1$ weighted by the confidence that edge is correct. We denote as \textit{anchors}  the nodes whose associated group element (i.e., sign) is known a-priori and note that one can easily incorporate anchor information in the above MPS algorithm. 
We initialize the algorithm (at iteration 0) as follows, for a non-anchor node $i$ we set $p_i^{+,0}=p_i^{-,0}=0.5$, while for an anchor node $i$ we set $p_i^{+,0}=1$ if $a_i=1$, or $ p_i^{-,0}=1$ if $a_i=-1$ (and enforce such constraints given by the anchors at all iterations, i.e., for anchor nodes, the probabilities $p_i^{+,r},p_i^{-,r}$ remain unchanged throughout the iterations). In other words, our proposed message passing algorithm propagates across the network the information given by the anchors. Finally, note that when there is no anchor information in the network, we simply pick a single node at random (the \textit{root} node), fix its sign and consider it an anchor (in this case the final solution will be given up to a global sign change, as mentioned in Section \ref{sec:sync}).

\begin{figure}[h]
\begin{center}
\subfigure[$ $]{\includegraphics[width=0.235\columnwidth]{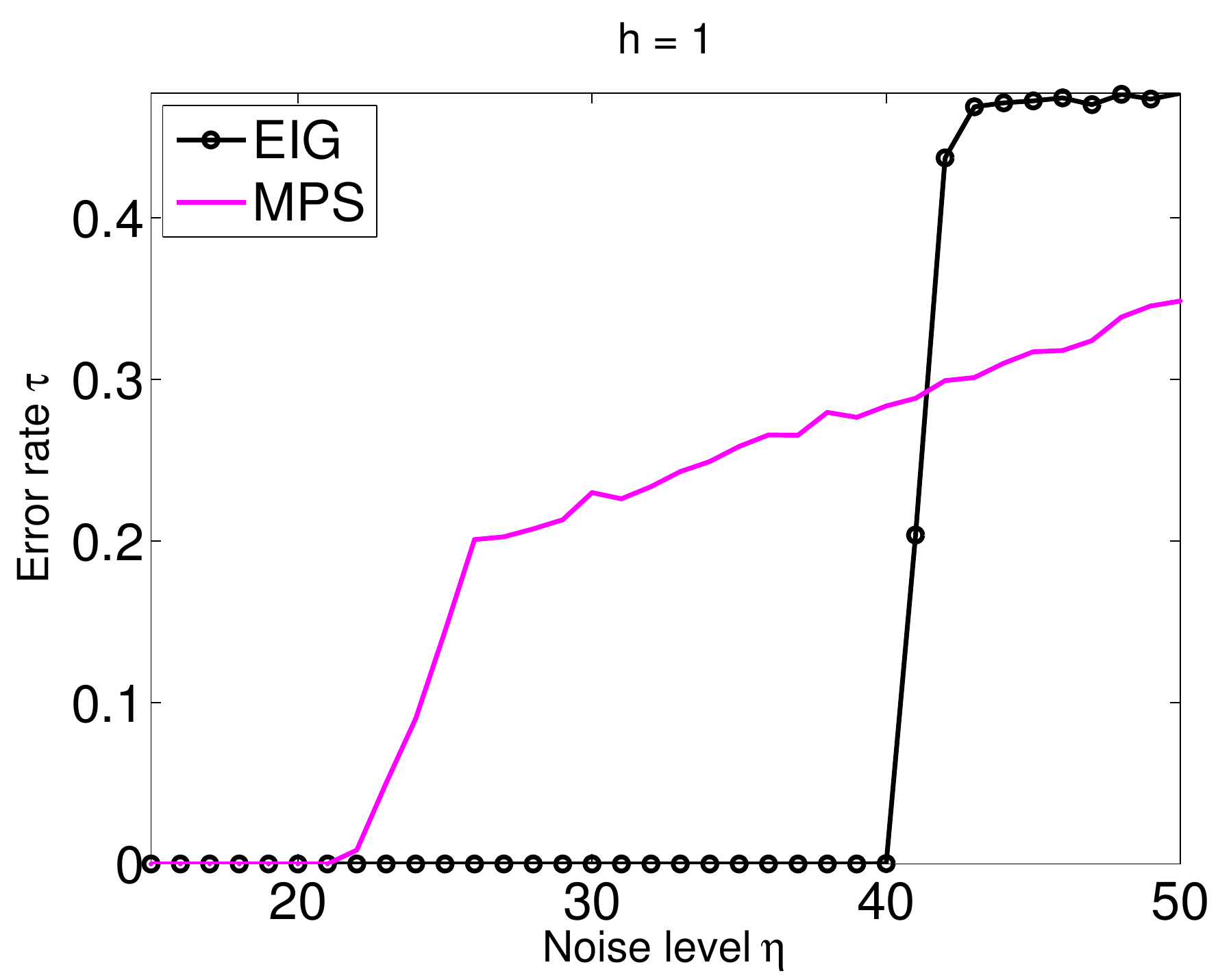}}
\subfigure[$ $]{\includegraphics[width=0.235\columnwidth]{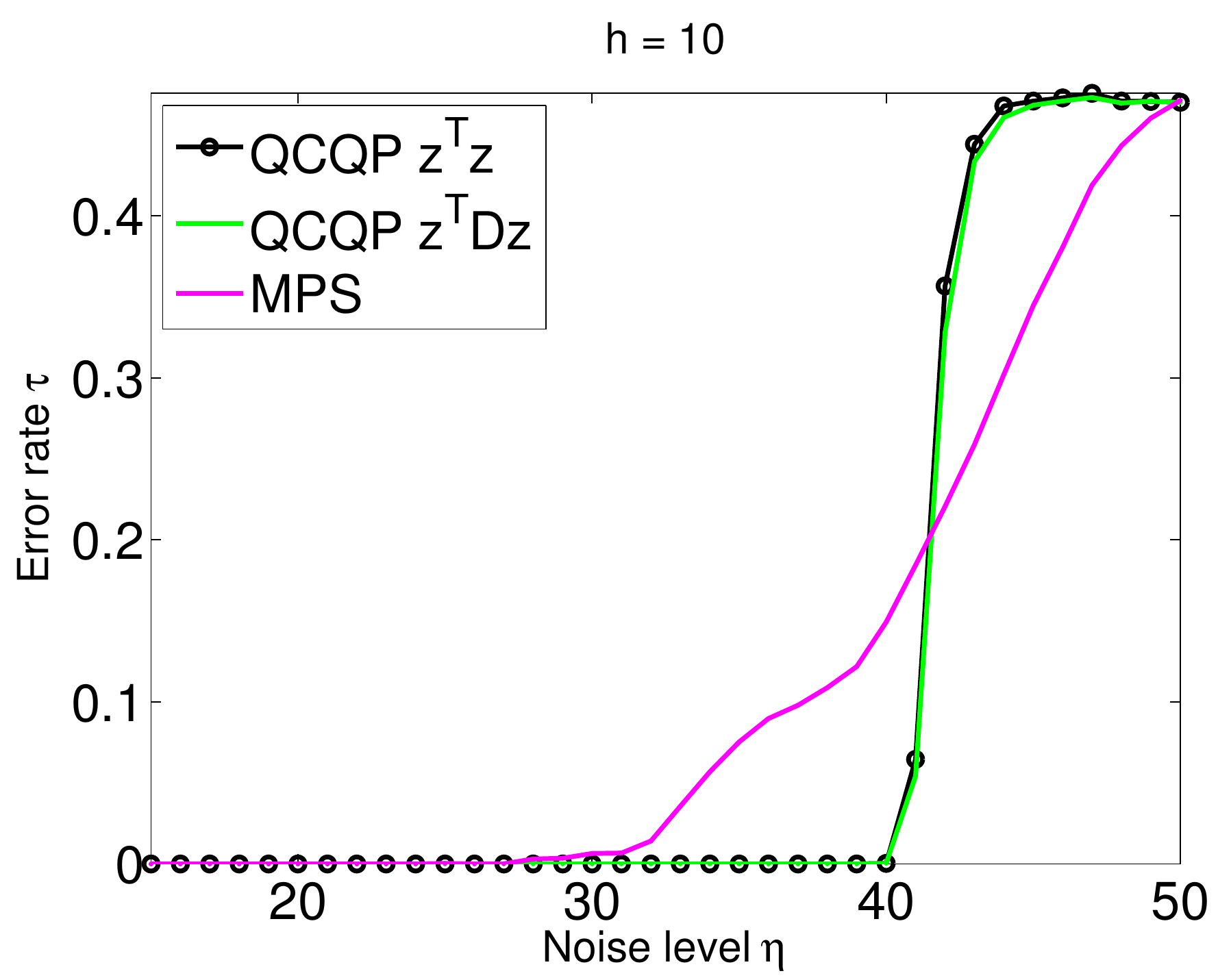}}
\subfigure[$ $]{\includegraphics[width=0.235\columnwidth]{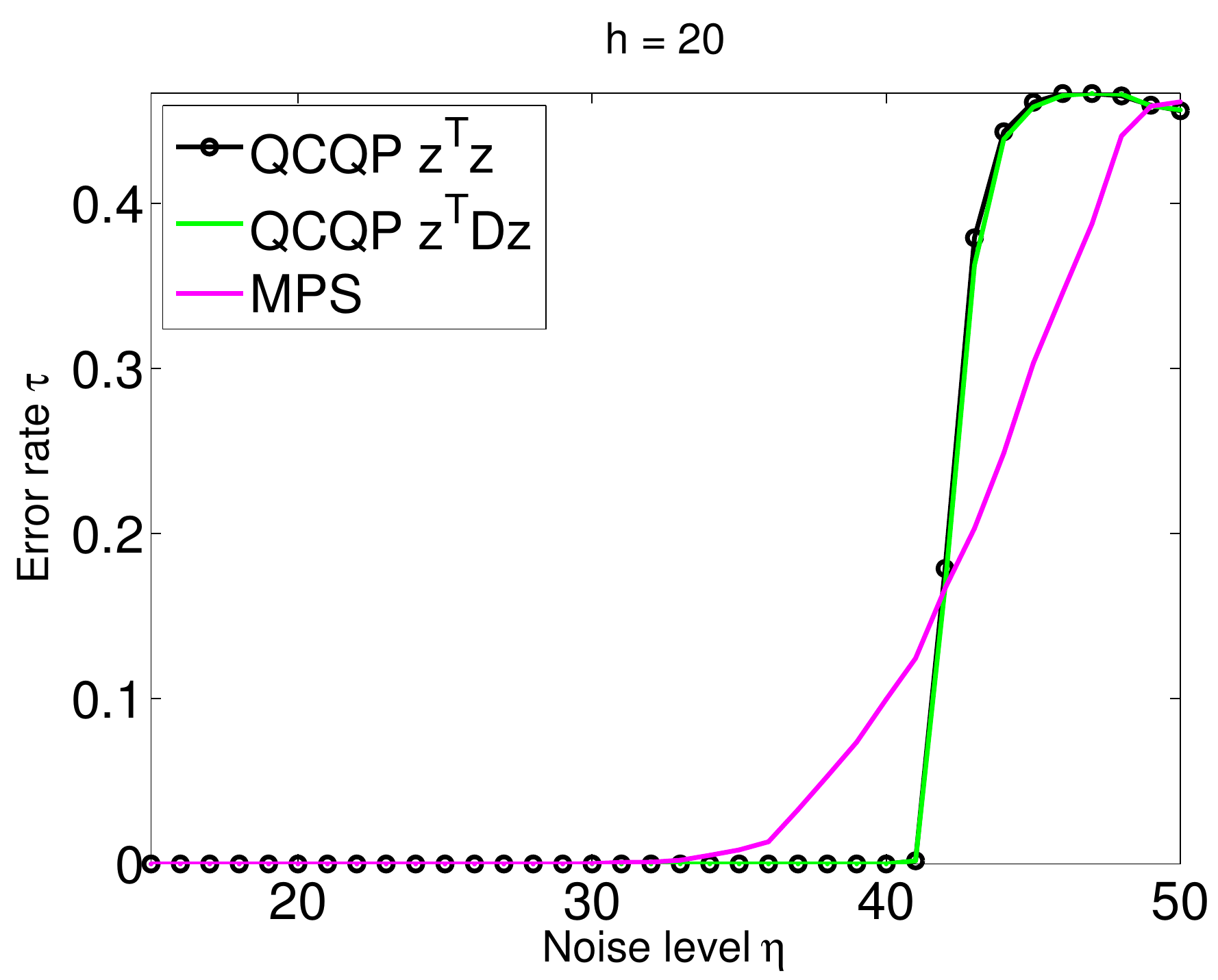}}
\subfigure[$ $]{\includegraphics[width=0.235\columnwidth]{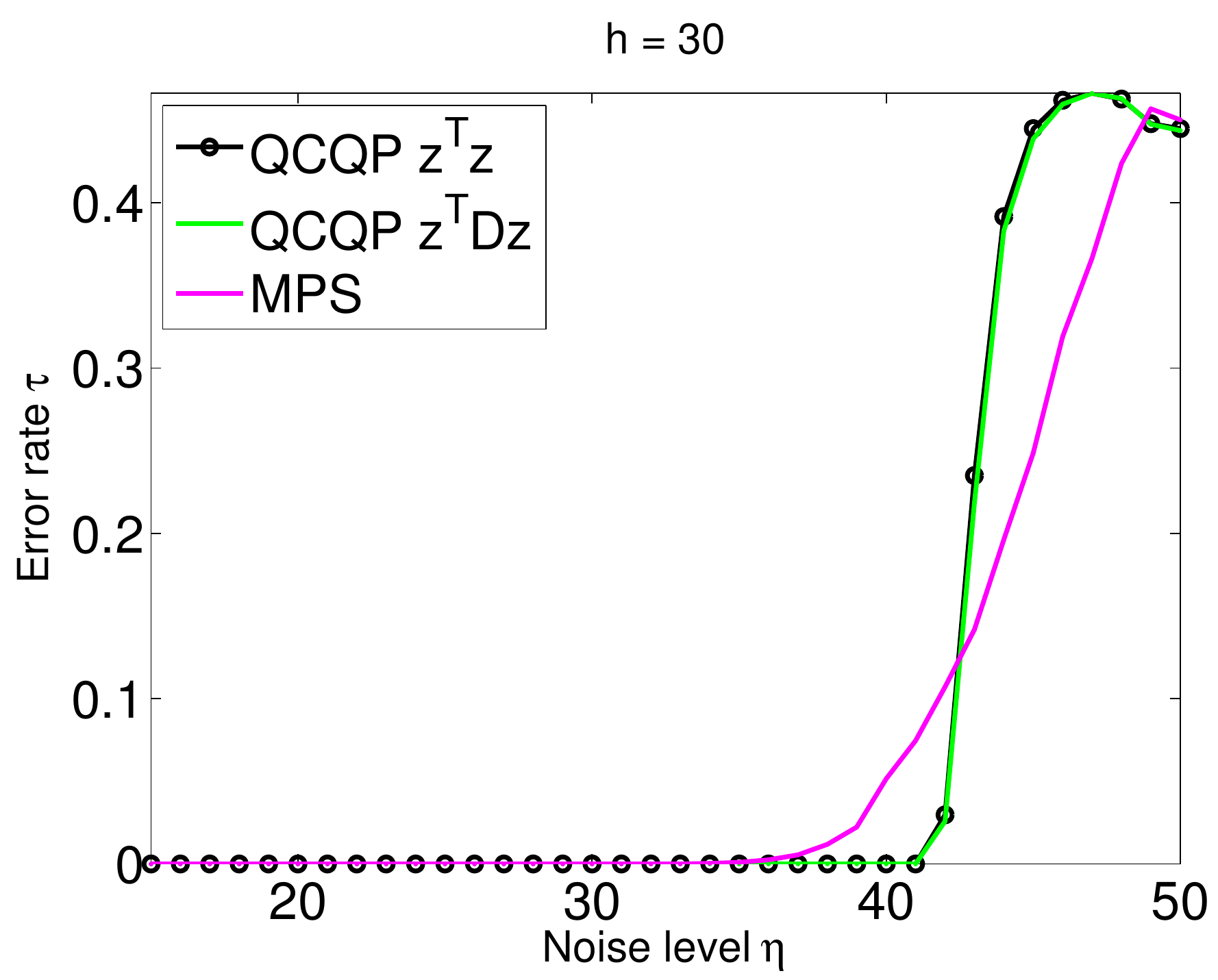}}
\end{center}
\caption{ Comparison of the eigenvector/QCQP and MPS algorithms when $G$ is the complete graph on $n$ vertices $G=K_{100}$, and the underlying subgraph $H$ of \textit{bad} edges (i.e., noisy edges) is approximately a regular graph of degree $d$ (x-axis). We vary the number of anchors $h$ in the range $\{1,10,20,30\}$, and average the result of each experiment over 100 runs.
}
\label{fig:Gcomp_HReg}
\end{figure}

We tested the above MPS algorithm for three different families of graphs and noise models.
In the first model shown in Figure \ref{fig:Gcomp_HReg}, we plot the recovery rates for the case when $G$ is the complete graph on $n=100$ vertices $G=K_{100}$, and the subgraph $H$ of  \textit{bad} edges, i.e., edges with incorrect measurements, is (approximately) a regular graph of degree $d$. 
The probability of an erroneous measurement is given by 
$\eta=1-p = \frac{d}{n-1}$, since every node has approximately $d$ bad incident edges out of a total of $n-1$ incident edges. Thus, we only vary $d$ in the range $[15,50]$ that would correspond to a noise level $\eta$ in the range $[15\%,50\%]$. The $x$-axis in Figure \ref{fig:Gcomp_HReg} represents the degree $d$ of the bad graph $H$, and the $y$-axis is the recovery error $\tau$, i.e., the percentage of nodes whose sign was incorrectly estimated.

We vary the number of anchors in the range $ h=\{1,10,20,30\}$, and remark that the number of anchors has an impact on the performance of any synchronization algorithm since anchors carry already known accurate information.
We compare the results of the MPS algorithm with those achieved by the eigenvector synchronization methods (when $h=1$) and the two versions of the quadratically constrained quadratic program (QCQP) based approach for synchronization with anchors (when $h > 1$). We refer the reader to Appendix B for a review of the QCQP relaxations which are able to incorporate anchor information. 
The results of this first experiment, averaged over $100$ runs, indicate that the eigenvector method clearly outperforms the MPS algorithm for low levels of noise (especially for a small number of anchors), but the MPS algorithm consistently returns more accurate solutions for very high levels of noise. Note that the marginal advantage one method has over the other, in the lower or higher end of the noise spectrum, diminishes as the number of anchors increases.

\begin{figure}[h]
\begin{center}
\includegraphics[width=0.235\columnwidth]{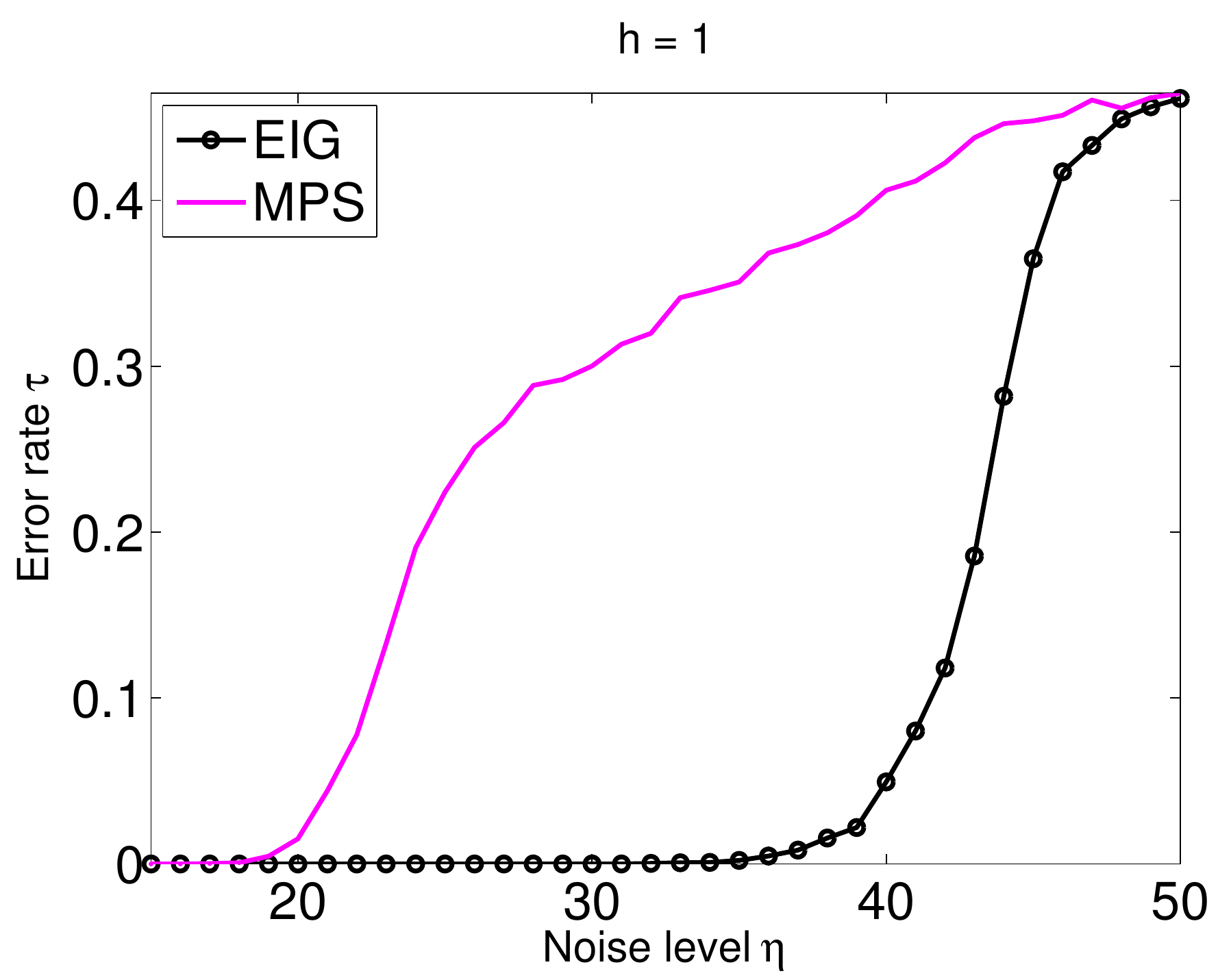}
\includegraphics[width=0.235\columnwidth]{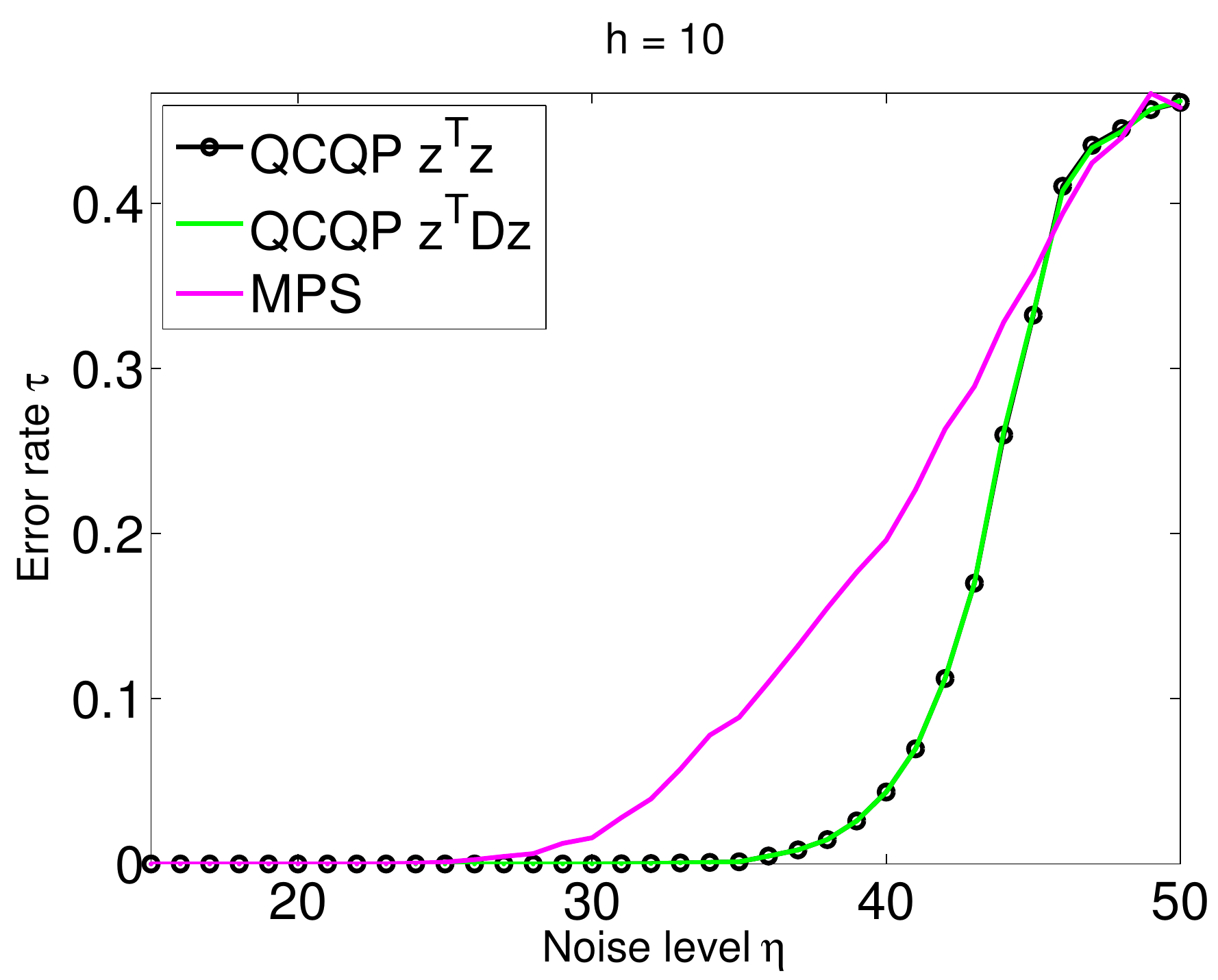}
\includegraphics[width=0.235\columnwidth]{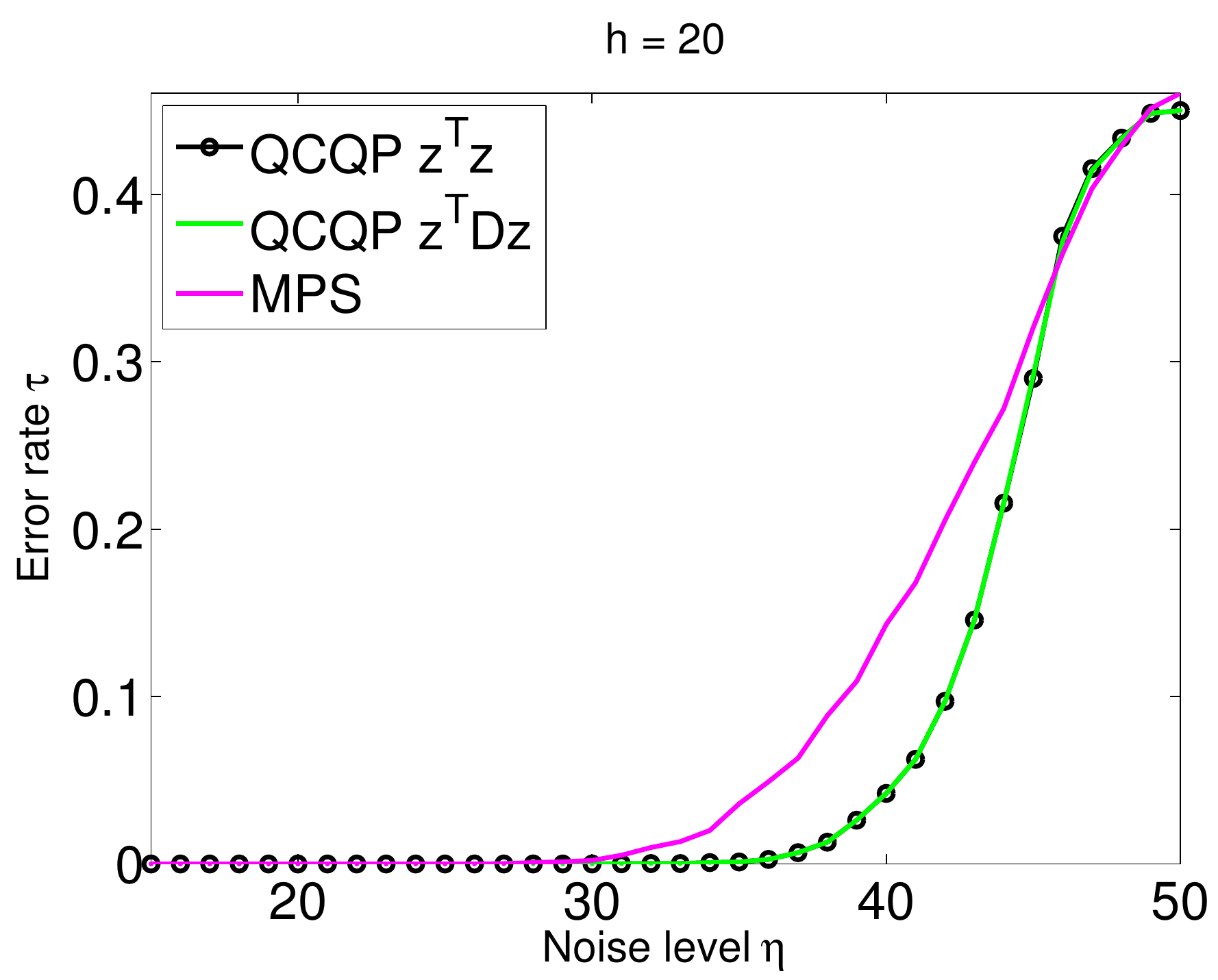}
\includegraphics[width=0.235\columnwidth]{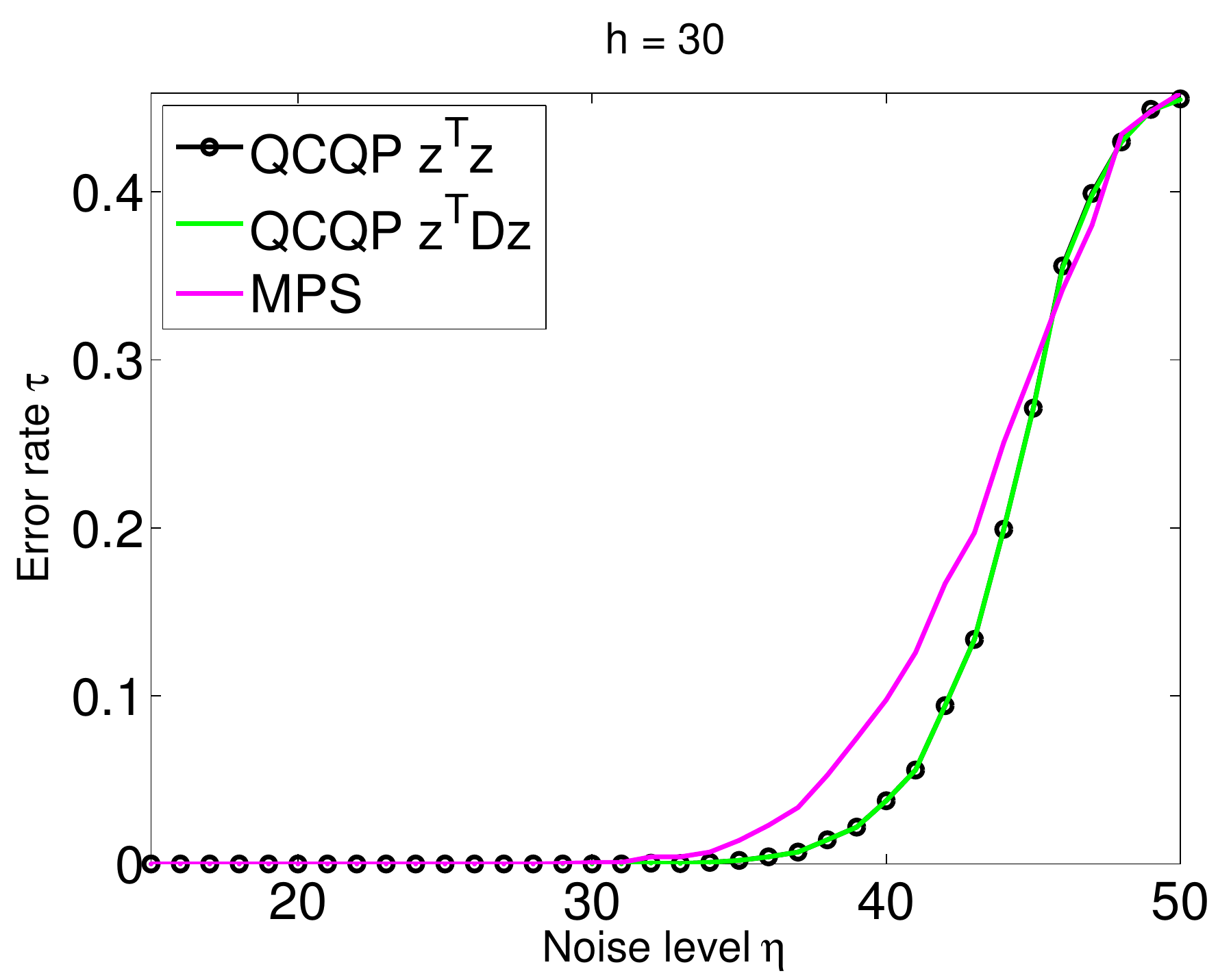}
\end{center}
\caption{ Comparison of the eigenvector/QCQP and MPS algorithms when $G$ is the complete graph on $n=100$ nodes $G = K_{100}$, and the underlying subgraph of noisy edges $H$ is an Erd\H{o}s-R\'{e}nyi random graph $G(n,\alpha)$ of expected degree $d$, where 
$ \alpha = \frac{d}{n-1}$. We vary the bad degree $d$ ($x-$axis), experiment with different number of anchors $h=\{1,10,20,30\}$, and average the result of each experiment over 100 runs.
}
\label{fig:Gcomp_HRandom}
\end{figure}

In Figure \ref{fig:Gcomp_HRandom}, we perform a similar experiment for the case when $G=K_n$ again, but this time the subgraph $H$ of \textit{bad} edges is a random Erd\H{o}s-R\'{e}nyi graph. In this noise model, the performance of the MPS algorithm is strictly inferior to that of the eigenvector methods and QCQP formulations, across all levels of noise in the measurements and all numbers of anchors.

\begin{figure}[h]
\begin{center}
\subfigure[$ m_{pa}=10$]{
\includegraphics[width=0.235\columnwidth]{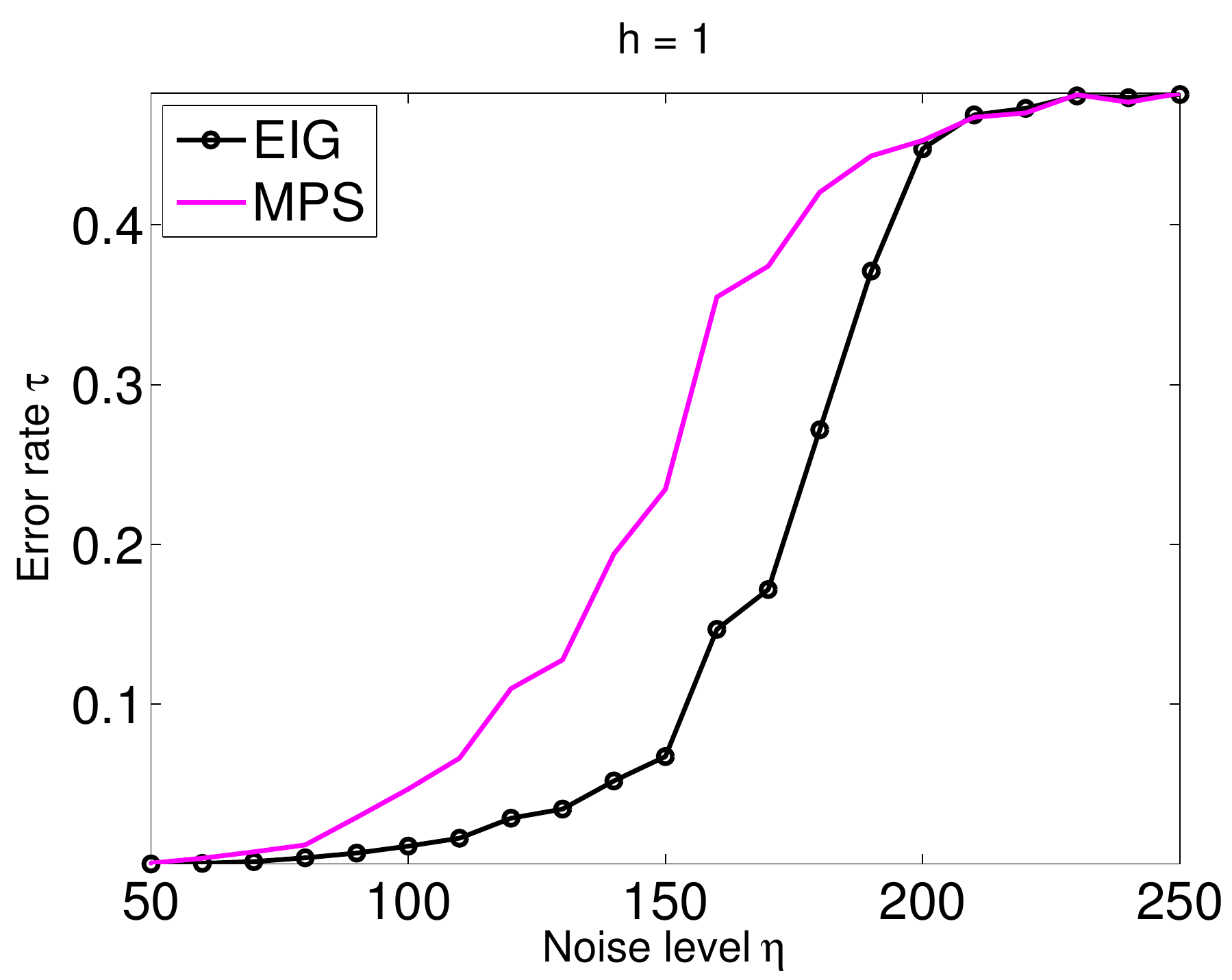}
\includegraphics[width=0.235\columnwidth]{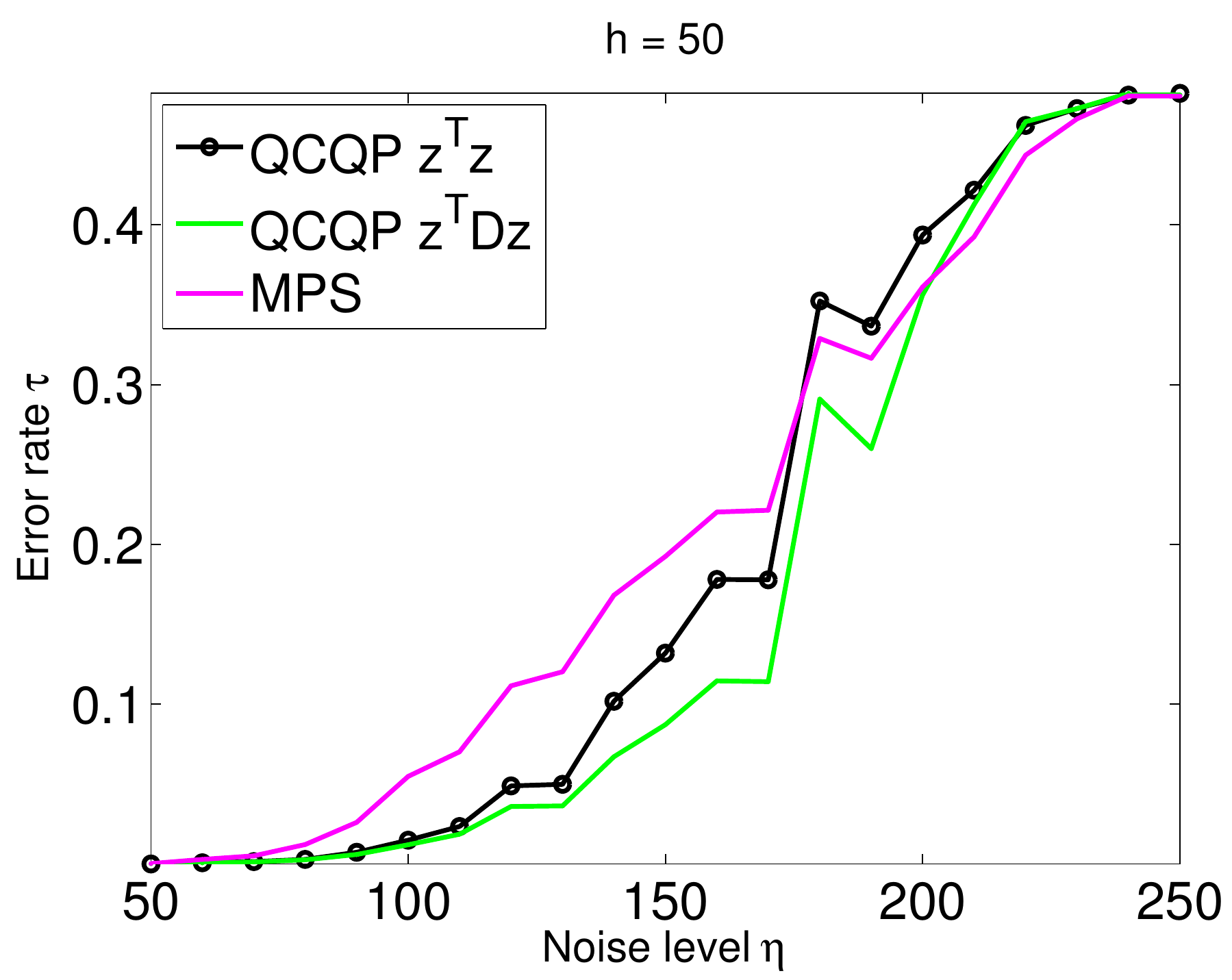}
\includegraphics[width=0.235\columnwidth]{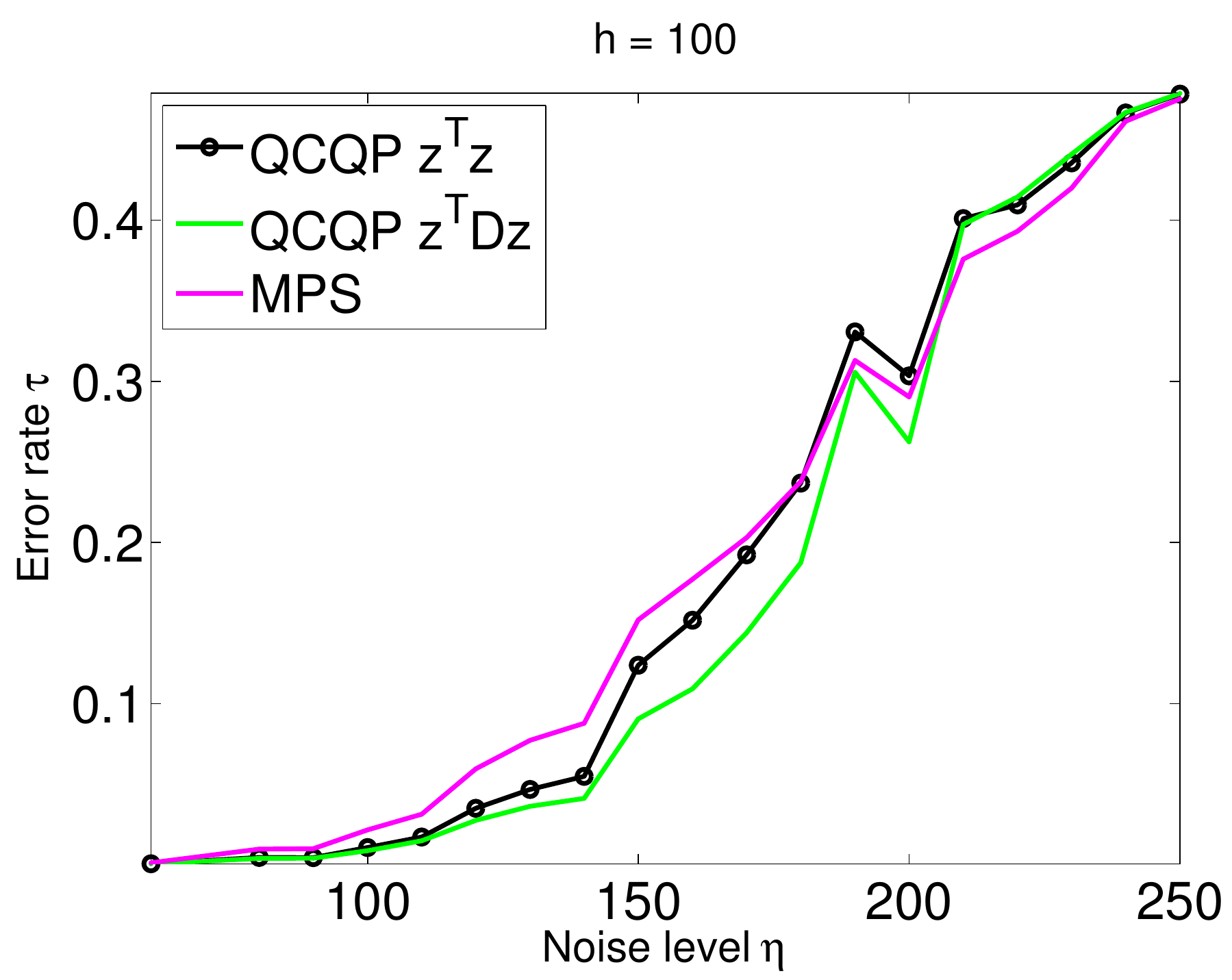}
\includegraphics[width=0.235\columnwidth]{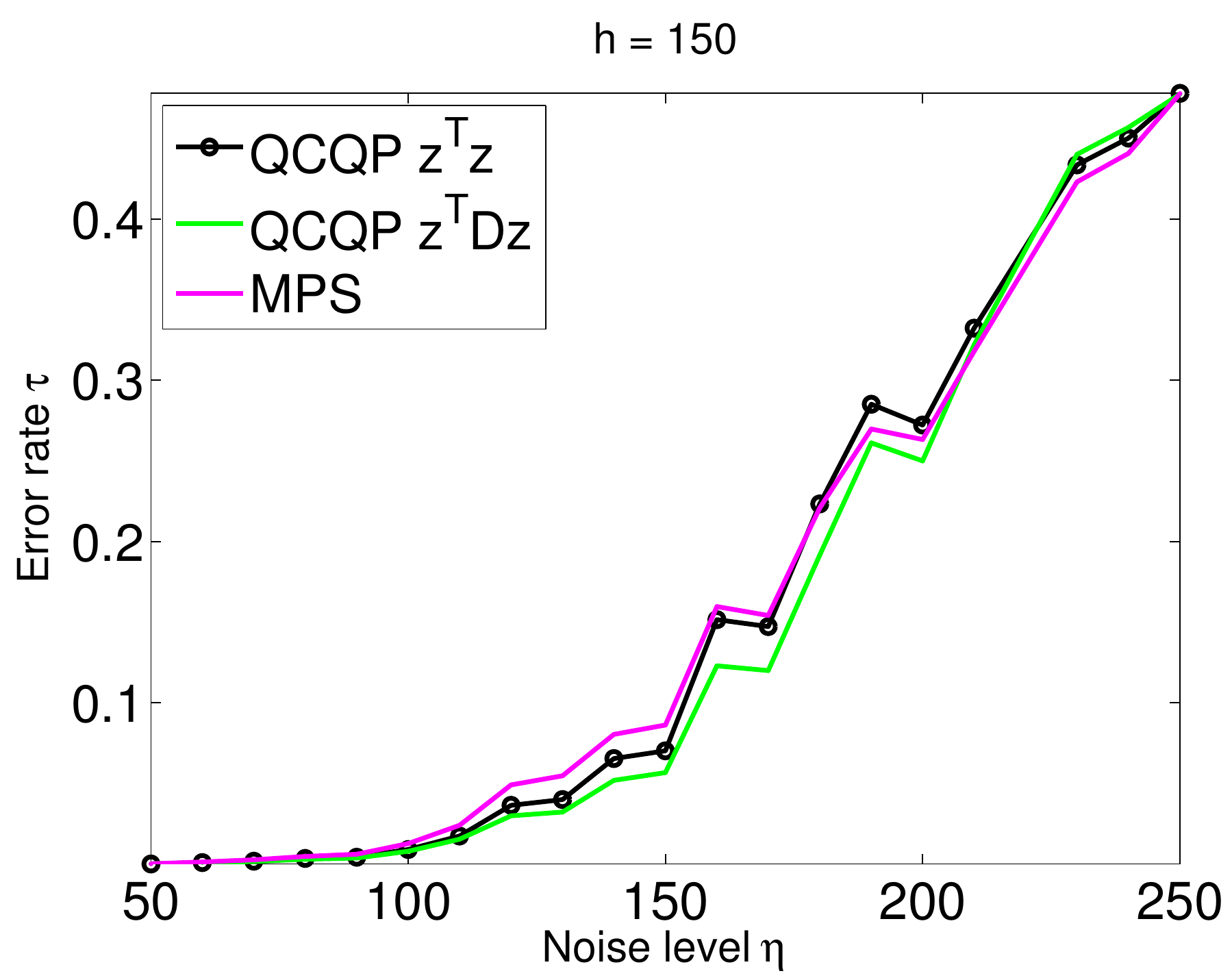}
}
\subfigure[$ m_{pa}=30$]{
\includegraphics[width=0.235\columnwidth]{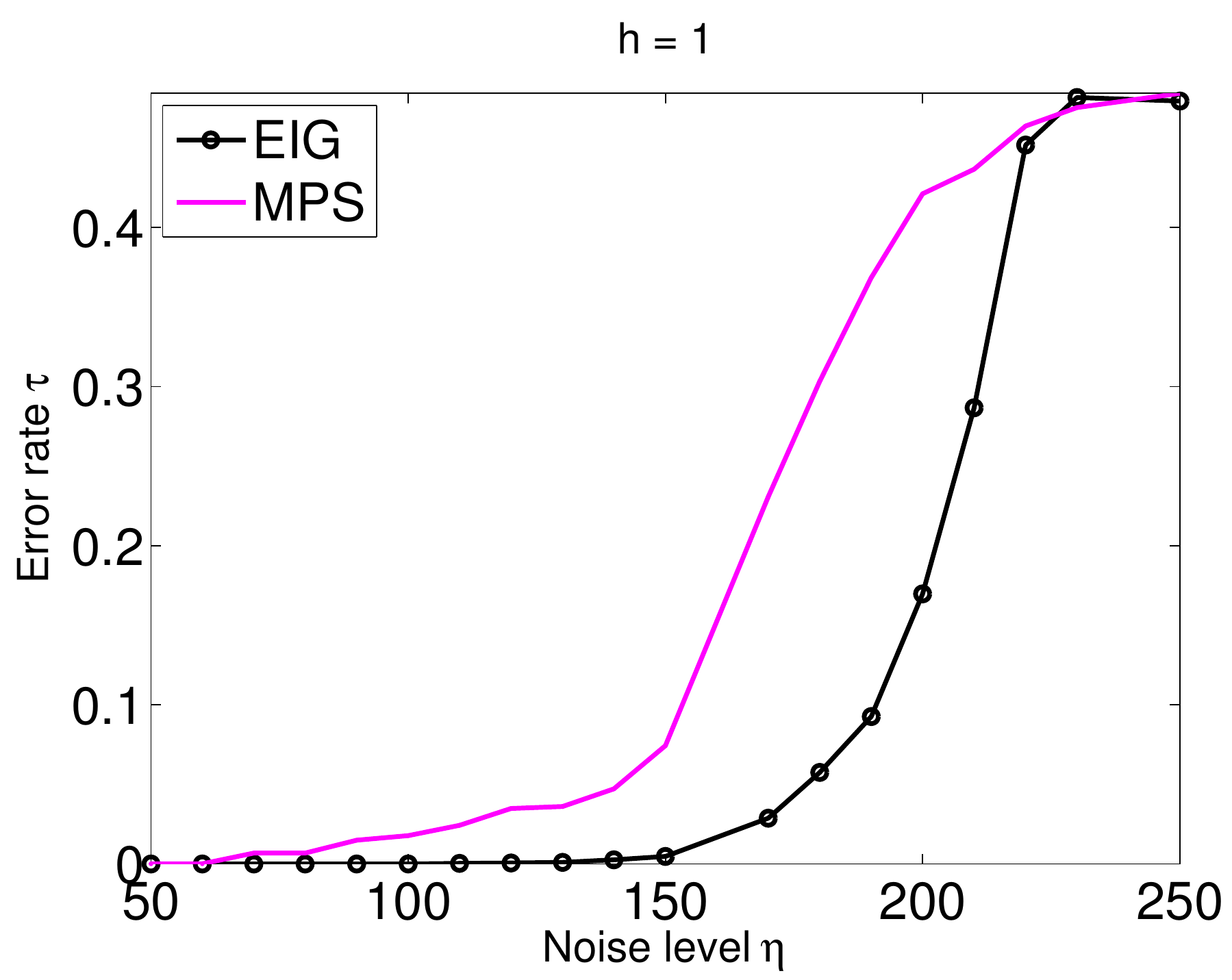}
\includegraphics[width=0.235\columnwidth]{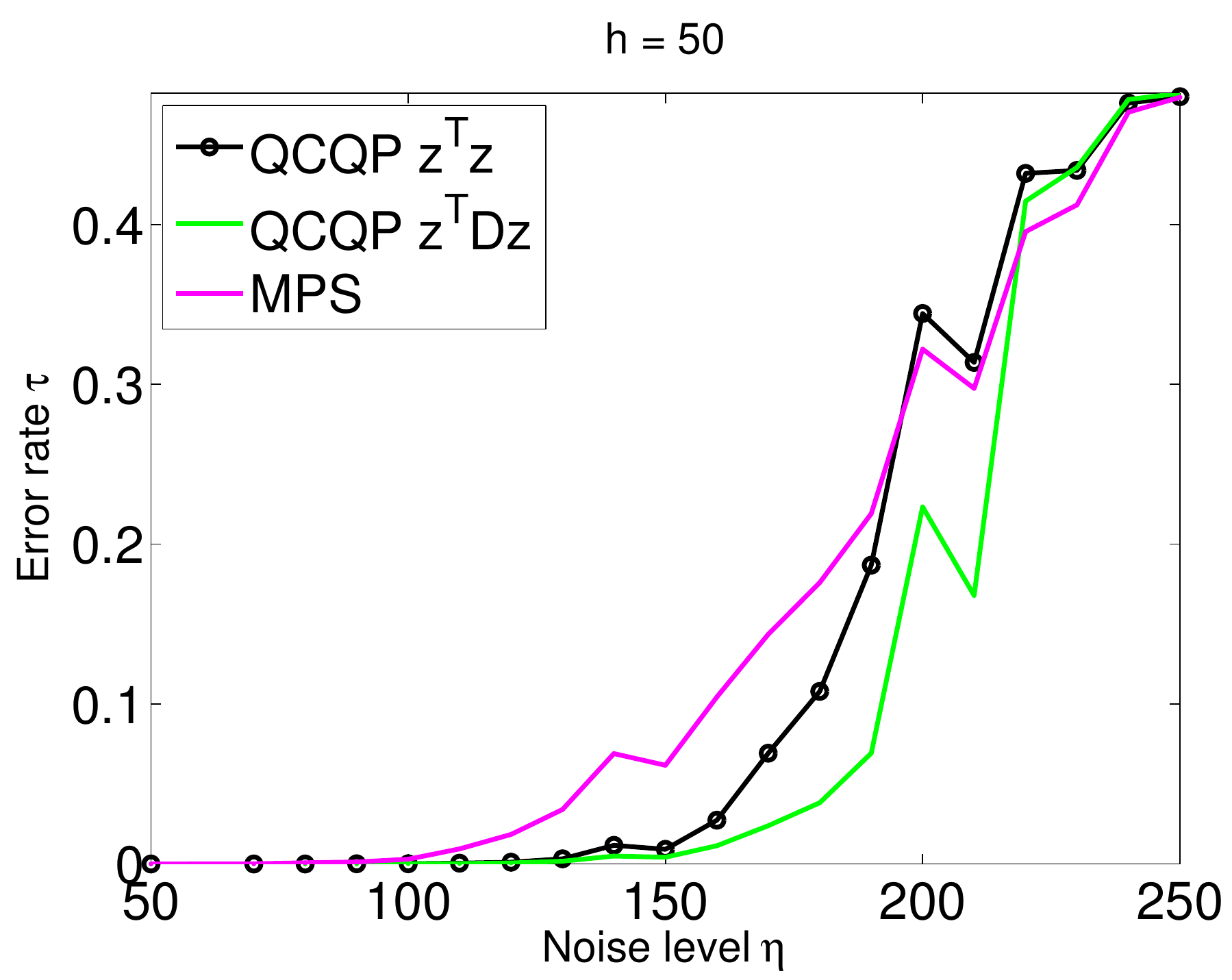}
\includegraphics[width=0.235\columnwidth]{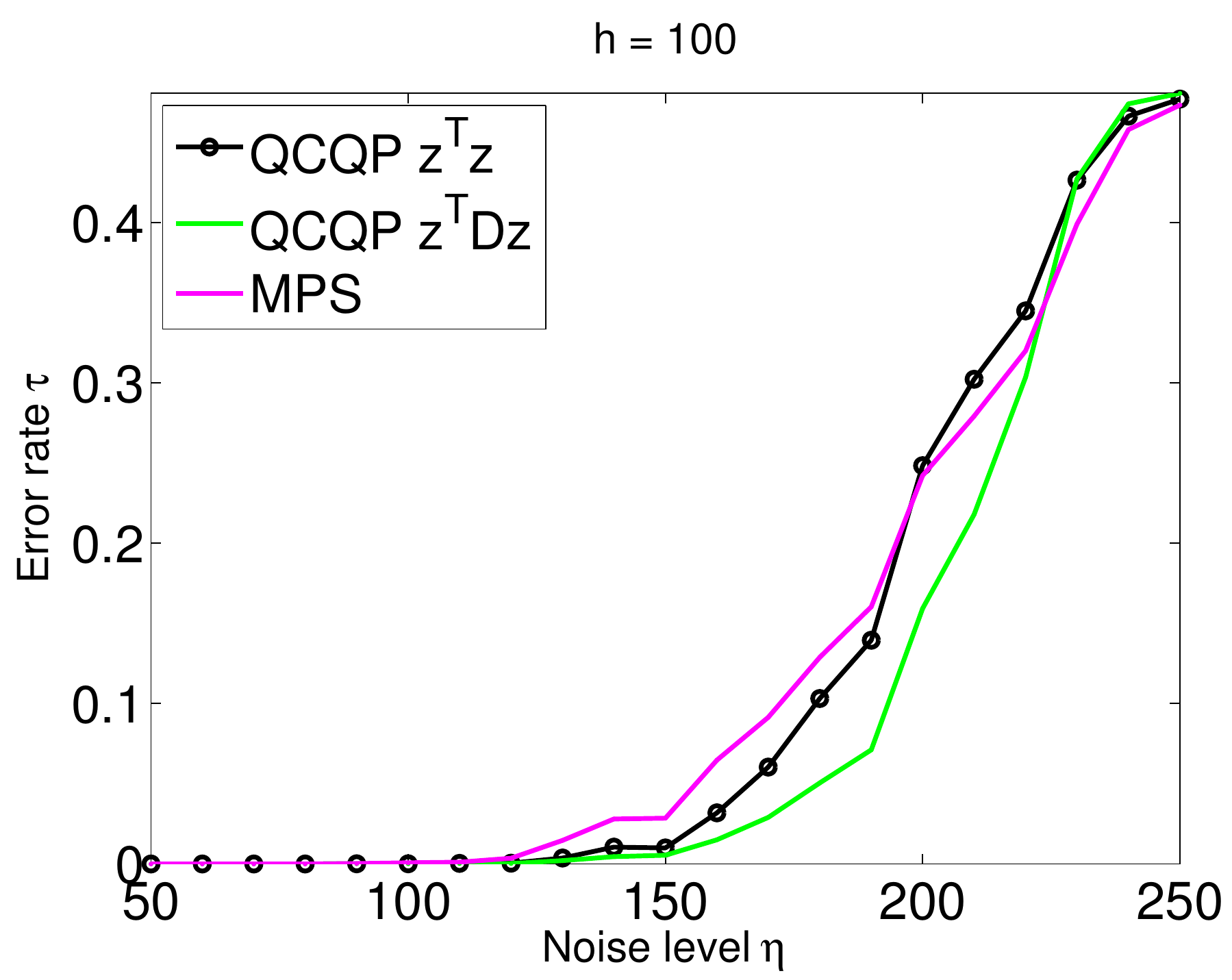}
\includegraphics[width=0.235\columnwidth]{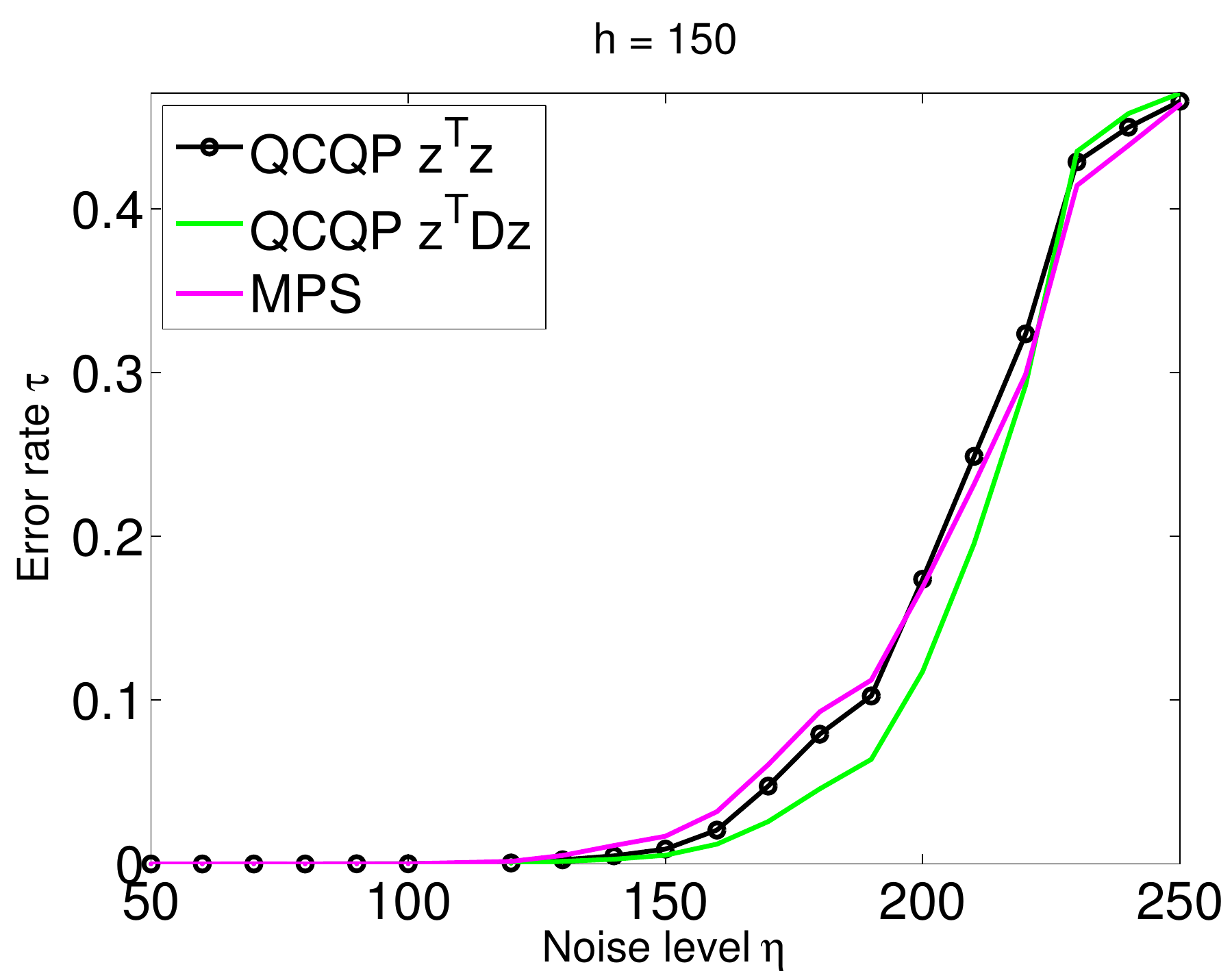}
}
\end{center}
\caption{ Comparison of the eigenvector/QCQP and MPS algorithms when $G$ is a graph on $n=100$ nodes which follows the preferential attachment model where a new node is incident with $m_{pa}$ already existing nodes chosen with a probability proportional to their degree. 
The underlying subgraph of noisy edges $H$ is an Erd\H{o}s-R\'{e}nyi random graph $G(n,\alpha)$ of average degree $d$. We vary the bad degree $d$ ($x-$axis), and average the result of each experiment over 100 runs.
}
\label{fig:GPrefAtt_HRandom}
\end{figure}

We plot in Figure \ref{fig:GPrefAtt_HRandom} the recovery rates when $G$ is a scale-free graph generated using the popular preferential attachment model, while the subgraph $H$ of \textit{bad} edges is a random Erd\H{o}s-R\'{e}nyi graph.
Following the preferential attachment model   \cite{Barabasi99emergenceScaling}, we grow a graph by adding new vertices one by one, with each new node having exactly $m_{pa}$ edges connecting it to already existing nodes, chosen proportionally to their degrees. In this scenario, the MPS algorithm performs worse than the eigenvector method, but as the number of anchors increase the performance of the MPS algorithm almost matches the one of the QCQP formulations.


\begin{figure}[h]
\centering
\includegraphics[width=0.239\columnwidth]{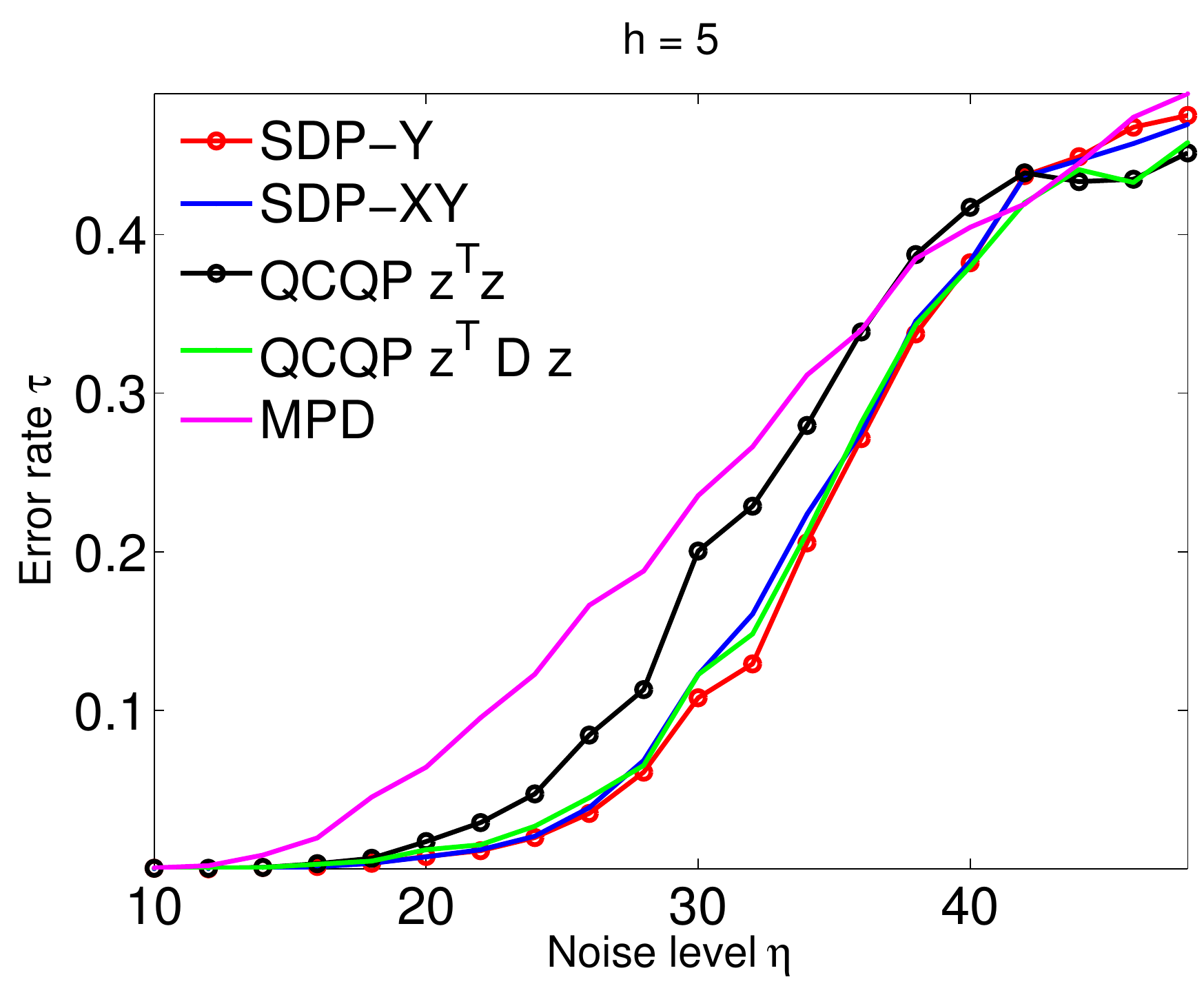}
\includegraphics[width=0.239\columnwidth]{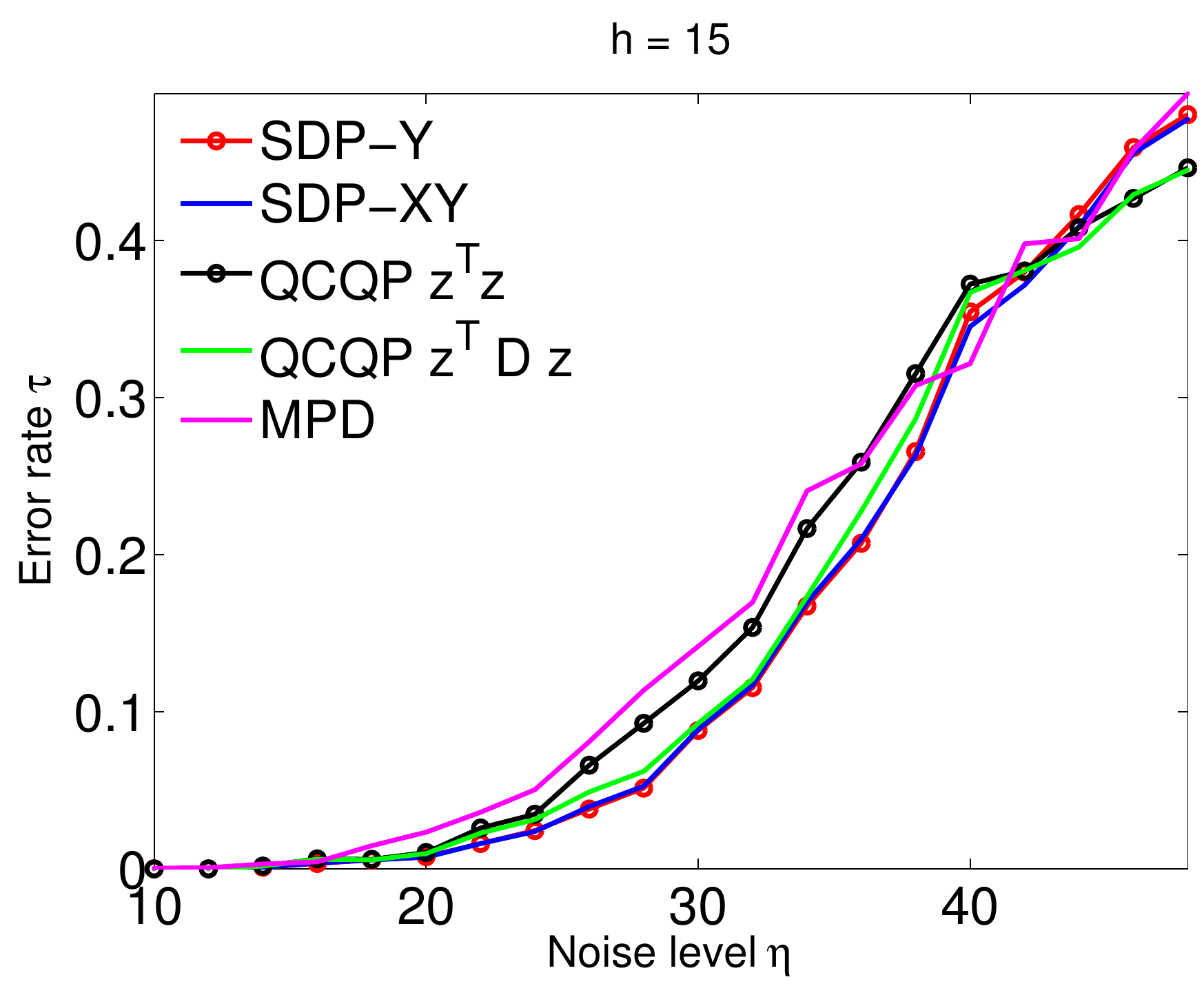}
\includegraphics[width=0.239\columnwidth]{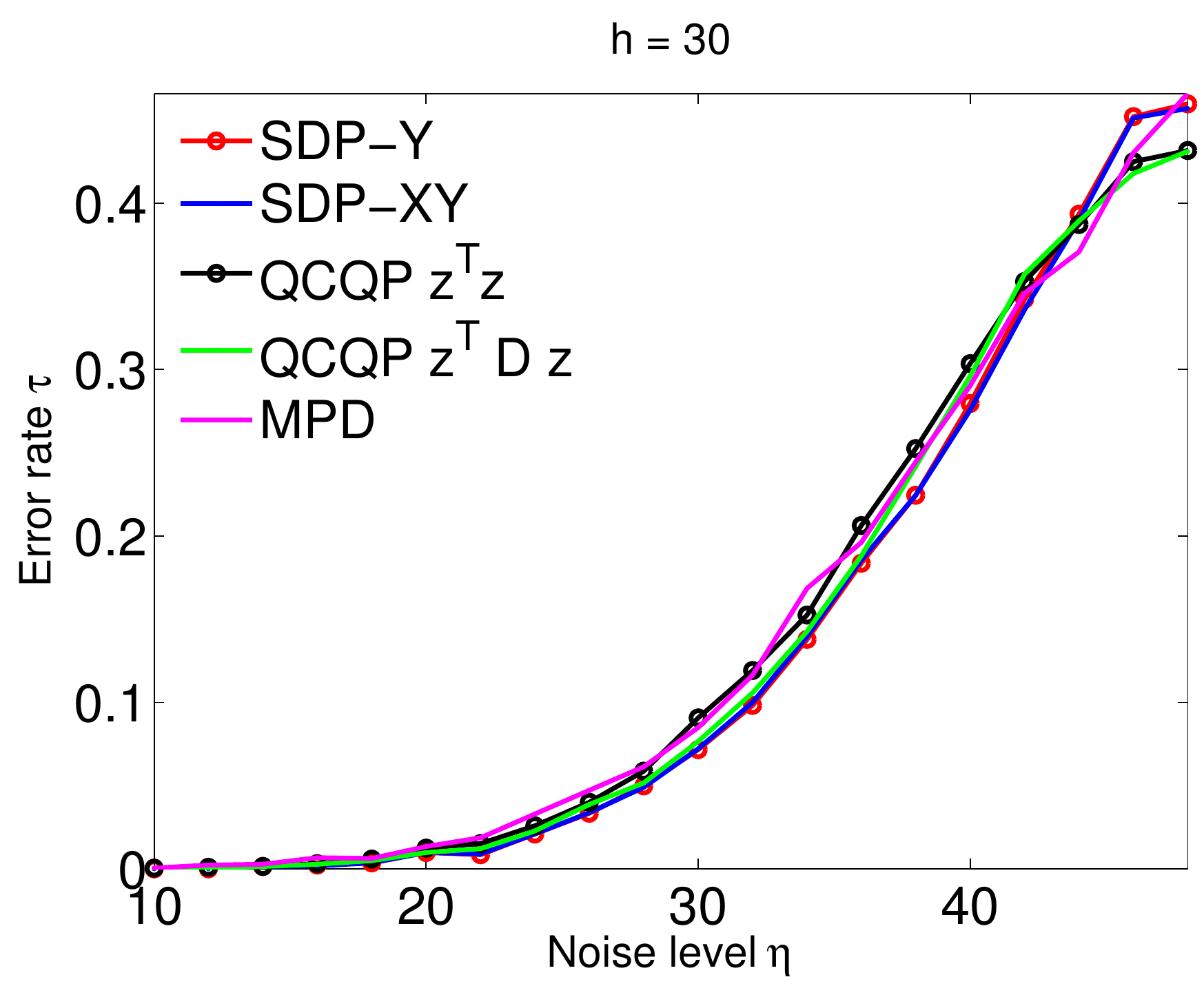}
\includegraphics[width=0.239\columnwidth]{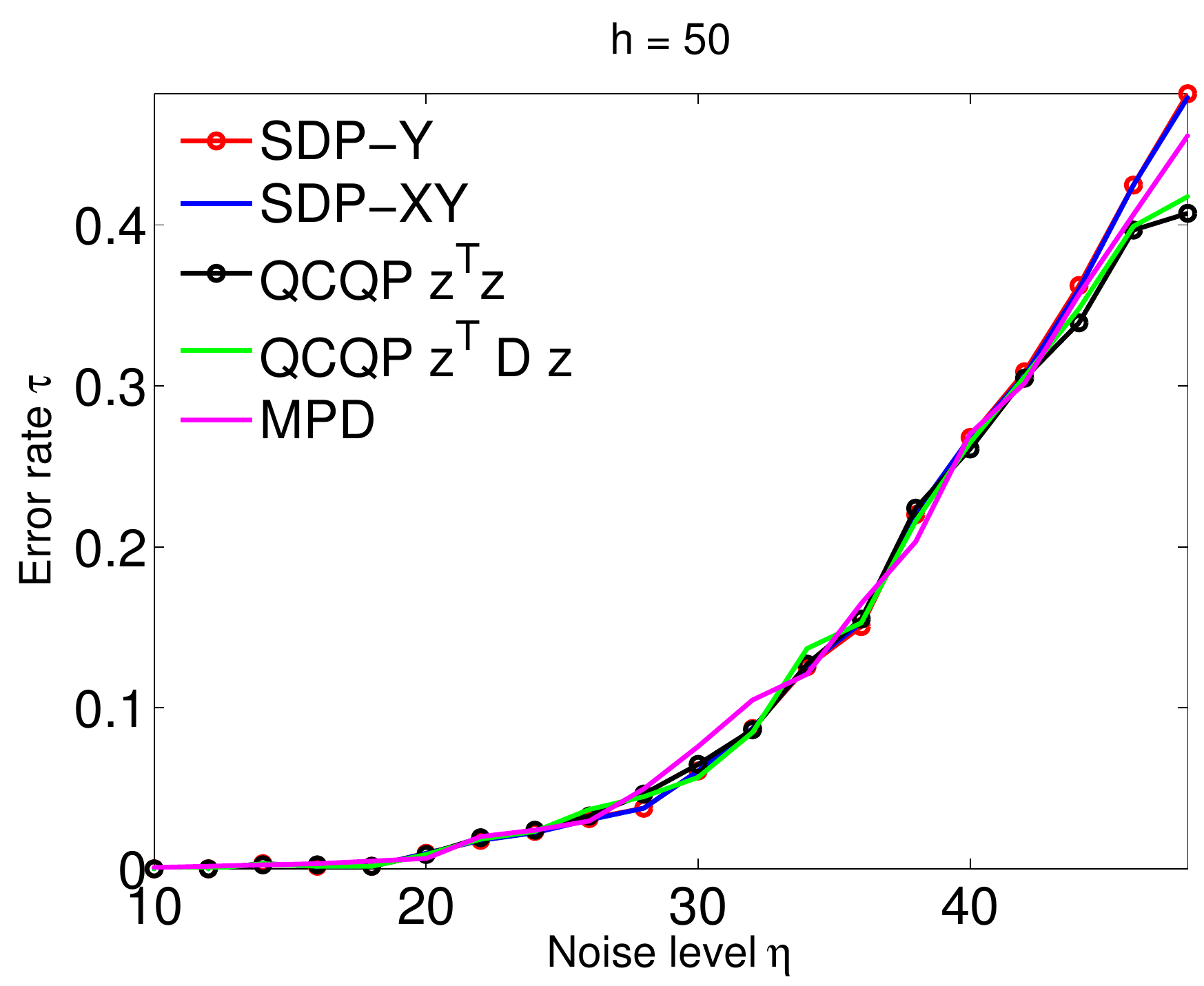}
\caption{Comparison, in terms of robustness to noise, of the five algorithms  proposed for synchronization with anchor information: the two QCQP formulations using the two different constraints: $z^Tz=s$ and $z^T D z = \Delta$ as they appear in equations (\ref{maximization_zTz}) and (\ref{maximization_zTDz}), the two SDP-based formulations SDP-Y and SDP-XY as formulated in
(\ref{SDP_maxAnch}) and (\ref{SDP_max_XY}), and finally the message passing synchronization algorithm MPS. We denote by $s$ the number of sensors $s=n-h$, and by $\Delta$ the sum of the degrees of all sensors. The adjacency graph of available pairwise measurements is an Erd\H{o}s-R\'{e}nyi  graph $G(n,\alpha)$ with $n=75$ and $\alpha=0.2$, and the subgraph of corrupted edges is a random subgraph of $G$. Also, $h$ denotes the number of anchors, chosen uniformly at random from the $n$ nodes. Results are averaged over 50 runs.}
\label{fig:comp_sync_anch}
\end{figure}

As a final set of experiments, we also compare the results of the above proposed MPS algorithm with the quadratically constrained quadratic program (QCQP) and SDP-based methods for synchronization with anchors introduced in \cite{asap3d} and reviewed in Appendix B, when the adjacency graph of available pairwise measurements is an Erd\H{o}s-R\'{e}nyi graph $G(n,\alpha)$ with $n=75$ and $\alpha=0.2$. Furthermore, the corrupted edges are chosen at random from the existing edges with probability $\eta$.  We vary the number of anchors $ h=\{5,15,30,50\}$, chosen uniformly at random from the $n$ nodes, and we average the results over 50 runs. The results in Figure \ref{fig:comp_sync_anch} show that the performance of the MPS algorithm is very similar to the one of its competitors, especially for a large number of anchors.

An analysis of the above message passing algorithm in terms of its convergence and robustness to noise is a project in itself and is beyond the scope of this paper. We believe that there are two  main reasons that make the formulation of SYNC($\mathbb{Z}_2$) as a message passing algorithm appealing. First, in the presence of anchors, MPS may give superior results when compared to the eigenvector formulation, over certain noise regimes and models. 
Second, the MPS algorithm enjoys the flexibility of incorporating additional constraints that may not be accommodated by any of the other formulations.
As an example, we point out the k-SYNC problem from Section 6.1, where we easily incorporate into the message passing algorithm constraints specific to the k-SYNC problem (\ref{medianMPS}), where all nodes within a given partition correspond to the same unknown group element.
Note that the SDP-based formulation, which can easily accommodate the constraints given by anchors or partitions (as described in Section \ref{sec:sync_partition}) is computationally expensive when compared to the MPS algorithm, although also more robust. 

As an overall conclusion, the numerical simulations detailed in this section suggest the  computationally appealing MPS algorithm is perhaps more suitable for certain instances (underlying graph models) of the synchronization problem
when we are aware that there is a large amount of noise in the available measured data. Otherwise, we expect the eigenvector method to yield more robust results, though it is an interesting question whether there exist families of graphs for which the MPS method yields better solutions that the eigenvector method, even at low levels at noise.

We point out that the above proposed message passing algorithm bears similarities with other propagation models considered in the literature in the past decades, but it is also distinct in the sense that it is a (deterministic) algorithm that solves a problem where edges, whenever present, carry additional information on the pairwise interaction between pairs of nodes.
The popular voter model is a stochastic model for opinion formation among interacting agents of a finite population \cite{liggett1997}, in which any given agent flips its state to a new state at a rate that is proportional to the number of neighbors that posses that new state, and one of the main questions of interest is consensus of the finite system to a given state. 
Another line of work similar to our proposed MPS approach is given by the very popular information cascades models \cite{Bikhchandani1992_Cascade_Model,Banerjee_Model_Herd_Behavior}, where agents make decisions sequentially based on a mix of own private information but also that of interacting agents, and the probability that a particular choice is the best one is computed via Bayes's Rule, as MPS also does. However, at a give node, MPS aggregates information from its neighbors based on both the belief of its neighbors and the beliefs on the incident edges that capture the pairwise interaction measurements and thus encode the group structure available in the group synchronization problem.
Others such models include the popular Sznajd model for the dynamics of opinions and consensus in a population \cite{sznajd2000opinion,Generalized_Sznajd}, the Deffuant model for social influence, where randomly chosen agents interact to mix opinions only if the distance between their respective opinions are shorter than a prescribed threshold \cite{Deffuant,Albert_statisticalmechanics,Fortunato2004_Deffuant}.


Finally, we point out an encouraging analogous situation in the case of the \textit{planted clique problem}, where a recent improvement over spectral methods is due to a belief propagation algorithm. 
In the planted clique problem, one chooses a random graph 
$G(n, \frac{1}{2})$ and a random  subset $Q$ of vertices of size $K$ and force it to be a clique by connecting every pair of vertices of $Q$ by an edge. The goal is to give a polynomial time algorithm for finding the planted clique almost surely for various values of $K$. 
Until recently, the best polynomial-time algorithms require
that $ K \geq c \sqrt{n}$ for some constant $c$, first achieved by \cite{Alon98findinga} through a spectral technique. The recent result of \cite{montanariClique} shows that the planted clique can be identified in nearly linear time if the clique size is at least $ (1+ \epsilon) \sqrt{n/e}$, for any $\epsilon > 0$ (thus below the spectral threshold $\sqrt{n}$), and does so via an algorithm derived from belief propagation, a heuristic method for approximating posterior probabilities in graphical models.



\section{Synchronization for Partition Networks (k-SYNC)} \label{sec:sync_partition}

The problem of synchronization over the group $\mathbb{Z}_2$ with partition constraints  can be stated as follows. To each node $v_i \in V$ of the graph $G$, there corresponds an unknown group element $z_i \in \mathbb{Z}_2$.  Given a partition of the vertex set $V$ into $k$  pairwise disjoint non-empty subsets
$\mathcal{A}_{1}, \ldots, \mathcal{A}_{k}$ with 
\begin{equation}
  \mathcal{A}_{i} \cap  \mathcal{A}_{j} = \emptyset, \;\; \forall \; i\neq j \in \{1,\ldots,k \} \;\;\;\;\;\;\;\;\;\;\;\; \mbox{and}  \;\;\;\;\;\;\;\;\;\;\;\;
  \bigcup_{1\leq j \leq k} \mathcal{A}_{j} = V(G) 
\end{equation}
such that $z_{j_1}= \ldots = z_{j_t} \in \mathbb{Z}_2$ for $v_{j_1},\ldots, v_{j_t} \in \mathcal{A}_{j}$ (with $|\mathcal{A}_{j}| = t$), the task is to recover the unknown group elements $ z_1, \ldots, z_n \in \mathbb{Z}_2$ from an incomplete set of (possibly noisy) pairwise group measurements $z_i z_j^{-1}$, for $ (i,j) \in E$ and $ v_i \in \mathcal{A}_{u}, v_j \in \mathcal{A}_{v}$, and $u \neq v$.
In other words, we partition the graph nodes into $k$ disjoint subsets, where each subset corresponds to some unknown group element $ \theta_l \in \mathbb{Z}_2, l=1,\ldots,k$. Given a sparse noisy sample of pairwise measurements between pairs of nodes that belong to two different partitions, the task is to recover the unknown group elements $\theta_l, l=1,\ldots,k$, and thus implicitly of all $n$ nodes since $z_{j_1}= \ldots = z_{j_t} = \theta_l \in \mathbb{Z}_2$, if $x_{j_1},\ldots, x_{j_t} \in \mathcal{A}_{l}$.

\begin{figure}[h]
\begin{center}
\subfigure[Network]{\includegraphics[width=0.34\columnwidth]{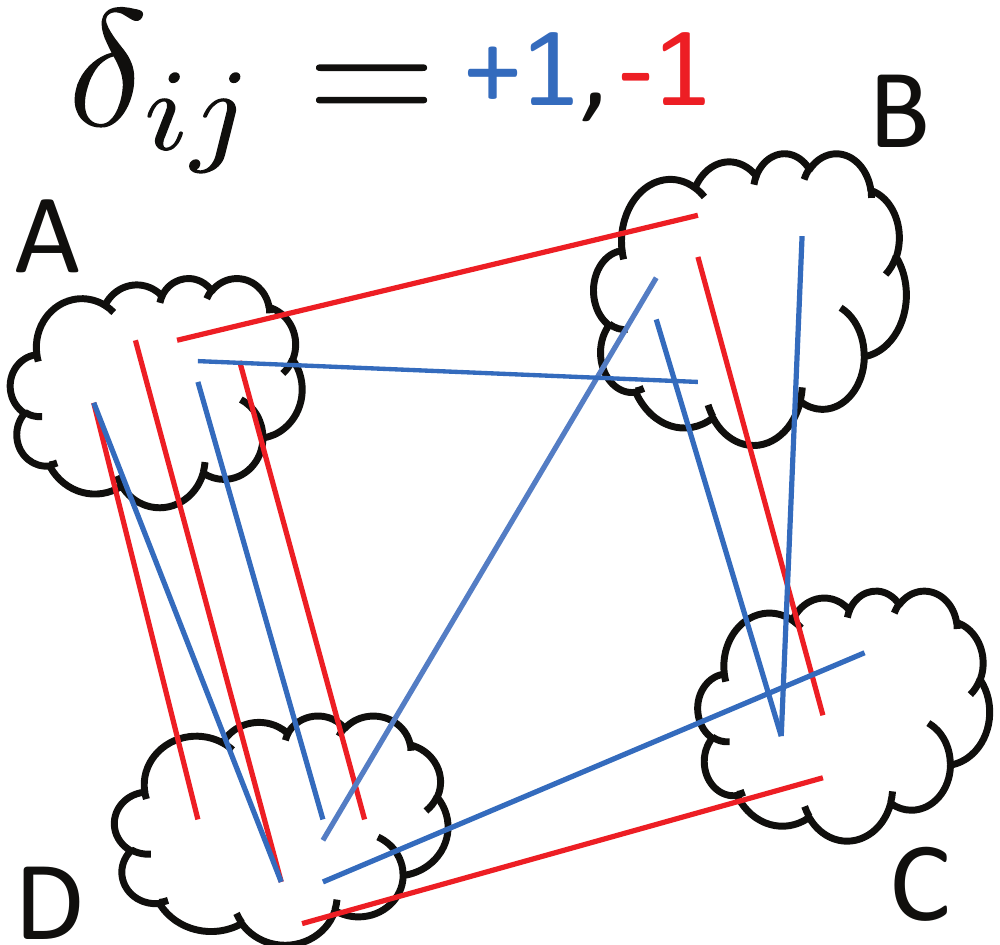}}
\subfigure[Graph of good edges]{\includegraphics[width=0.29\columnwidth]{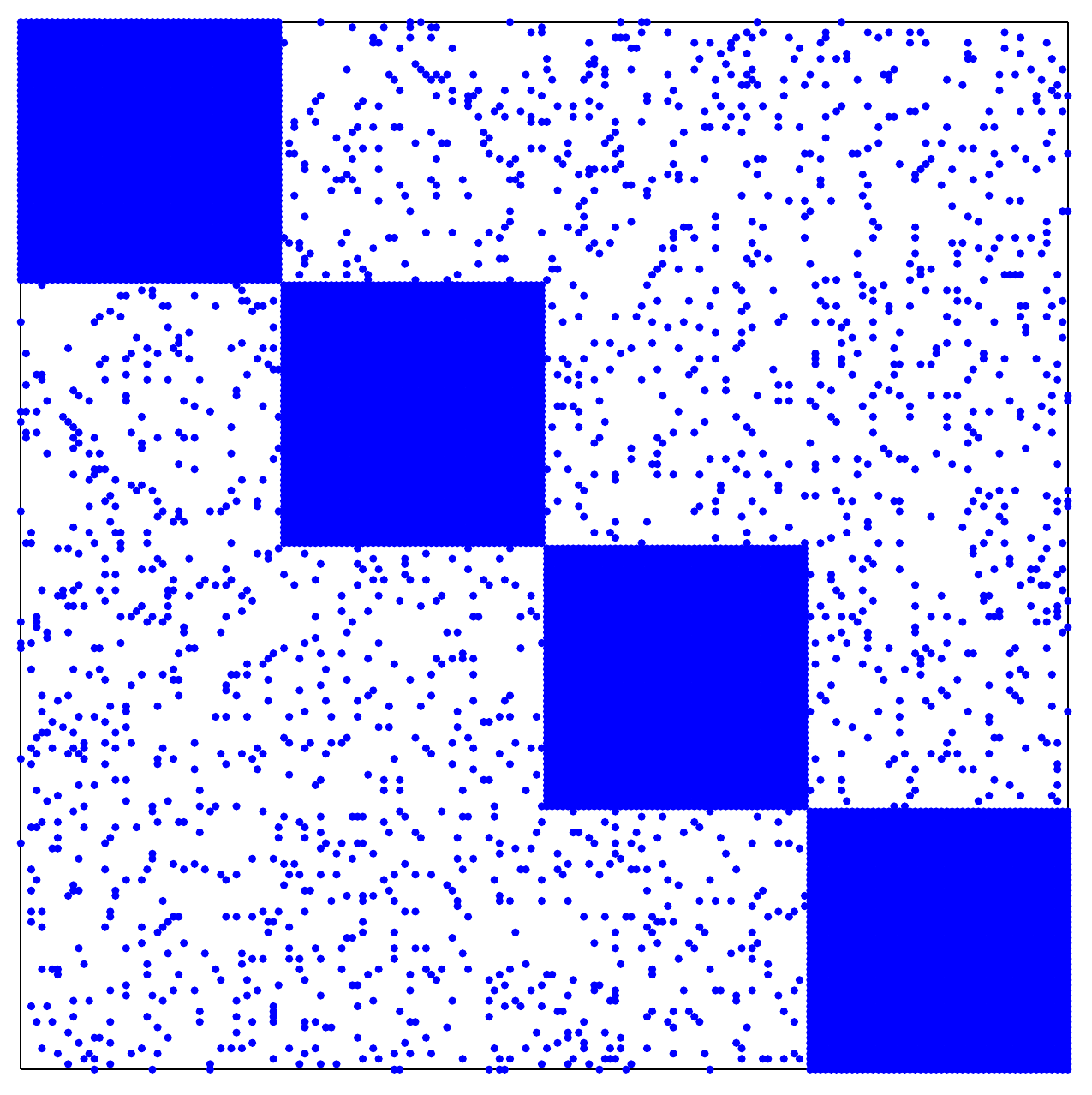}}
\subfigure[Graph of bad  edges]{\includegraphics[width=0.29\columnwidth]{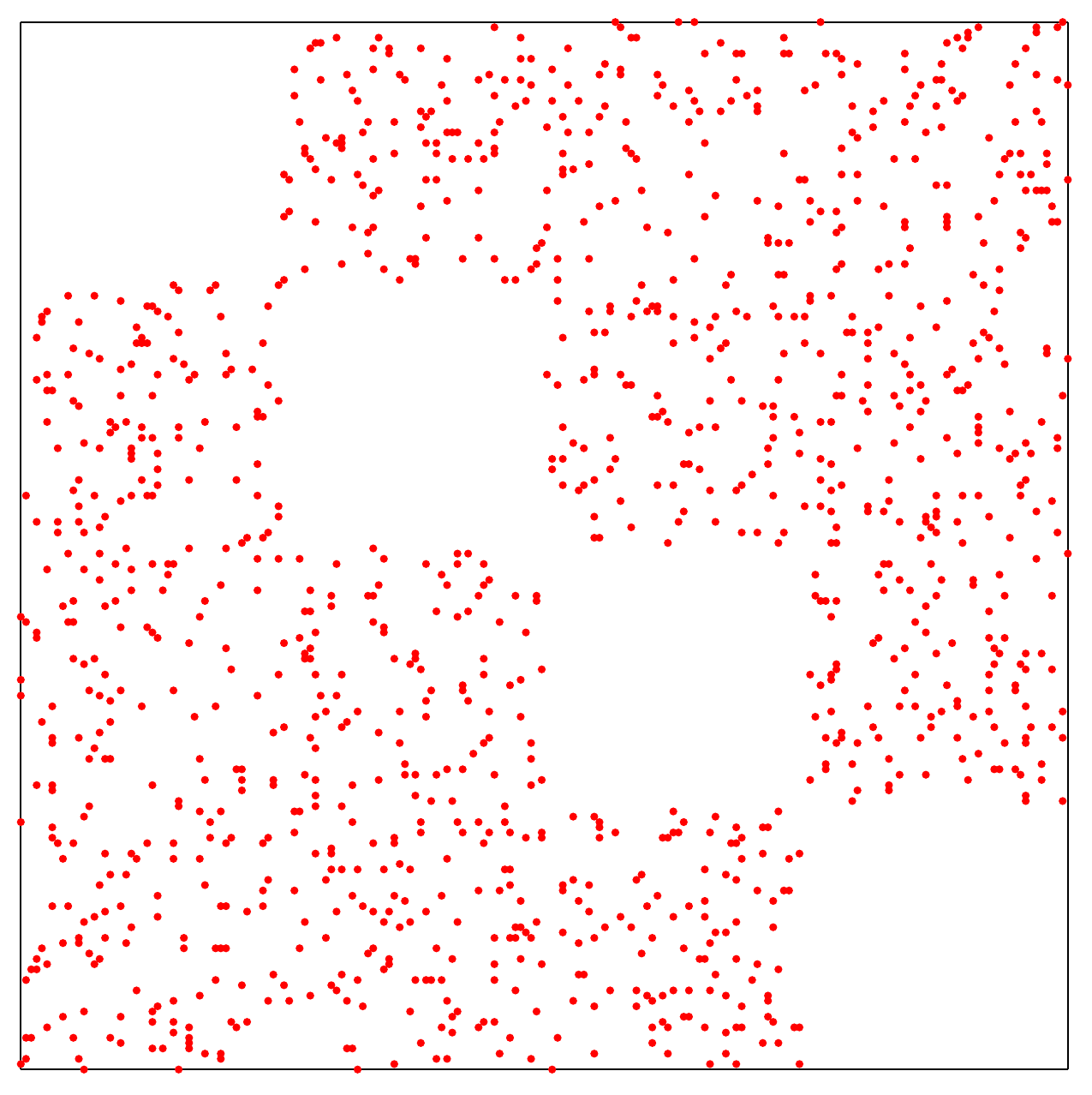}}
\end{center}
\caption{(a) An example of a multipartite network with four partitions $\{A,B,C,D\}$, where the blue, respectively red, edges across the partitions denote the correct, respectively incorrect, pairwise measurements between pairs of points that belong to different partitions. $\delta_{ij}=1$ denotes a correct measurements, while $\delta_{ij}=-1$ denotes a corrupted measurement. (b) The adjacency graph of good edges.  Note that all edges with endpoints within the same partition always denote correct measurements $\delta_{ij}=1$. (c) The adjacency graph of bad edges $\delta_{ij}=-1$.}
\label{fig:exampleMultipartite}
\end{figure}

Figure \ref{fig:exampleMultipartite}(a) illustrates an example of a multipartite network with $k=4$ partitions.  
We show in Figure \ref{fig:exampleMultipartite} (b) and (c)  the adjacency matrices of the subgraphs of good (correct) and bad (incorrect) edges, and note that all edges with endpoints within the same partition always carry correct pairwise  measurements that are equal to $+1$ since nodes within the same partition $\mathcal{A}_i$ denote the same group element $\theta_i$, and $\theta_i^2=1, \forall \theta_i \in \mathbb{Z}_2$.

We denote by \textbf{k-SYNC} the above synchronization problem over $\mathbb{Z}_2$ with the  additional constraints that nodes within the same partition all correspond to the same unknown group element. In the context of the Congress data, the additional information is that the same senator may serve on multiple (not necessarily consecutive) Congresses. We denote by $m_i$ the multiplicity of each distinct senator $i$, i.e., the number of Congresses she or he has served on. The average multiplicity of all senators serving in the first 41 Congresses is $\overline{m} = 5.3$. Since it is reasonable to assume that a senator will maintain the same political orientation across different Congresses, we would like to use this information to robustly detect the party affiliation of each senator across time. From the perspective of the k-SYNC problem, the network with $n=4196$ nodes in the Congress data may be partitioned into $k=735$ non-overlapping subsets of nodes, one for each unique senator. The nodes within each partition block $\mathcal{A}_u$ correspond to all mandates of senator $u$ across time.

\subsection{Spectral and SDP algorithms for the k-SYNC problem.}

The partition-constrained synchronization problem can be formulated as follows
\begin{equation}
	\begin{aligned}
	& \underset{ \boldsymbol{z} = (z_1, \ldots, z_n)  \in \mathbb{Z}_2^n}{\text{maximize}}
	& & \boldsymbol{z}^T Z  \boldsymbol{z} &  \\
	& \text{subject to}
	&  &   \boldsymbol{z}^T \boldsymbol{z} = n      \\
	&  &  &    z_i z_j = 1, \;\; \forall \; i,j \in \mathcal{A}_u,\;\; \forall u=1,\ldots,k.
	\end{aligned}
\label{max_Const}
\end{equation}
Note that if the number of partitions $k$ is large and close to $n$, then the additional constraints contribute with less information, and are less helpful in increasing the robustness to noise of any algorithm one may use. On the other hand, if $k$ is small relative to $n$, then we are facing an easier task since there are less group elements to the recovered and a large amount of redundant information in the network. Note that the case $k=n$ corresponds to the usual synchronization problem discussed in Section \ref{sec:sync} when no additional information is available. Unfortunately, since the above quadratic program (\ref{max_Const}) can no longer be cast as an eigenvector problem, we consider again a relaxation via semidefinite programming. The objective function in (\ref{max_Const}) can be written as
\begin{equation}
  \sum_{i,j=1}^{n} z_i {Z}_{ij} z_j = Trace(Z \Upsilon)
\end{equation}
where $\Upsilon$ is the $n \times n$ symmetric rank-one unknown matrix with entries $\pm 1$
\begin{equation}
\Upsilon_{ij} = \left\{
     \begin{array}{rl}
 z_i  z_j^{-1} & \;\; \text{ if } i \in \mathcal{A}_u, j \in \mathcal{A}_v, u \neq v \\
 1             & \;\; \text{ if } i,j \in \mathcal{A}_u \;\; u=1,\ldots,k.\\
      \end{array}
   \right.
\label{upsilondef_congress}
\end{equation}
Note that $\Upsilon$ has ones on its diagonal $ \Upsilon_{ii} =1, i=1,\ldots,n$, and the partition information gives another layer of hard constraints in the form of block submatrices with all ones entries, since the pairwise measurement between a pair of nodes that belong to the same partition is always 1. The SDP formulation is now
\begin{equation}
	\begin{aligned}
	& \underset{\Upsilon \in \mathbb{R}^{n \times n}}{\text{maximize}}
	& & Trace(Z \Upsilon) \\
	& \text{subject to}
	& &   \Upsilon_{ij} = 1 & \;\; \text{ if }  i,j \in \mathcal{A}_u,  \;\; u=1,\ldots,k \\
		& & &   \Upsilon \succeq 0,
	\end{aligned}
 \label{SDP_max_a}
\end{equation}
where the maximization is taken over all semidefinite positive real-valued matrices  $\Upsilon \succeq 0$. Since $\Upsilon$ is not necessarily a rank-one matrix, the SDP-based estimator is given by the best rank-one approximation to  $\Upsilon$, which can be computed via an eigen-decomposition. 
From a computational perspective, solving such SDP problems is computationally feasible only for relative small values of $n$  (typically several thousand unknowns, up to $n=10-15,000$), though there exist distributed methods for solving such convex optimization problems, such as the popular Alternating Direction Method of Multipliers (ADMM) \cite{Boyd_ADMM} which can handle large-scale problems arising in statistics and machine learning \cite{ADMM_Sparse_Low_Rank}. The remainder of this section describes several algorithms for solving the k-SYNC problem with partition constraints.

\textbf{Algorithm EIG-k-SYNC.} Unfortunately, due to the additional constraint, the above maximization in (\ref{max_Const}) can no longer be
cast as an eigenvector problem. One obvious approach to solving (\ref{max_Const}) is to ignore the second constraint and solve the resulting problem using the eigenvector method. If the data is not very noisy and the underlying graph $G$ is not very sparse, or $k$ is large compared to $n$, then we expect the eigenvector method to perform well, even without taking advantage of the additional constraint information. We denote by EIG-k-SYNC the approach of solving k-SYNC without taking the partition constraints into account.

\textbf{Algorithm MVEIG-k-SYNC.} The next natural step is to enforce the additional constraints after running EIG-k-SYNC, by a simple majority voting rule such as
\begin{equation}
    \bar{z}_{A_l} = \bar{z}_{l_1} = \bar{z}_{l_2} = \ldots = \bar{z}_{l_t} = \mbox{Majority}( z_{l_1}, z_{l_2},\ldots, z_{l_t}),
\end{equation}
where $\mbox{Majority}( z_{l_1}, z_{l_2},\ldots, z_{l_t}) =+1$ if at least half of the elements have value $+1$, and $-1$ otherwise. We denote by MVEIG-k-SYNC the synchronization algorithm which first runs EIG-k-SYNC followed by the majority voting scheme applied as a post-processing step. As expected, MVEIG-k-SYNC performs better than EIG-k-SYNC, but is not very robust at high levels of noise. Its disadvantage is that it is not able to integrate the partition constraints during the eigenvector computation.

\textbf{Algorithm PART-k-SYNC.} Next, we consider another eigenvector formulation which synchronizes the \textit{partition} graph (to be made clear shortly) of size $k$, as opposed to synchronizing the original graph $G$ of size $n$. This approach  is able to integrate the constraints across partitions at the price of losing information between the initial individual nodes belonging to different partitions. To each partition block $\mathcal{A}_i$ we associate a node $a_i$, whose value in the final solution will give the sign to all nodes in partition $\mathcal{A}_i$. For each pair of partition blocks $\mathcal{A}_i$ and $\mathcal{A}_j$, denote by $E(\mathcal{A}_i, \mathcal{A}_j,+)$ (respectively, $E(\mathcal{A}_i, \mathcal{A}_j,-)$) the number of existing edges with a $+1$ (respectively, $-1$) measurement, that connect a node in $\mathcal{A}_i$ with a node in $\mathcal{A}_j$. 
We let $E(\mathcal{A}_i, \mathcal{A}_j) = E(\mathcal{A}_i, \mathcal{A}_j,+)  + E(\mathcal{A}_i, \mathcal{A}_j,-)$, and note that $E(\mathcal{A}_i, \mathcal{A}_j)$ is the number of edges connecting partition blocks $\mathcal{A}_i$ and $\mathcal{A}_j$. We define the partition graph $G^\mathcal{A}=(V^\mathcal{A},E^\mathcal{A})$ to be the graph of size $k$ with node set $V^\mathcal{A} = \{a_1,a_2,\ldots,a_k\}$ and edge set $E^\mathcal{A}$, where  $G^\mathcal{A}_{ij}=1$ if $E(\mathcal{A}_i,\mathcal{A}_j) > 0$ (and we say the two partitions are adjacent), and $G^\mathcal{A}_{ij}=0$ otherwise.
For every pair of adjacent partitions, we would like to compute the pairwise group measurement, i.e., decide whether the two partitions have the same sign or no. To that end, we build the matrix $W$ of size $k \times k$, where
\begin{equation}
  \overline{W}_{ij} = \left\{
     \begin{array}{rl}
  +1 & \;\; \text{ if} (i,j) \in E^\mathcal{A} \text{ and }  E(\mathcal{A}_i, \mathcal{A}_j,+)   >  E(A_i, A_j,-) \\
  -1 & \;\; \text{ if} (i,j) \in E^\mathcal{A} \text{ and }  E(\mathcal{A}_i, \mathcal{A}_j,+)  <  E(\mathcal{A}_i, \mathcal{A}_j,-) \\
   0 & \;\; \text{ if} (i,j) \notin E^\mathcal{A} \text{ or } E(\mathcal{A}_i, \mathcal{A}_j,+) = E(\mathcal{A}_i, \mathcal{A}_j,-).
     \end{array}
   \right.
\label{H_ij_def_a}
\end{equation}
Yet, a better approach would be to weight the pairwise measurement between two adjacent partitions, and interpret this as a measure of confidence on the measurement. Thus, for the synchronization of the partition graph we use the following similarity matrix
\begin{equation}
  W_{ij} = \left\{
     \begin{array}{rl}
+\frac{ E(\mathcal{A}_i, \mathcal{A}_j,+)}{E(\mathcal{A}_i, \mathcal{A}_j)} & \;\; \text{ if} (i,j) \in E^\mathcal{A} \text{ and } E(\mathcal{A}_i, \mathcal{A}_j,+)>E(\mathcal{A}_i, \mathcal{A}_j,-) \\
  - \frac{ E(\mathcal{A}_i, \mathcal{A}_j,-)}{E(\mathcal{A}_i, \mathcal{A}_j)}  & \;\; \text{ if} (i,j) \in E^\mathcal{A} \text{and }  E(\mathcal{A}_i, \mathcal{A}_j,+)  <  E(\mathcal{A}_i, \mathcal{A}_j,-) \\
   0 & \;\; \text{ if} (i,j) \notin E^\mathcal{A} \text{ or } E(\mathcal{A}_i, \mathcal{A}_j,+) = E(\mathcal{A}_i, \mathcal{A}_j,-).
     \end{array}
   \right.
\label{H_ij_def_b}
\end{equation}
We are now ready to synchronize the partition graph whose pairwise relations are given by matrix $W$, and infer the sign of each partition. One criticism of this approach is that the good measurements between pairs of nodes get ``diluted" when converting them to measurements between pairs of partitions. Also, if the sign of a partition node $a_i$ is inferred incorrectly, so will be the sign of all nodes in partition $\mathcal{A}_i$.

\textbf{Algorithm SDP-k-SYNC.} We denote by  SDP-k-SYNC the algorithm which solves the SDP formulation in (\ref{SDP_max_a}), and remark that  SDP 
problems are expensive to solve compared to the eigenvector-based methods and our proposed message passing algorithm, since the complexity of solving an SDP problem scales as O($n^3$), i.e., cubic in the number of nodes in the graph  \cite{Vandenberghe94sdp,Porkolab97onthe}.

\textbf{Algorithm MPS-k-SYNC (Message Passing Synchronization).} 
Finally, we also consider the message passing synchronization formulation introduced in Section \ref{sec:MPD}. In light of the structural information available in the k-SYNC problem, we have modified the algorithm in the following two ways. First, at each iteration, we reinforce the edge weights within the partitions to be 1, i.e., $w_{ij}^{+,r}=1$, since $z_i z_j=1$ whenever both $i$ and $j$ belong to the same partition $\mathcal{A}_u$. Second, at each iteration, we update the node probabilities within each partition $\mathcal{A}_u$ to reflect the average belief of all nodes within $\mathcal{A}_u$,  since they all have the same sign. More precisely, if $\mathcal{A}_u$ is a partition of size $l$, with nodes $\mathcal{A}_u=\{y_1,\ldots,y_l\}$, then we update the individual node probabilities to the average belief of all nodes within the partition. However, to obtain a more accurate estimate for the value of the current belief on the overall sign of a partition block $\mathcal{A}_u$, we choose to compute the median, rather than the mean, of the individual nodes probabilities contained in $\mathcal{A}_u$
\begin{equation}
p_{y_i}^{+,r} \mapsto \operatorname{median}(p_{y_1}^{+,r},\ldots,p_{y_l}^{+,r}),\forall i=1,\ldots,l.
\label{medianMPS}
\end{equation}
We update $p_{y_i}^{-,r}$ similarly, and note that the equality $p_{y_i}^{+,r} + p_{y_i}^{-,r}=1,  \forall y_i \in \mathcal{A}_u$  remains valid. If the noise level in the original pairwise measurements $z_iz_j$ is small, then we expect that at each iteration $r$, the probabilities $p_{y_1}^{+,r}, \ldots, p_{y_l}^{+,r}$ will have small variance and take values  close to $0$, or all close to $1$ (and $p_{y_1}^{-,r}, \ldots, p_{y_l}^{-,r}$ will take values close to $1$, respectively $0$). However, for higher levels of noise, many of the above probabilities will be very inaccurate (i.e., very far from their true value of $0$ or $1$), and we choose the median value as a more robust approximation for the aggregated belief of all nodes within a partition.

\subsection{Numerical results for the k-SYNC problem}
This subsection details the empirical results when using all the above algorithms for solving the k-SYNC problem for two synthetically generated networks and the real Congress data set previously introduced in Section \ref{sec:syncCongress}. 
In the Congress data, the two communities (the Republican and Democrats) are roughly the same size, and this is also the case for the synthetic data. For the latter case, each senator is a republican or democrat with equal probability, thus the sizes of the two parties is  approximately equal.

\begin{figure}[h]
\begin{center}
\includegraphics[width=0.32\columnwidth]
{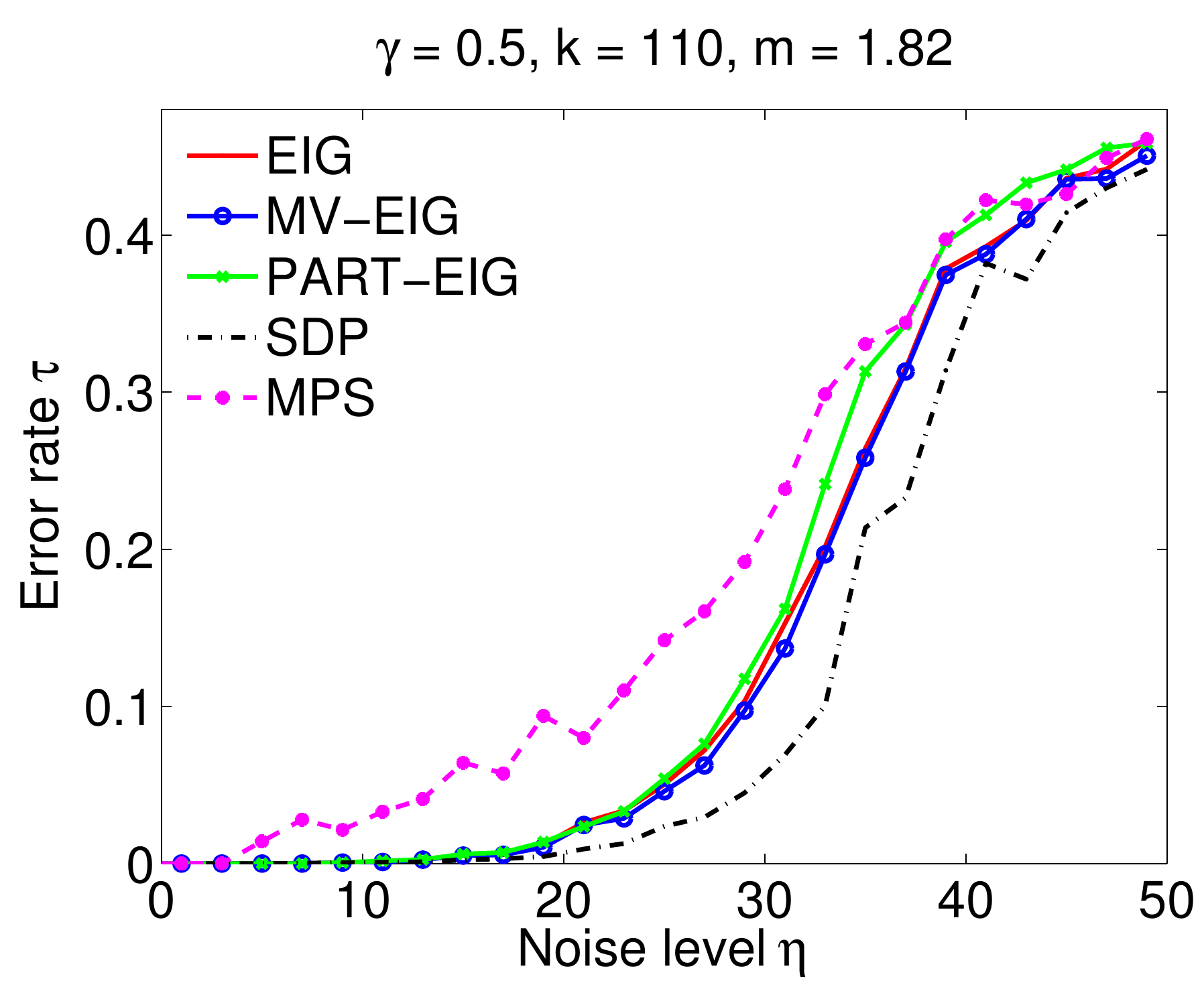}
\includegraphics[width=0.32\columnwidth]
{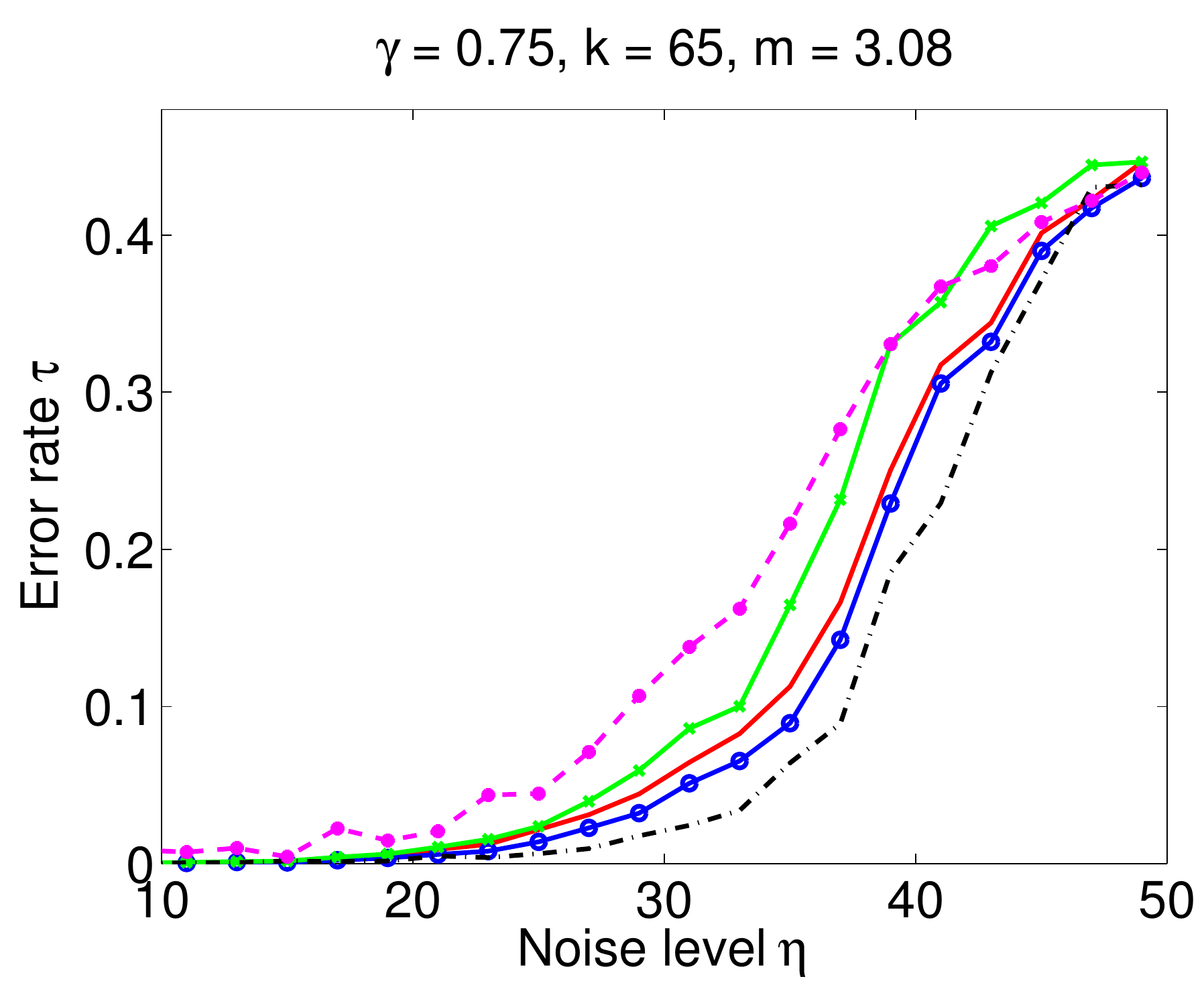}
\includegraphics[width=0.32\columnwidth]
{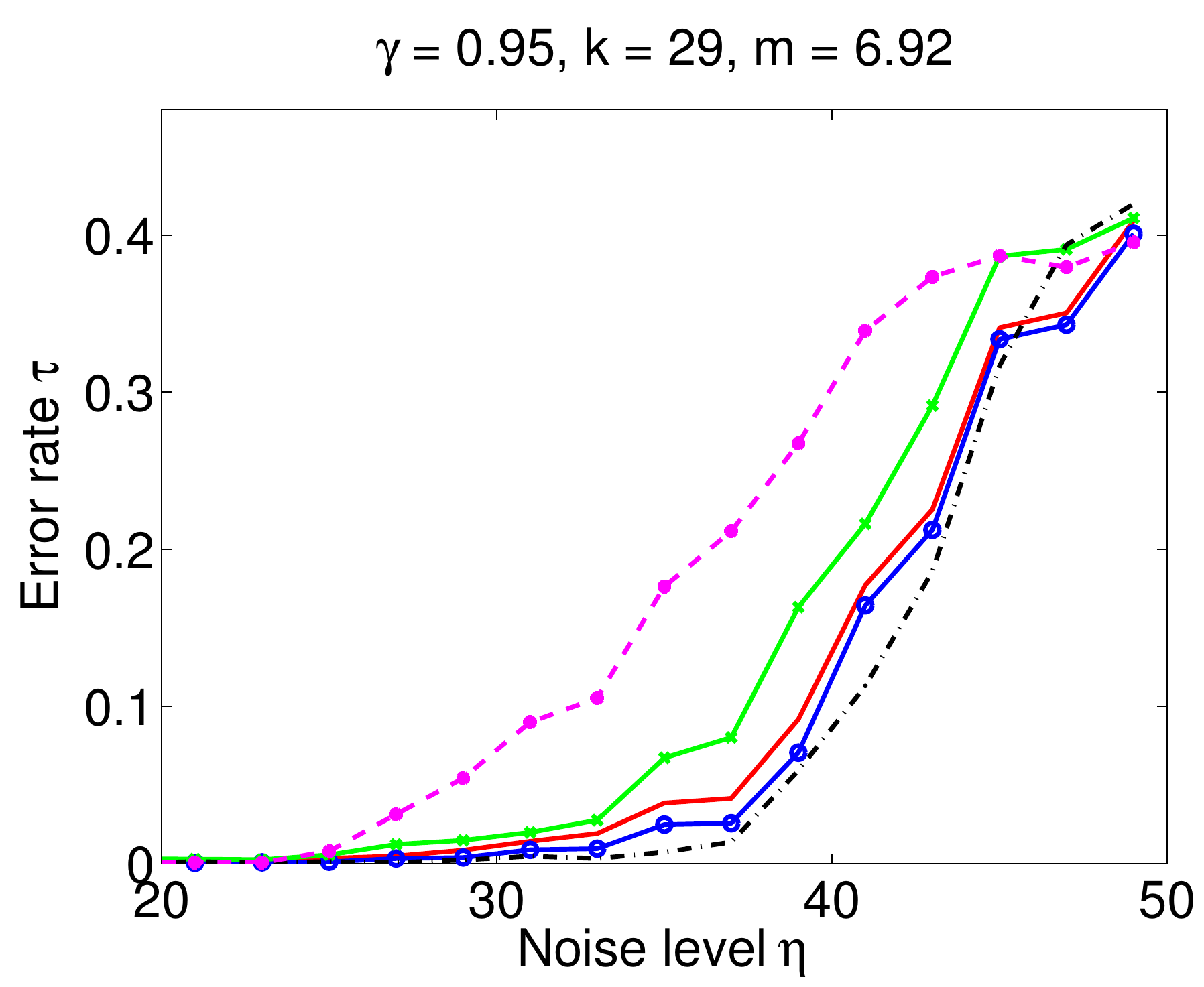}
\end{center}
\caption{Model: 10 Congresses, each with 20 senators (each one equally likely to be a $\pm1$, i.e., a Republican or a Democrat), and each congress is a $G(n,\alpha)$ with $n=200$ and $\alpha=0.50$. A senator from Congress $i$ gets to serve in
Congress $i+1$ with probability $\gamma$. For each $\gamma$, $k$ is the resulting number of partitions (results are averaged over 25 runs) and $\nu$ the
frequency of occurrence of each senator. Note that $n = k \nu = 200$. The $x$-axis is the
noise level (the probability that a measurement $z_i z_j$ is flipped) and the $y$-axis is the
recovery error (the percentage of nodes whose sign is estimated incorrectly).
}
\label{fig:withPersistenceProb}
\end{figure}

\textbf{Synthetic Congress model I.} The first synthetic  example we consider closely models the Congress voting data set. We fix the number of Congresses $C=10$, and the number of senators $n=20$ in each Congress. A senator that participates in any given Congress also serves on the next Congress with probability $\gamma$ (that we refer to as the ``persistence probability"). The number of unique senators that serve across the 10 Congresses is on average $k$, and their average multiplicity $m$ (average number of terms a senator serves on). Throughout the experiments, we let the ground truth variables $z_i$ take value $\pm1$ with equal probability. 
The plots in Figure \ref{fig:withPersistenceProb} show the recovery errors (averaged over 25 runs) as we increase the noise level, for three different values of the persistence probability $\gamma=\{0.5,0.75,0.95\}$. The ranking of the algorithms is consistent across the three networks. SDP is the best performer, closely followed by MV-EIG, EIG. The PART-EIG algorithms performs very similar to the latter two ones when the network is sparse ($\gamma=0.5$), but worse for the other two denser networks. MPS is  the worst performer, especially when the measurement graph $G$ is sparse.

\begin{figure}[h]
\begin{center}
\includegraphics[width=0.32\columnwidth]{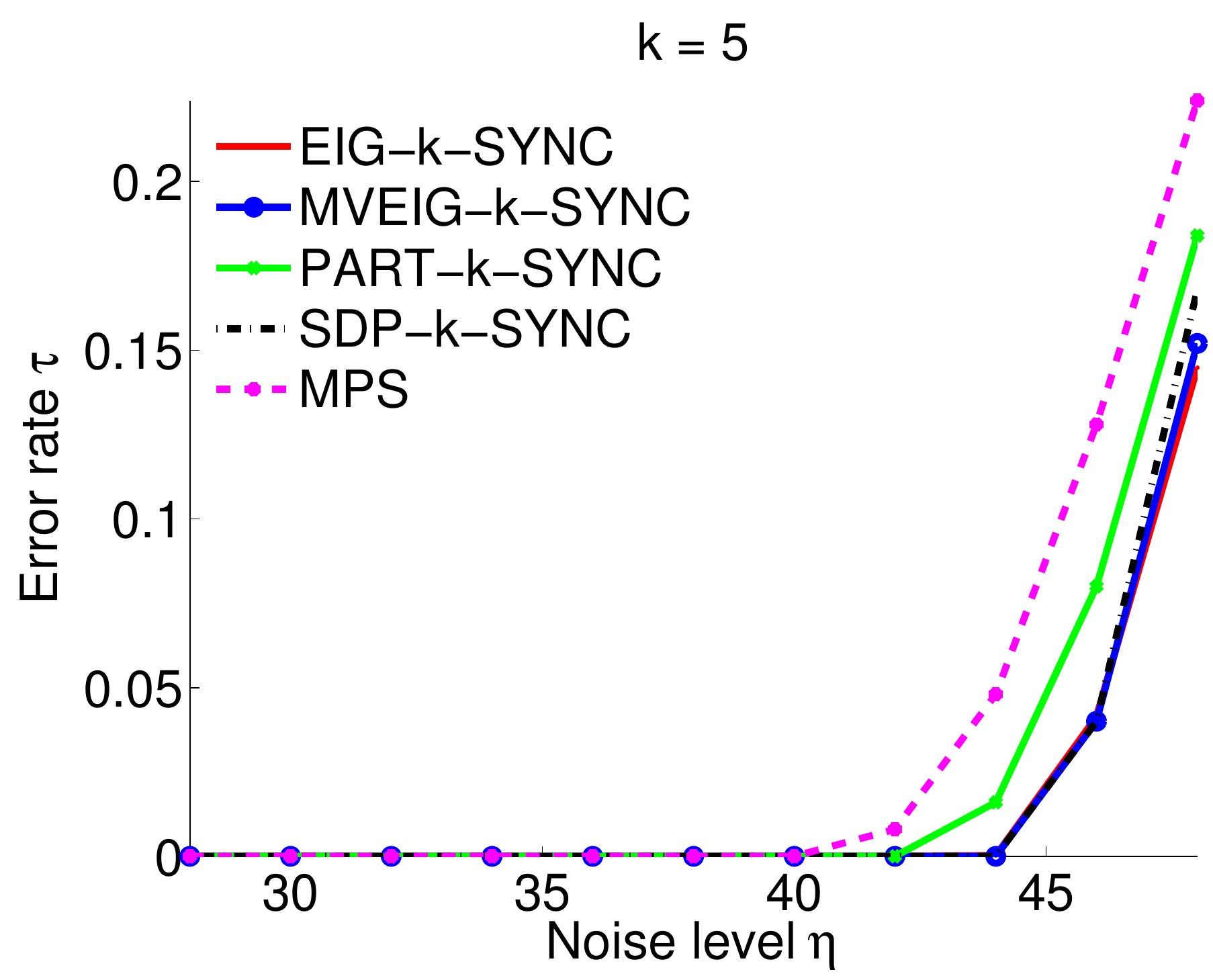}
\includegraphics[width=0.32\columnwidth]{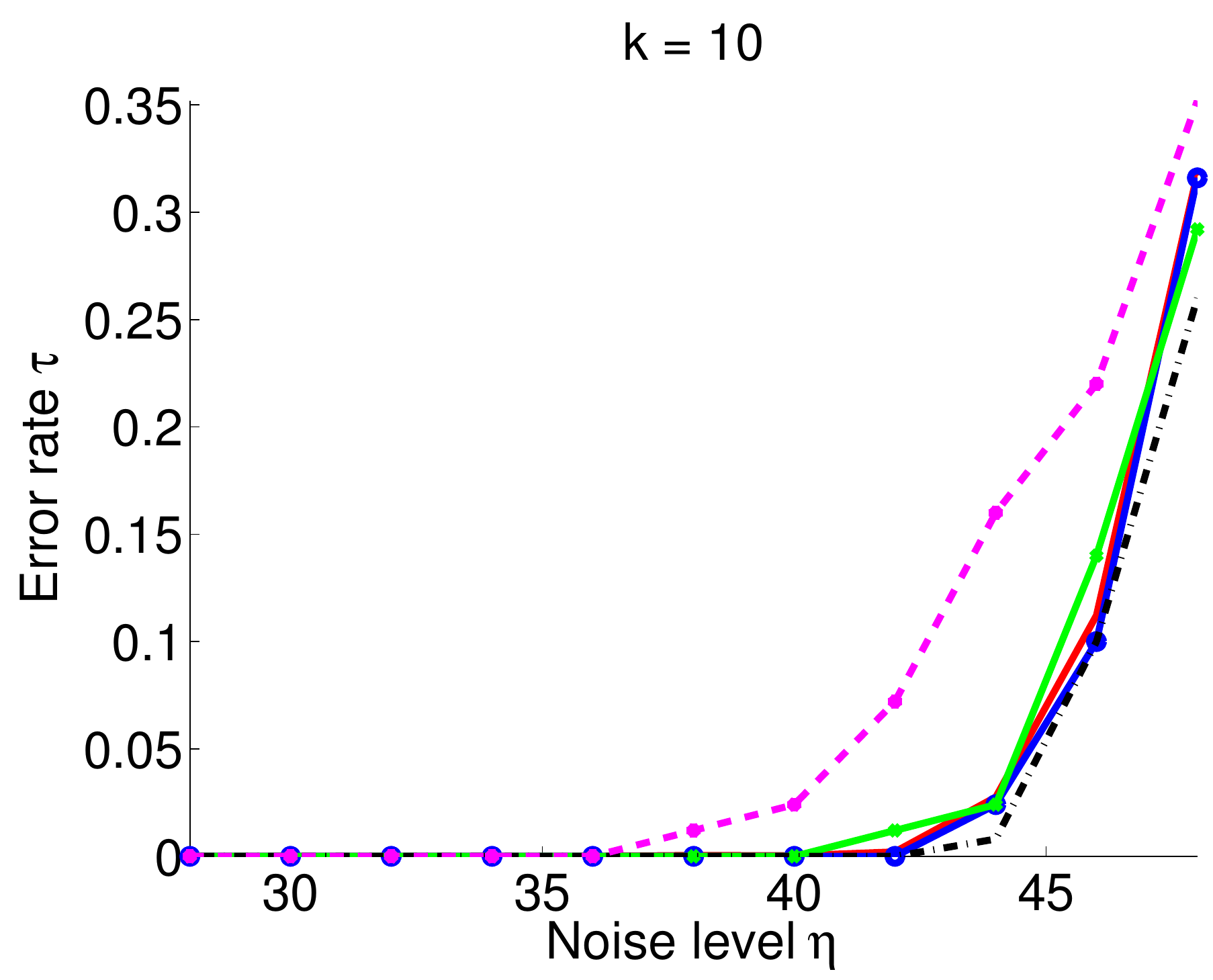}
\includegraphics[width=0.32\columnwidth]{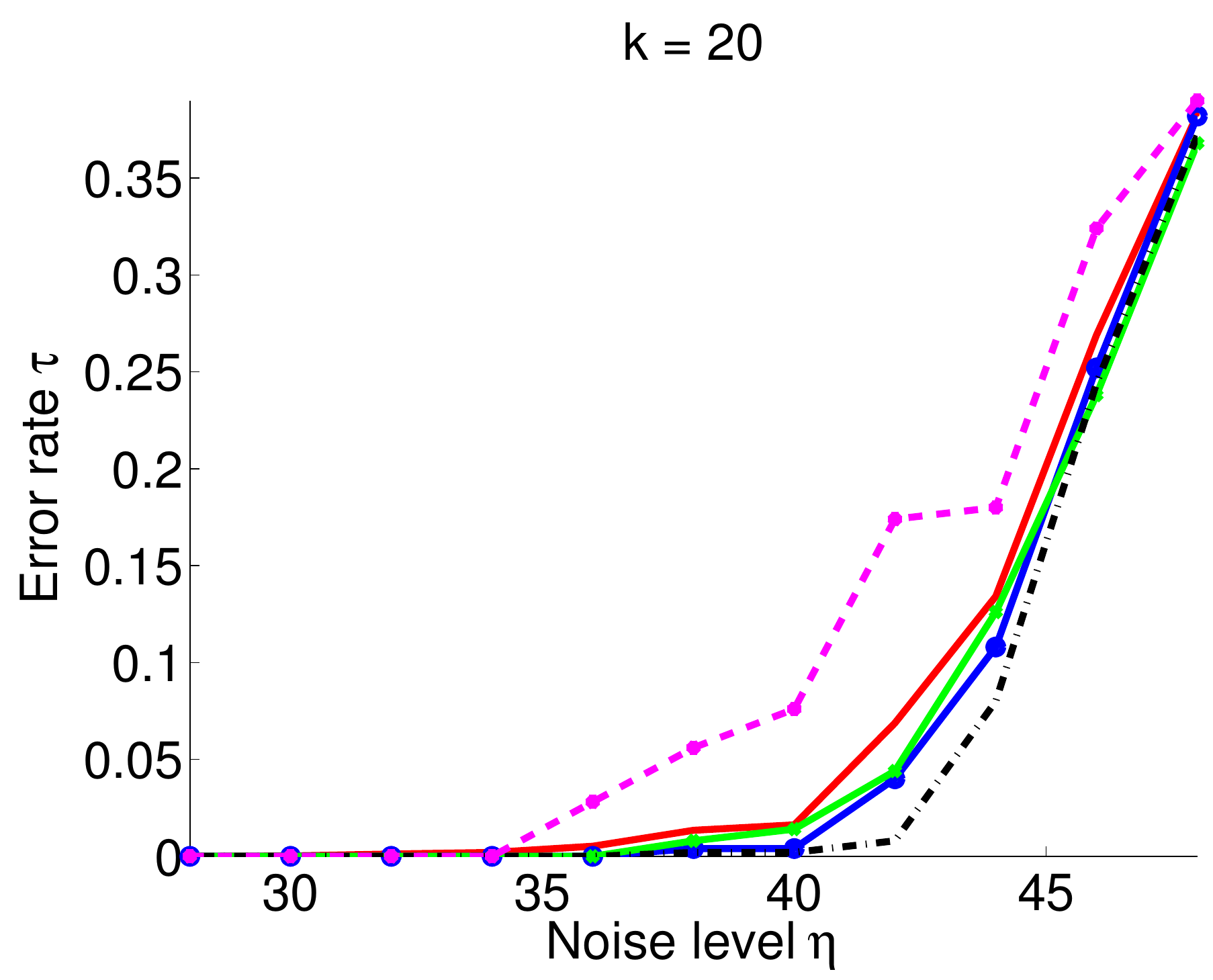}
\includegraphics[width=0.32\columnwidth]{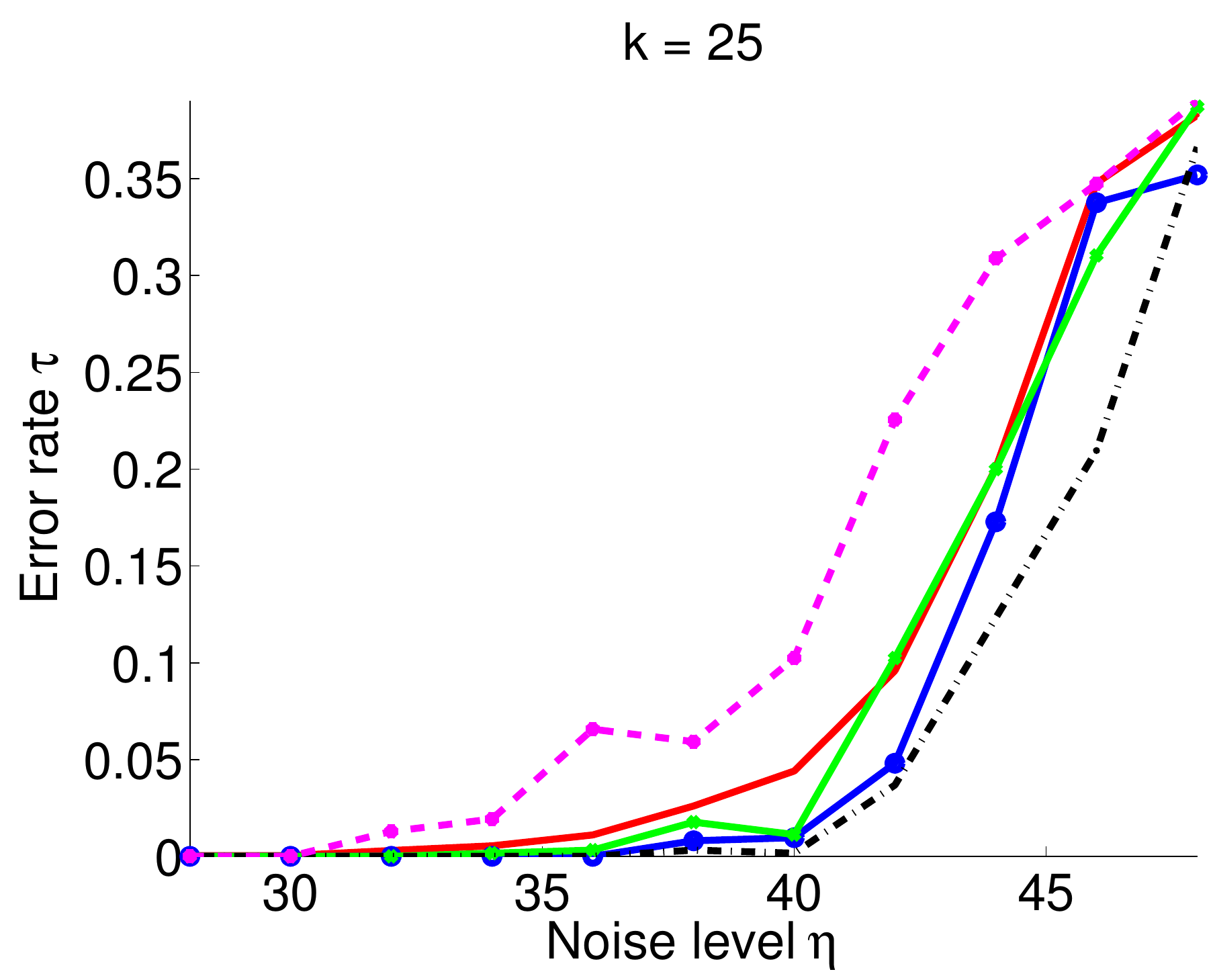}
\includegraphics[width=0.32\columnwidth]{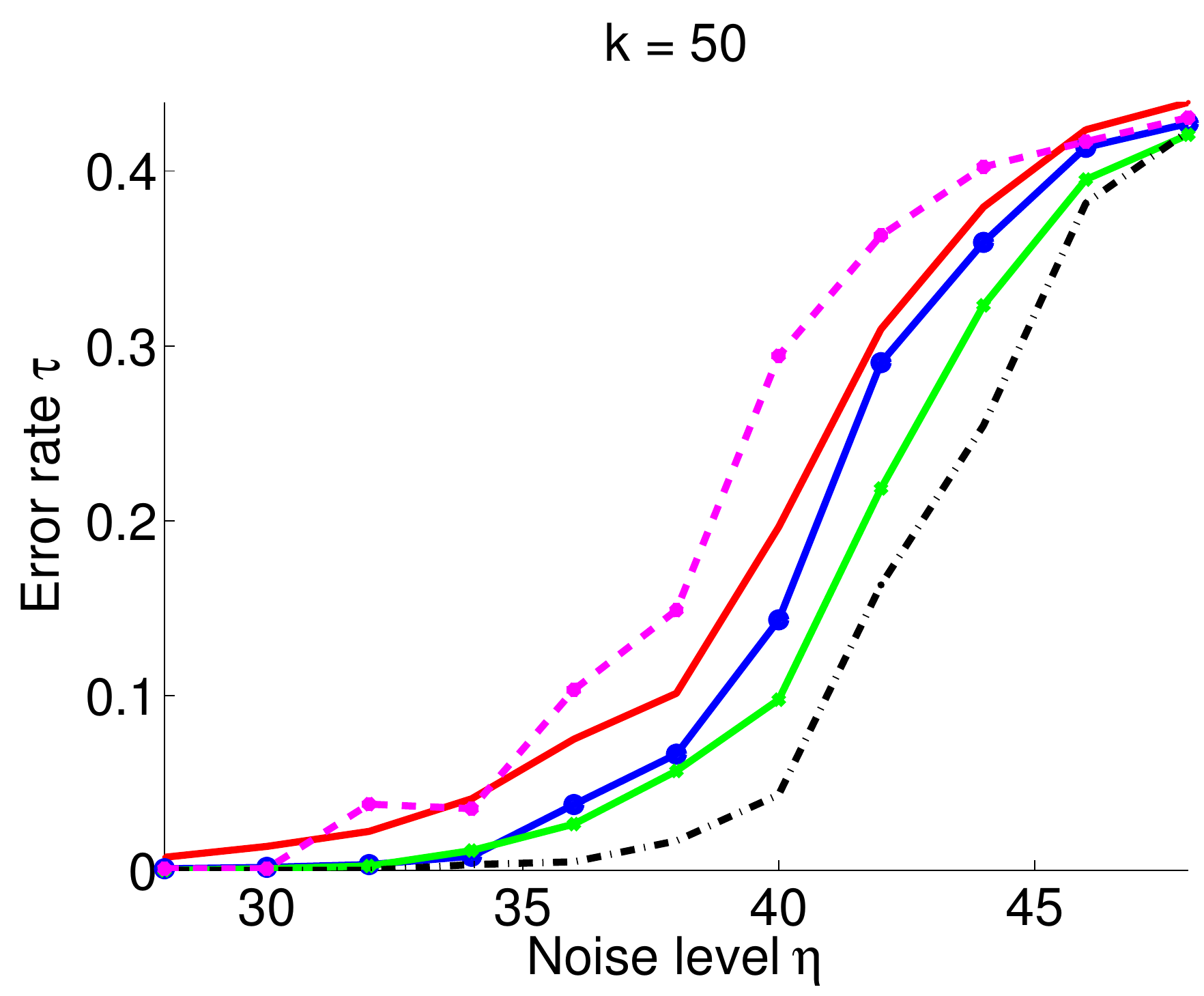}
\includegraphics[width=0.32\columnwidth]{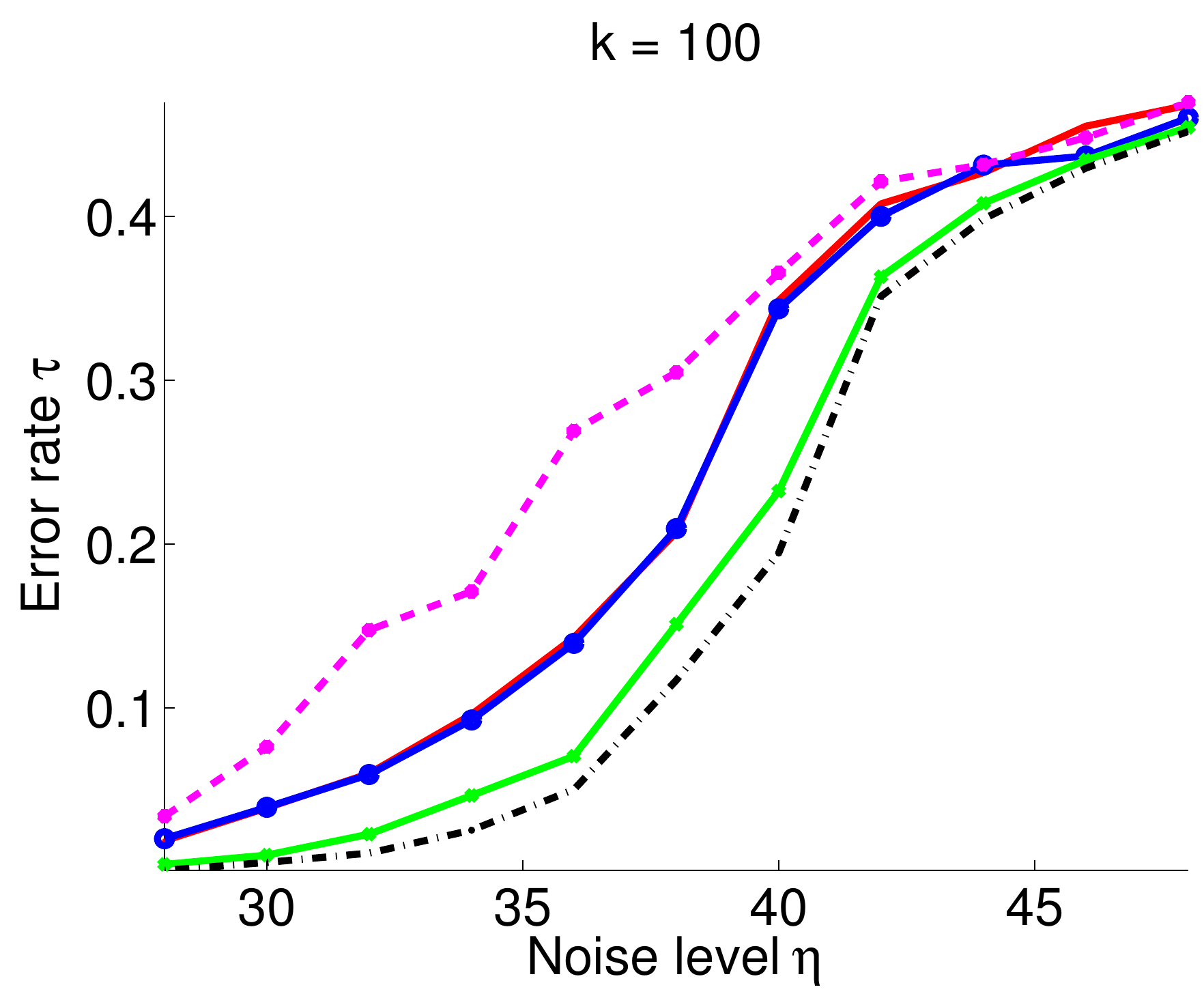}
\end{center}
\caption{ A network of $n=200$ nodes partitioned into $k$ blocks of equal size $m=\frac{200}{k}$. 
Each block corresponds to a unique senator, and all nodes within the same partition correspond to multiple terms that senator has served on. The graph is no longer a ``time-series" graph, but an Erd\H{o}s-R\'{e}nyi $G(n,\alpha)$ with $n=200$ and $\alpha=0.10$. Note that in this model the "line" structure of the network is no longer present. 
}
\label{fig:fixedMult}
\end{figure}

\textbf{Synthetic benchmark  II.} In Figure \ref{fig:fixedMult} we consider a second set of experiments on a different network model,  where we assume that all partition sets have the same size $m=\frac{200}{k}$, nodes within the same partition correspond to the same element $z_i \in \mathbb{Z}_2$, and for any pair of nodes that belong to two different partitions there is available information on their measure of similarity/dissimilarity. 
We point out that this synthetic model is not appropriate for a Congress model since it is unrealistic to assume that one can extract a similarity measure for two senators in two different (perhaps very far apart in time) Congresses. However, it still remains a valid benchmark for a synchronization algorithm on a partitioned network.  

The adjacency graph of the network is a random Erd\H{o}s-R\'{e}nyi graph $G(n,\alpha)$ with $n=200$ and $\alpha=0.10$, with the additional constraint that nodes that belong to the same partition are always connected. In other words, a pair of nodes $(i,j)$ from distinct partitions share an edge with probability $\alpha$ and have $Z_{ij}=\pm1$ depending on their measure of similarity and whether the edge has been corrupted with noise or no, while nodes $(i,j)$ within the same partition are connected with probability 1 and have $Z_{ij}=1$. Note that the graph is no longer a ``time series" graph as in the previous example, and the ``line" structure no longer exists. In our experiments, we vary the number of partitions in the range $k=\{5,10,20,25,50,100\}$. When $k$ is small (and hence $m$ large) all algorithms perform very well as most of the edges in the graph have noiseless measurements, and only edges across different partitions can be corrupted by noise. On the other hand, for large values of $k$, the sizes of the complete subgraphs of noiseless edges that span nodes corresponding to the same partition is much smaller, and a larger percentage of edges are subject to having their measurement corrupted by noise. As expected, in the latter scenario, the noise robustness of all algorithms degrades, with SDP and MPS being the best, respectively worst, performers.

\textbf{Real Congress data set.} Finally, we apply the aforementioned algorithms to the voting Congress data set. As previously described in Section \ref{sec:syncCongress}, it is very often the case that a senator serves on
multiple Congresses. Note that usually  senators serve on consecutive Congresses, but this is not always the case. Furthermore, for the purpose of this experiment, we restrict out attention only to the first 20 consecutive Congresses, due to the computationally expensive SDP computations. We start as described in Section \ref{sec:syncCongress}, by linearly mapping the voting agreement patterns (percentage of laws on which two senators agree) to values  in the interval $[-1,1]$ using the transformation (\ref{transformation}). For increased robustness, we hard-threshold the entries of matrix $\overline{W}$ by zeroing out those whose absolute value is smaller than a chosen threshold $\theta=0.45$. Our choice for the value of $\theta$ was made with two considerations in mind. A small value of $\theta$ would allow for a larger number of noisy measurements, while a large value would filter out many noisy edges. However, a large value for  $\theta$ would disconnect the underlying graph associated to the matrix $\overline{W}$ of pairwise measurements. We have also experimented with smaller values of the parameter $\theta$, but have consistently obtained less accurate results across all algorithms.

We have tested the five algorithms described at the beginning of this Section for solving the k-SYNC problem on the real Congress data set. We conducted experiments in an incremental approach, by expanding the network to include an additional Congress at each step. In other words, we start by synchronizing the first Congress, where each node of the network corresponds to a unique senator. In the next experiment, we add the second Congress, and already there are senators who have served on both Congress. When considering multiple Congresses, we denote the multiplicity of a senator to be the number of distinct Congresses on which she or he served. More specifically, for a given voting network on a total of $n$ nodes that span $r$ consecutive Congresses, each of size $S$, with $n=r S$, we denote the unique senators by $u_1,\ldots u_k \in \mathbb{Z}_2$ and their corresponding multiplicities by $m_1,\ldots,m_k$, with $m_1+m_2 + \ldots + m_k = n$.

\begin{figure}[h]
\begin{center}
\includegraphics[width=0.33\columnwidth]{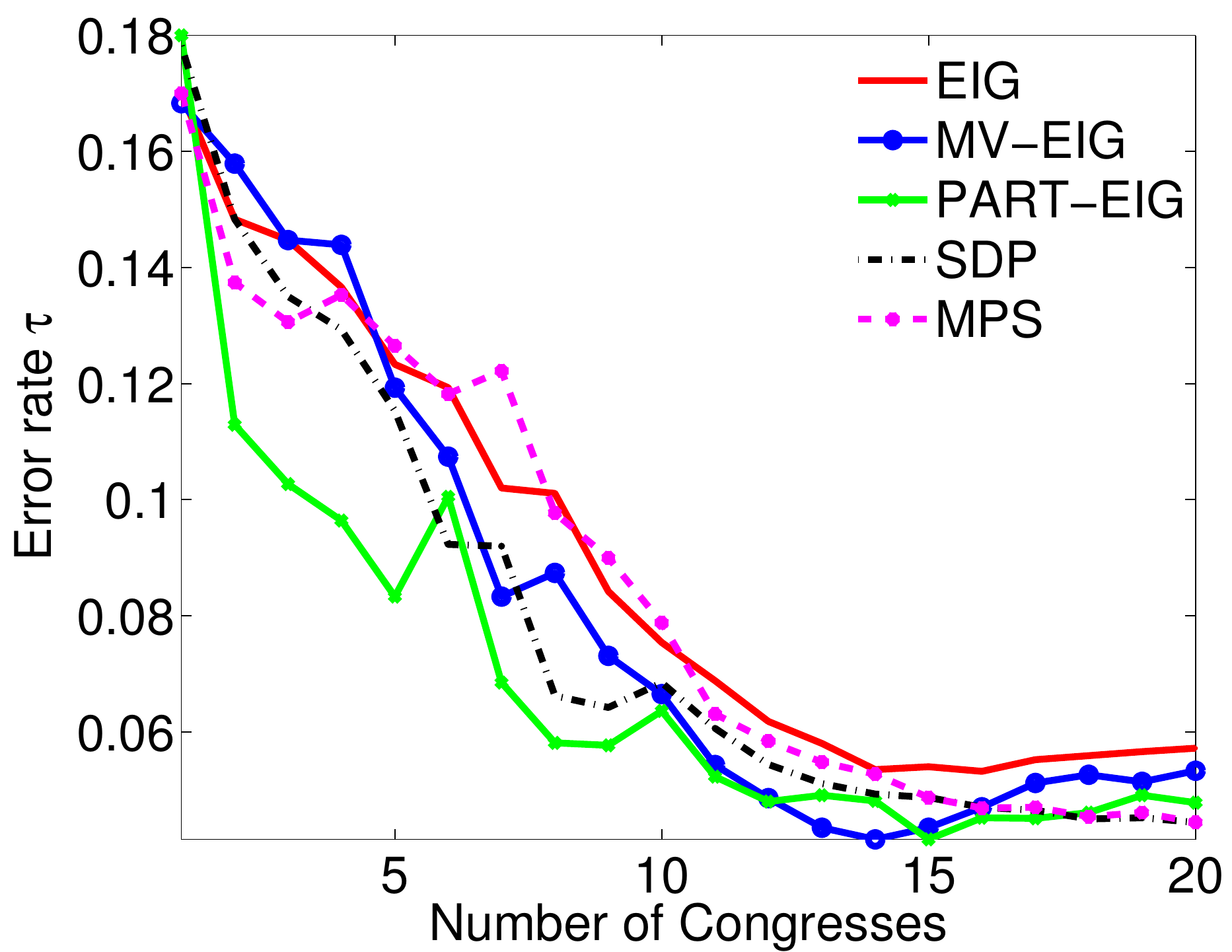}
\includegraphics[width=0.32\columnwidth]{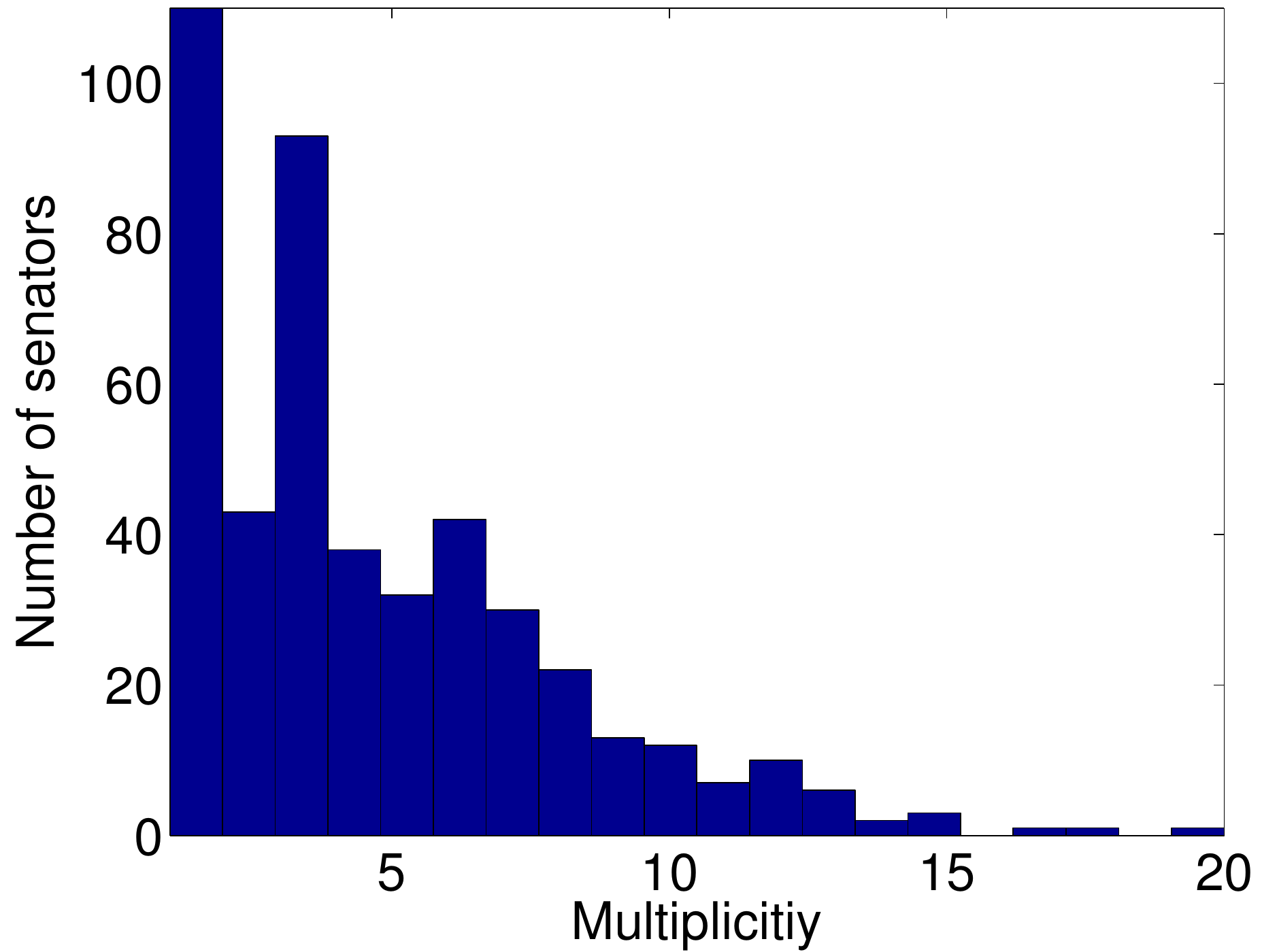}
\includegraphics[width=0.32\columnwidth]{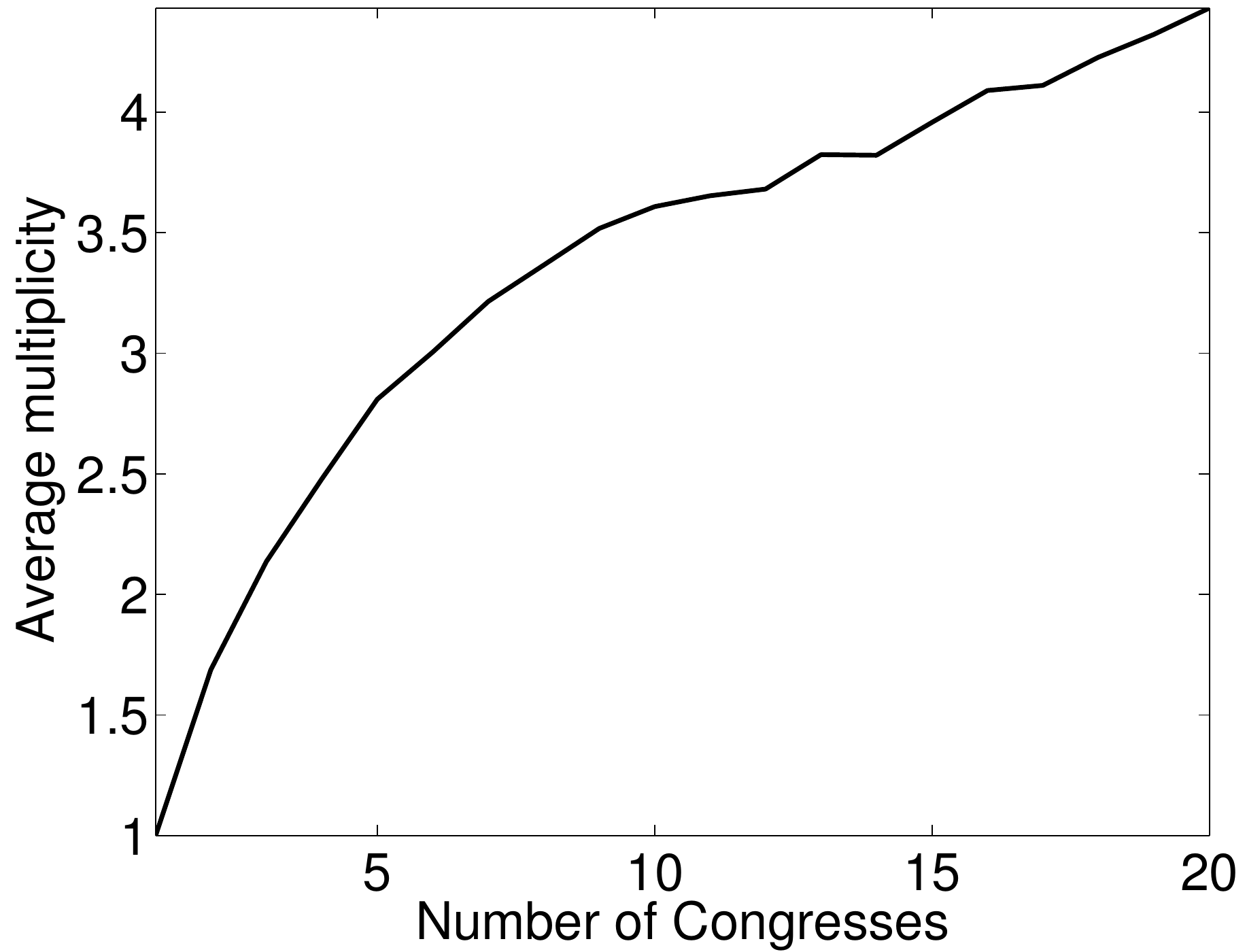}
\end{center}
\caption{ Left: the error rates as we increase the number of Congresses, across all algorithms proposed for k-SYNC. Middle: histogram of the multiplicity of all senators, at the end of the 20 Congresses we have considered in this experiment. Right: The average multiplicity of a senators, as the number of Congresses increases.
}
\label{fig:ksyncRealCong}
\end{figure}

As we add more Congresses to the network, more and more senators will have participated in multiple Congresses. The middle plot in Figure \ref{fig:ksyncRealCong} shows the histogram of the number of (non-necessarily consecutive) terms each of the $k=735$ unique senators has served on, across all 20 Congresses we considered. Note that a senator serves on average on 5.6 terms, and very few senators serve on more than 15 Congresses. The right plot in Figure \ref{fig:ksyncRealCong} shows the average multiplicity of the senators during the first 20 consecutive Congresses. Note that, as time goes on, the average multiplicity increases which means that senators tend to serve on more Congresses and only a few new senators join the Congress. This last figure also explains the descending slope of the error rates for all algorithms. Any senator $u_i$ that has served on multiple terms adds more ``redundancy" to the network, in the sense that we have available a larger number of voting patterns for senator $u_i$, which makes it easier to decide to which party she or he belongs to.

\section{Summary and open related problems}  \label{sec:summary}
\normalsize


In this paper we have investigated the synchronization problem over the group $\mathbb{Z}_2$, where the goal is to recover the unknown group elements from noisy pairwise measurements of their ratios.  We applied the eigenvector synchronization method to the U.S. Congress roll call voting data set, to robustly  identify the two political parties across time, taking into account the additional constraints due to senators serving on multiple Congresses. We used tools from random matrix theory, in particular recent results on perturbation of rank-one matrices, to give an analysis of the robustness to noise of the eigenvector method when the underlying graph of pairwise measurements is the Erd\H{o}s-R\'{e}nyi random graph. Furthermore, we also proposed a message passing synchronization algorithm that outperforms the existing eigenvector synchronization algorithm only for certain classes of graphs and noise regimes, and also enjoys the flexibility of incorporating additional  constraints that may not be easily accommodated by any of the other spectral or SDP-based methods. Finally, we have proposed and compared the performance of several algorithms based on spectral and SDP relaxations, and message passing, for the synchronization problem over $\mathbb{Z}_2$  when one has the additional information that certain subsets of nodes (that altogether form a partition of the network) represent the same unknown group element.

There are a number of interesting open questions left for future investigation.  One such question relates to the noise sensitivity analysis of the eigenvector synchronization method for various noise models, and when the underlying graph is no longer complete or Erd\H{o}s-R\'{e}nyi, but a random geometric graph or any other graph that has local structure. This calls for more advanced tools from random matrix theory
in order to understand the behavior of the largest eigenvalue of low-rank perturbations of large structured random matrices.

Another interesting direction to consider is an application of our methods to other data sets of similar nature. For example,  in finance, the given data may be readily available in the form of correlation matrices indexed by time. Consider for example an instance where one has available a correlation matrix (based on daily prices) between a class of $n$ assets, for each year during the past $N$ years. For example, one may be interested in tracking the evolution of a certain index which, at any given time, tracks a basket composed of $n$ stocks that have a natural separation into two categories (for example stocks belonging to two different industries). At the end of the year,  the index is \textit{rebalanced}, meaning that certain stocks will drop out of the index and others will join the index, such that the number of constituents in the index remains constant $n$.  Instances of such data are ubiquitous in the financial world,  where companies belonging to a given sector behave as they belong to a another sector (for example, an airline company which has high exposure to the financial sector, would be closer in behavior to the financial sector than to its own airline sector). Such an approach could be used to detect  misclassifications in existing hierarchical divisions such as  the standard GICS (Global Industry Classification System) or ICB (Industry Classification Benchmark)  \cite{GICS_ICB,GICSaffiliation}.

An analysis of the convergence and robustness to noise of the proposed message passing algorithm, and the investigation of a possible phase transition behavior are other venues of research, beyond the scope of this paper. A much more ambitious goal would be to further expand the proposed algorithm to groups other than $\mathbb{Z}_2$ (for example, SO(d) or O(d)) and compare it with the existing spectral and SDP relaxations.

Furthermore, understanding the role of the lower-order eigenvectors in the synchronization problem and the extent to which one may extract useful information from them is another question of interest. Empirical evidence in the case of $\mathbb{Z}_2$ suggests that subsequent eigenvectors localize on the ``inconsistent" parts of the network, i.e., subgraphs whose edges contain the largest amount of noise. In other words, our preliminary investigation suggests that such lower order eigenvectors  point out local cuts in the underlying subgraph of noisy edges.

Continuing the analogy with the graph cut problem, another interesting question concerns a flow-based approach for solving the synchronization problem. SDP, spectral and flow-based methods are the three main classes that cover the most popular graph partitioning relaxation algorithms, usually with tradeoffs between cut quality, balance and computational cost. However, the group synchronization problem is currently solved via SDP and spectral relaxations, and no flow-based approach currently exists.


The eigenvector method itself cannot accommodate available additional information, such as the existence of anchors or partition information. However, it would be interesting to explore whether the traditional power method can be modified to incorporate such additional constraints at each iteration (for example, reinforcing at each round that the entries corresponding to the available anchors have the desired sign), and to analyze the convergence and robustness to noise of this approach.

One may also consider an online version of synchronization,  for a graph whose set of nodes increases by one at each iteration. At each update a new vertex $v_i$ is added to the graph $G_t$, with $V(G_{t+1}) = V(G_t) \cup \{v_i\} $, and $v_i$ is connected to $k$ randomly chosen existing vertices in $G_t$, with the new edge connections holding both accurate and noise corrupted pairwise measurements between the new node and the rest of the graph. 
A trivial approach to classifying the polarity of the new node would be to consider all its neighbors, and assume their polarity is fixed and already available (from a previous solution for $G_t$). Each such neighbor $v_j$ contributes with a vote on the possible parity of node $v_i$, i.e., the vote would be $Z_{ij} z_j$. Considering all the $k$ votes from the neighbors of $v_i$, one can then consider the majority vote and decide on the parity of the new node $v_i$ accordingly. In other words, the estimate would be given by 
\begin{equation}
\hat{z}_i = \mbox{median}( Z_{ij_1} \hat{z}_{j_1}, \ldots,  Z_{i j_k}  \hat{z}_{j_k} )
\end{equation}
 Alternatively, one may explore a Nystr\H{o}m extension-based method, to integrate consistently on the fly the newly arrived information with the existing one, without having to recompute from scratch a global solution. While the addition of a new node, or several such new nodes, may not affect the stability of the overall accuracy of the current assignment, one may expect that if enough new nodes are added to the network, the current solution (obtained via the above Nystr\H{o}m extension procedure) may be significantly different than the solution one would obtain by recomputing from scratch a new global solution of $G_{t+1}$ taking into account the newly arrived information (either via the eigenvector synchronization algorithm or the SDP approach if additional constraints exist). Thus, it is natural to ask how often does one need to recompute a global solution for such an online algorithm and still preserve the accuracy of the overall solution. For example, the percentage of \textit{unhappy} edges (edges for which $z_iz_jZ_{ij} = -1$) may be a useful proxy for deciding when a global recalculation for $G_{t+1}$ is required.

Another question worth investigating is a combinatorial characterization of \textit{synchronizable} graphs, i.e., graphs for which the synchronization problem yields a perfect solution under certain combinatorial sufficient and necessary conditions on the underlying subgraphs of good and bad edges. Perhaps a first step in the above direction would be to provide such a combinatorial characterization for certain families of graphs. A trivial example in this direction is the following. If every bad edge $e_{t}=(i,j)$ of the graph $G$ is contained in a subgraph $H_t=K_4$ such that $|V(H) \cap V(G\backslash H)|=1$ and the remaining other 5 edges of $H_t$ are all correct, then we know for sure we can provide a perfect solution for the synchronization problem. This stems from the following two facts. First, we can decompose the objective function $f_G$ in (\ref{maxZ2}) associated to graph $G$, into two separate objective functions $f_H$ and $f_{G\backslash H}$ that can be solved  independently, and later merged into an global solution that remains optimal. Note that this observation remains true for any subgraph $H$ (not necessarily $K_4$) as long as the subgraphs $H$ and $G\backslash H$ have at most one node in common (the statement is trivially true is they do not overlap, as we can synchronize each connected component separately). 
Second, if exactly one edge of a complete graph on four vertices has been corrupted by noise, then there exists a unique solution (up to a global sign change) that maximizes the objective function in (\ref{maxZ2}) associated to a $K_4$ graph, and that unique solution is also noiseless.

Another possible venue to investigate is an information theoretic analysis of the synchronization problem over $\mathbb{Z}_2$, for various families of graphs and noise models. An analysis of the eigenvector synchronization method in the presence of noise was first explored by Singer \cite{sync}, in the case of the group SO(2), and uses tools from random matrix theory that allow for a precise matrix perturbation analysis that quantifies the robustness to noise of the method under a certain random noise model. Furthermore, it provided  an information theoretic analysis showing that the eigenvector method is asymptotically nearly optimal and achieves the information theoretic Shannon bound up to a multiplicative factor that depends only on the discretization error of the measurements.
We expect that a similar approach would yield an information theoretic lower bound for the permissible amount of noise that any synchronization algorithm over $\mathbb{Z}_2$ can tolerate,  while still being able to provide an accurate recovery. On a related note, Bandeira et al. \cite{afonso} proved very recently a Cheeger-type inequality via the graph Connection Laplacian operator, providing a deterministic worst case performance guarantee for the synchronization problem over the group O(d) of orthogonal transformations.

Finally, motivated by the size of many of the modern data sets, it is natural to consider a divide-and-conquer approach for the synchronization method. While the eigenvector algorithm scales well for sparse networks and runs in almost linear time in the number of egdes (with the number of iterations depending on the spectral gap), for very large networks with millions of nodes, a spectral approach is far from being computationally feasible. An alternative method one can propose would be to partition the graph into many non-overlapping subgraphs $S_1,S_2,\ldots,S_t$ (using for example the highly scalable Louvain method for community detection in graphs \cite{louvainMethod}), compute an optimal solution $\hat{S}_i$ for each subgraph via any of the available methods, and finally piece together all local solutions by running a final synchronization algorithm on the new graph $H$ whose nodes $h_i$ correspond to subgraphs $S_i$. The pairwise measurement between a pair of nodes $(h_i, h_j)$ is obtained by aggregating the information available in the edges of the initial graph $G$ between the subgraph $S_i$ and $S_j$. In such an approach, the final solution for the $n$ nodes of the initial graph $G$ would be given by $\hat{h}_1 \hat{S}_1, \ldots, \hat{h}_k \hat{S}_k$. In this setting, a natural question to ask would be how much worse the resulting solution can be, with respect to the expensive calculation of a global solution of the entire network.


\section*{Acknowledgments}
We gratefully acknowledge the SAMSI Institute for hosting the year-long program in Complex Networks, providing support, and facilitating discussions via workshops and research meetings. We are grateful to Amit Singer for his guidance and support via Award Number R01GM090200 from the NIGMS, and Award Number FA9550-09-1-0551 from AFOSR, and to Andrea Bertozzi for support via award number AFOSR MURI grant FA9550-10-1-0569. We thank Prakash Balachadnran, Mauro Maggioni, Michael Mahoney, Peter Mucha and Mason Porter for the many useful discussions during the SAMSI workshops and research meetings, and to the anonymous reviewers for their thorough and constructive comments.  We acknowledge Peter Mucha and Mason Porter for providing us with a processed version of the U.S. Congress roll call voting data set.


\clearpage

\section{Appendix A: Noise sensitivity analysis for synchronization over $\mathbb{Z}_2$} \label{sec:Appendix_rmtx}


This appendix gives an analysis of the eigenvector method for synchronization over $\mathbb{Z}_2$ when the underlying graph $G$ of pairwise measurements is an Erd\H{o}s-R\'{e}nyi random graph (or possibly a complete graph $K_n$), and the subgraph of noisy edges is a random subgraph of $G$.  
We follow closely the analysis detailed in previous work by Singer \cite{sync}, pertaining to group SO(2), and note that crucial to the analysis are recent random matrix theory results of  F\'eral and P\'ech\'e \cite{FeralPeche} on the largest eigenvalue of rank-one deformation of (real, symmetric) large random matrices. 

We denote by $G(n,\alpha)$ the initial Erd\H{o}s-R\'{e}nyi random graph of pairwise measurements on $n$ vertices, where each edge is present with probability $p$ independent of the other edges. The probabilistic model for the noisy measurements on the existing edges is given by a binomial model where each measurement $Z_{ij}$ is correct with probability $p$, and incorrect with probability $\eta = 1-p$.  We denote by $E_{good} \subseteq E$, respectively $E_{bad} \subseteq E$, the subset of correct, respectively incorrect, edge measurements, and by $\bar{G}$ the complement of the graph $G$, whose edge set we denote by $E(\bar{G})$. In light of the above notation, the available measurement matrix $Z$ is given by the following mixture model
\begin{equation}
Z_{ij} = \left\{
     \begin{array}{rll}
   z_i  z_j^{-1}  & \;\; \text{ if } (i,j) \in E_{good}   	& \text{with probability } p \alpha 	\\
 - z_i  z_j^{-1}  & \;\; \text{ if } (i,j) \in E_{bad}, 	& \text{with probability } (1-p) \alpha	\\
		     0   & \;\; \text{ if } (i,j) \in E_{\bar{G}}, 	& \text{with probability } 1-\alpha	\\
     \end{array}
   \right.
\label{noiseModelZ2}
\end{equation}
Following the approach used in \cite{sync}, the goal is to write the given measurement matrix $Z$ as a rank-one perturbation, and make use us recent results from the random matrix theory community on the largest eigenvalue of such rank-one deformation.  For convenience, we set the diagonal entries of $Z$ to $Z_{ii} = (2p-1) \alpha$, and choose to denote $z_j^{-1}$ by $z_j$ since in $\mathbb{Z}_2$ an element is its own inverse. Next, we write the symmetric matrix $Z$ as a random perturbation of a rank-one matrix, with the expected value of its elements given by
\begin{equation}
\mathbb{E}(Z_{ij}) = z_i z_j p \alpha - z_i z_j (1-p) \alpha = z_i z_j (2p-1) \alpha.
\label{expZij}
\end{equation}
Note that the expected value of $Z$ is a rank-one matrix
\begin{equation}
      \mathbb{E}(Z_{ij}) = n \alpha (2p-1) \alpha \boldsymbol{tt}^{T},
\label{theRankOneMtx}
\end{equation}
where $\boldsymbol{t}$ is the normalized vector $\parallel \boldsymbol{t} \parallel = 1$ with entries
\begin{equation}
 	t_i = \frac{ z_i}{\sqrt{n}}, \;\; i=1,\ldots,n.
\label{tdef}
\end{equation}
We are now ready to write the matrix $Z$ in the form of a random perturbation of a rank-one matrix
\begin{equation}
 	Z = \mathbb{E}(Z_{ij}) + R =  n(2p-1) \alpha  \boldsymbol{t} \boldsymbol{t}^T + R,
\label{decompZ}  
\end{equation}
where $R$ is a random matrix with $R_{ii} = 0$, whose elements $R_{ij} =Z_{ij} - \mathbb{E}(Z_{ij}) $ are zero-mean independent random variables. Using (\ref{expZij}), the entries of the random matrix $R$ are given by the following mixture model
\begin{equation}
  R_{ij} = \left\{
     \begin{array}{rll}
   z_i z_j  -  z_i z_j (2p-1) \alpha,  & \;\; (i,j) \in E_{good},    	& \text{with probability } p \alpha   \\
  -z_i z_j  -  z_i z_j (2p-1) \alpha,  & \;\; (i,j) \in E_{bad},      	& \text{with probability } (1-p) \alpha	\\
              -  z_i z_j (2p-1)  \alpha,  & \;\; (i,j) \in E_{\bar{G}},    	& \text{with probability } 1-\alpha	\\
     \end{array}
   \right.
\label{second_H_ij_def}
\end{equation}
Furthermore, the entries of $R$ have zero mean 
\begin{align*}
\mathbb{E}(R_{ij})  & = [ z_i z_j ( 1 - (2p-1) \alpha ]    p \alpha           +  [ -z_i z_j ( 1 + (2p-1) \alpha) ]  (1-p) \alpha + [  -  z_i z_j (2p-1)  \alpha]  (1-\alpha) \\
		                & =  z_i z_j \alpha [  (1 - 2p \alpha  + \alpha)  p    - (1 + 2p \alpha  - \alpha)  (1-p)                     -     (2p-1)   (1-\alpha)   ]  = 0, 	
\end{align*}
where the last equality follows after several algebra manipulations. The diagonal elements $R_{ij}$ have zero variance, while the off-diagonal elements have variance 
\begin{equation}
	Var(R_{ij}) = [1 - (2p-1) \alpha]^2  p \alpha    +  [ 1 + (2p-1) \alpha]^2  (1-p) \alpha  +  [ (2p-1)  \alpha]^2  (1-\alpha) =\alpha (1-\alpha + 4 p \alpha - 4 p^2 \alpha) \\
\end{equation}
using the fact that $(z_i z_j)^2 = 1$.
Note that in the case of a complete graph, $G=K_n$ and $\alpha=1$, the variance vanishes when $p=1$, since all edges are correct. This is also the case whenever $p=0$ and all edges are incorrect, meaning the sign of all equations is flipped, in which case we still recover the original solution. When $G$ is a random graph, the variance is $\alpha (1-\alpha)$, whenever $p=0$ or $p=1$.


Following the approach in \cite{sync}, the interpretation of $R$ as a random matrix now comes into play, with the distribution of its eigenvalues following Wigner's semi-circle law \cite{wigner1,wigner2}. Denoting
$ \theta=n(2p-1) \alpha$  and  $ \sigma^2 = n\alpha (1-\alpha + 4 p \alpha - 4 p^2 \alpha)$, 
an adaptation of Theorem 1.1 from \cite{FeralPeche}
guarantees that the largest eigenvalue $\lambda_1(Z)$ jumps almost surely outside the support $[ -2 \sigma, 2\sigma]$ of the semicircle law
as soon as $ \theta > \sigma$,
which translates to
\begin{equation}
	n(2p-1) \alpha > \sqrt{   n\alpha (1-\alpha + 4 p \alpha - 4 p^2 \alpha)   }.
\label{condgap}
\end{equation}
After squaring both sides and solving the associated quadratic, the above inequality holds true whenever
	 $p >  \frac{1}{2} + \frac{1}{2 \sqrt{\alpha(n+1)}} $
since we are considering the case when $ p >\frac{1}{2}$. In other words, as soon as the correct edge probability $p$ is larger than the threshold probability
\begin{equation}
 p^*  \approx  \frac{1}{2}+\frac{1}{2\sqrt{\alpha n}}.
\label{condpTh}
\end{equation}
we are guaranteed that the largest eigenvalue $\lambda_1(Z)$ will jump outside the support 
$ [-2\sigma,2\sigma] $ of the semi-circle law \cite{FeralPeche,peche}. Note that the results of \cite{FeralPeche,peche} on the distribution of the largest eigenvalue of perturbed rank-1 matrices hold for the case when the rank-1 matrix is assumed to be a constant matrix. In the above setting (\ref{decompZ}), this translates to the rank-1 matrix in (\ref{theRankOneMtx}) having constant entries, in other words it is required that the vector defined by (\ref{tdef}) to be the all ones vector, an assumption we can make without loss of generality since once can reduce the $tt^{T}$ matrix to the all ones matrix by conjugating with the diagonal matrix $Z$, whose diagonal entries are given by $Z_{ii}=z_i$. 
Figure \ref{fig:spectra_Compl_ErdosR_short}  
shows histograms of the eigenvalues of the matrix $Z$ of size $n=1000$ for the complete graph case (thus $\alpha=1$), as well as the corresponding error levels $\tau$, for different values of $p$. For this experiment,  $\frac{1}{2} + \frac{1}{2 \sqrt{n+1}} \approx 51.6$, and indeed the recovered solution is meaningful as long as $p$ is above this threshold.


The correlation between the ground truth solution and the top eigenvector $\boldsymbol{v_1}$ can be lower bounded by the spectral gap between the top eigenvalues of matrices $Z$ and $R$. If $ \boldsymbol{v_1}$ denotes the top eigenvector of $Z$, with corresponding eigenvalue  $\lambda_1^{(Z)}$,  $ Z \boldsymbol{v_1} = \lambda_1^{(Z)} \boldsymbol{v_1}$, the following can be concluded using the rank-one perturbation in  (\ref{decompZ})
\begin{equation}
	 \lambda_1^{(Z)} \boldsymbol{v_1} =  Z \boldsymbol{v_1} = \left[ n(2p-1) \alpha  \boldsymbol{t} \boldsymbol{t}^T + R \right] \boldsymbol{v_1}
\end{equation}
Left multiplication by $\boldsymbol{v_1}^T$ yields
\begin{equation}
	\lambda_1^{(Z)} =  \boldsymbol{v_1}^T \left[ n(2p-1) \alpha  \boldsymbol{t} \boldsymbol{t}^T + R \right] \boldsymbol{v_1}  =    n \alpha (2p-1)  \langle \boldsymbol{v_1}, \boldsymbol{t}  \rangle ^2  +   \boldsymbol{v_1}^T R  \boldsymbol{v_1} 
\end{equation}
Since $ \boldsymbol{v_1}^T R \boldsymbol{v_1} \leq  \lambda_1^{(R)} $, where $ \lambda_1^{(R)} $ denotes the top eigenvector of the noise random matrix $R$.
 \begin{equation}
	\langle \boldsymbol{v_1},\boldsymbol{t}  \rangle ^2  \geq    \frac{      \lambda_1^{(Z)}   -  \lambda_1^{(R)}      }   { n \alpha (2p-1) },
 \end{equation}
in other words, the larger the gap between   $ \lambda_1^{(Z)}   -  \lambda_1^{(R)} $, the higher the correlation between the recovered solution and the ground truth.

\section{Appendix B: Synchronization with Anchors (ANCH-SYNC)} \label{sec:Appendix_sync_anchors}

In the context of the synchronization problem, \textit{anchors} are nodes whose corresponding group element in $\mathbb{Z}_2$ is known a priori. We will refer to the non-anchor nodes as \textit{sensors}, following the terminology from sensor network localization. For a given graph $G=(V,E)$ with node set $V$ ($|V|=n$) corresponding to a set of $n$ group elements composed of anchors $A=\{a_1,\ldots,a_h\}$ with $a_i \in \mathbb{Z}_2$ and sensors $S=\{s_1,\ldots,s_l\}$ with $ s_i \in \mathbb{Z}_2$, with $n = h + l$, and edge set $E$ of size $m$ corresponding to an incomplete set of $m$ (possibly noisy) pairwise group measurements $s_i s_j^{-1}$ with $s_i,s_j \in S$ or $a_i s_j^{-1}$ with $a_i \in A, s_j \in S$, the goal is to provide accurate estimates $\hat{s}_1,\ldots,\hat{s}_l \in \mathbb{Z}_2$ for the unknown sensor group elements $s_1,\ldots,s_l$.

The synchronization problem in the presence of anchors, that we shall refer from now on as ANCH-SYNC, can no longer be cast as an eigenvector problem. In recent work \cite{asap3d}, we introduced several methods for incorporating anchor information in the  synchronization problem over $\mathbb{Z}_2$, in the context of the molecule problem from structural biology. The first approach for solving ANCH-SYNC that we proposed in \cite{asap3d} relies on casting the problem as a quadratically constrained quadratic program (QCQP), while a second one  relies on an SDP formulation. We briefly summarize below the above two approaches, and refer the reader to Section 7 of \cite{asap3d} for additional details. The purpose of this section is to compare their performance with the message passing synchronization algorithm introduced in Section \ref{sec:MPD}.

The QCQP method follows a similar approach to equations (\ref{maxZ2}), (\ref{relmaxZ2}) and (\ref{LS_DZ}) that motivated the eigenvector synchronization method. Unfortunately, maximizing the quadratic form $\boldsymbol{x}^{T}Z \boldsymbol{x}$ under the anchor constraints $x_i = a_i, i \in \mathcal{A}$, is no longer an eigenvector problem. In order to incorporate the additional anchor information  we combined under the same objective function a quadratic term that corresponds to the contribution of the sensor-sensor pairwise measurements, and  a linear term that represents the contribution of the anchors-sensor pairwise measurements. Writing the solution vector in the form $\boldsymbol{x} = [ \boldsymbol{s} \; \boldsymbol{a}]^T$ that denotes both sensors and anchors, in \cite{asap3d} we formulated the synchronization problem as a least squares problem, by minimizing the quadratic form in (\ref{LS_DZ}) written as
\begin{equation}
 \left
[ \begin{array}{cc} \boldsymbol{s}^T  &  \boldsymbol{a}^T  \\  \end{array} \right]
\left[ \begin{array}{cc}
D_S - S & - U  \\
- U^{T} & D_V - V \\
\end{array} \right]
\left[ \begin{array}{c}
				\boldsymbol{s}  \\
				\boldsymbol{a} \\
				\end{array} \right]
 = \boldsymbol{s}^T (D_S - S)  \boldsymbol{s} - 2 \boldsymbol{s}^T U \boldsymbol{a} + \boldsymbol{a}^T (D_V - V) \boldsymbol{a},
\label{blockform}
\end{equation}
with
$$ Z = \left[ \begin{array}{cc}
S & U  \\
U^{T} & V \\
\end{array} \right],\;\;\;\;\;\;\;\;\;\;\;\;\; D =  \left[ \begin{array}{cc}
D_S & 0  \\
0 & D_V \\
\end{array} \right],  $$
where $S_{l \times l}$, $U_{l \times h}$ and $V_{ h \times h}$ denote the sensor-sensor, sensor-anchor, respectively anchor-anchor measurements, and $D$ is a diagonal matrix with $D_{ii} = \sum_{j=1}^{n} |Z_{ij}|$. Note that $V$ is a matrix with all nonzero entries, since the (correct)  measurement between any two anchors is readily available, and the vector $(Ua)_{l \times 1}$  can be interpreted as the anchor contribution in the estimation of the  sensors.
Since $ \boldsymbol{a}^T (D_V - V) \boldsymbol{a} $ is a (nonnegative) constant,
we are interested in minimizing the integer quadratic form $\boldsymbol{z}^T (D_S - S)\boldsymbol{z} - 2\boldsymbol{z}^TU\boldsymbol{a}$. Unfortunately, the non-convex constraint $\boldsymbol{z} \in \mathbb{Z}_2^l$ renders the problem NP-hard, and thus we introduce the relaxation to a quadratically constrained quadratic program (QCQP) from equation (\ref{maximization_zTz}), whose solution can be shown to be $\boldsymbol{z^*} = ( D_S - S + \lambda I )^{-1} (U \boldsymbol{a})$ \cite{asap3d}.

\noindent\begin{minipage}{.38\linewidth}
\begin{equation}
	\begin{aligned}
	& \underset{\boldsymbol{z} = (z_1, \ldots, z_l)}{\text{minimize}}
	& & \boldsymbol{z}^T (D_S - S)  \boldsymbol{z} - 2 \boldsymbol{z}^T U \boldsymbol{a}  &  \\
	& \text{subject to}
	& &  \boldsymbol{z}^T  \boldsymbol{z} = l
	\end{aligned}
\label{maximization_zTz}
\end{equation}
\end{minipage}%
\begin{minipage}{.6\linewidth}
\begin{equation}
	\begin{aligned}
	& \underset{\bar{\boldsymbol{z}} }{\text{minimize}}
	& & \bar{\boldsymbol{z}}^T D_S^{-1/2} (D_S - S) D_S^{-1/2} \bar{\boldsymbol{z}} - 2 \bar{\boldsymbol{z}}^T D_S^{-1/2} U \boldsymbol{a}  &  \\
	& \text{subject to}
	& &  \bar{\boldsymbol{z}}^T \bar{\boldsymbol{z}} = \Delta.
	\end{aligned}
\label{maximization_zTDz}
\end{equation}
\end{minipage}

In \cite{asap3d}, we also considered  a similar formulation where we replaced the constraint
 $ \boldsymbol{z}^T \boldsymbol{z} = l$ in (\ref{maximization_zTz}) by $\boldsymbol{z}^T D_S \boldsymbol{z} = \Delta$,
where $\Delta = \sum_{i=1}^l d_i $ is the sum of the degrees of all sensor nodes. Note that the change of variable $\bar{\boldsymbol{z}}=D_S^{1/2} \boldsymbol{z}$ yields the optimization problem shown in (\ref{maximization_zTDz}), which is very similar to the one in (\ref{maximization_zTz}).


We have seen in Section \ref{sec:sync} that an alternative approach to solving SYNC($\mathbb{Z}_2$) relies on semidefinite programming. In light of the optimization problem (\ref{SDP_max}), the SDP relaxation of (\ref{maxZ2}) in the presence of anchors is  shown in equation (\ref{SDP_maxAnch}), where the maximization is taken over all semidefinite positive real-valued matrices $\Upsilon \succeq 0$  with
\begin{equation}
\Upsilon_{ij} = \left\{
     \begin{array}{rl}
x_i  x_j^{-1} & \;\; \text{ if } i,j \in \mathcal{S} \\
x_i  a_j^{-1} & \;\; \text{ if } i \in \mathcal{S}, j \in \mathcal{A} \\
a_i  a_j^{-1} & \;\; \text{ if } i,j \in \mathcal{A}. \\
     \end{array}
   \right.
\label{upsilondefAnch}
\end{equation}
Note that $\Upsilon$ has ones on its diagonal $ \Upsilon_{ii} =1, \forall i=1,\ldots,n$, and the anchor information gives another layer of hard constraints.
Since $\Upsilon$ is not necessarily a rank-one matrix, the SDP-based estimator is given by the best rank-one approximation to the submatrix corresponding to the sensor-sensor measurements $\bar{\Upsilon}_{ \{1,\ldots,l\} \times \{1,\ldots,l \} }$, which we compute via an eigendecomposition.

Alternatively, to reduce the number of unknowns in (\ref{SDP_maxAnch}) from $n=l+h$ to $l$, one may consider the relaxation (\ref{SDP_max_XY})
where we relax the non-convex constraint $\Upsilon = \boldsymbol{x} \boldsymbol{x}^T$ (which guarantees that  $\Upsilon$ is indeed a rank-one solution) to $\Upsilon \succeq \boldsymbol{x} \boldsymbol{x}^T$, via Schur's lemma. This last matrix inequality is equivalent \cite{boyd94} to the last constraint in the SDP formulation in (\ref{SDP_max_XY}). As before, we obtain estimators $\hat{z}_1, \ldots,\hat{z}_l$ for the sensors by setting $\hat{z}_i= \text{sign}(x_{i}), \forall i=1,\ldots,l$.

\noindent\begin{minipage}{.48\linewidth}
\begin{equation}
	\begin{aligned}
	& \underset{\Upsilon \in \mathbb{R}^{n \times n}}{\text{maximize}}
	& & Trace(Z \Upsilon) \\
	& \text{subject to}
	& & \Upsilon_{ii} =1, i=1,\ldots,n \\
	& & &  \Upsilon_{ij} = a_i  a_j^{-1}, \;\; \text{ if } i,j \in \mathcal{A} \\
		& & &   \Upsilon \succeq 0
	\end{aligned}
\label{SDP_maxAnch}
\end{equation}
\end{minipage} 
\begin{minipage}{.48\linewidth}
 \begin{equation}
		\begin{aligned}
		& \underset{ \Upsilon \in \mathbb{R}^{l \times l}; \boldsymbol{x} \in \mathbb{R}^{l} }{\text{maximize}}
		& & Trace(S \Upsilon)  + 2 \boldsymbol{x}^T U \boldsymbol{a} \\
		& \text{subject to}
		& & \Upsilon_{ii} =1, \forall i=1,\ldots,l  \\
		& & &  \left[ \begin{array}{cc}
				\Upsilon  & \boldsymbol{x}  \\
				 \boldsymbol{x}^T & 1 \\
				\end{array} \right]  \succeq 0
		\end{aligned}
	\label{SDP_max_XY}
\end{equation}
\end{minipage}

In Figure \ref{fig:comp_sync_anch} we compare the performance of the MPS algorithm to the four algorithms introduced in \cite{asap3d} and summarized above. In the synthetic model we used in our simulations, the graph of available pairwise measurements is an Erd\H{o}s-R\'{e}nyi graph $G(n,\alpha)$ with $n=75$ and $\alpha=0.2$ (i.e., a graph with $n$ nodes, where each edge is present with probability $\alpha$, independent of the other edges). Figure \ref{fig:comp_sync_anch} shows the results of our  numerical experiments when we vary the number of anchors $h=\{5,15,30,50\}$. The set of anchors $A \subset V(G)$, with $|A|=h$, is chosen uniformly at random from the $n$ nodes. As the number of anchors $h$ increases, compared to the number of sensors $s=n-h$, the performance of the five algorithms is essentially the same. Only when the number of anchor nodes is small (for example when $h=5$), the SDP-Y formulation shows superior results, together with SDP-XY and QCQP with constraint $z^T D z = \Delta$, while the QCQP with constraint $\boldsymbol{z}^T \boldsymbol{z}=s$ and the message passing algorithm
perform less well. In practice, one would choose the QCQP formulation with constraint $\boldsymbol{z}^T D \boldsymbol{z} = \Delta$ or the message passing algorithm since the SDP-based methods are computationally expensive as the size of the problem increases.

\bibliographystyle{siam}
\bibliography{main_bib_sync}




\end{document}